\pdfoutput=1
\documentclass[11pt]{article}
\usepackage[utf8]{inputenc}
\usepackage[T1]{fontenc}  
\usepackage{natbib}
\usepackage[usenames,dvipsnames,svgnames,table]{xcolor}
\usepackage{graphicx}
\usepackage[font={small}]{caption}
\usepackage[a4paper,left=40mm,right=35mm,top=30mm,bottom=25mm]{geometry}
\usepackage[hidelinks]{hyperref}
\usepackage{algorithm} 
\usepackage{algpseudocode}
\usepackage{authblk}
\usepackage{amssymb}
\usepackage{mathtools}
\usepackage{siunitx}            % must be loaded pre-cleveref
\usepackage[capitalise]{cleveref}
\usepackage[section]{placeins}
\usepackage{stackengine}
\DeclareSIUnit\yr{a}

\newcommand{\mathtiny}[1]{\textnormal{\tiny #1}}
\newcommand{\fenics}{FEniCS}
\def\CC{{C\nolinebreak[4]\hspace{-.05em}\raisebox{.4ex}{\tiny\bf ++}}}

\begin{document}\pagestyle{plain}
% TITLE
\title{Galerkin Least-Squares Stabilization in Ice Sheet Modeling - Accuracy, Robustness, and Comparison to other Techniques}
\date{\vspace{-5ex}}
\author[1]{Christian Helanow}
\affil[1]{Stockholm University, Department of Physical Geography, SE-106 91 Stockholm, Sweden}
\author[2]{Josefin Ahlkrona}
\affil[2]{Uppsala University, Department of Information Technology. Current affiliation: University of Kiel, the  Mathematical Seminar, Westring 383, D-24118 Kiel, Germany}
\maketitle
%------------------------------------------

\begin{abstract}
\noindent
\small{%
We investigate the accuracy and robustness of one of the most common methods used in glaciology for the discretization of the $\mathfrak{p}$-Stokes equations: equal order finite elements with Galerkin Least-Squares (GLS) stabilization. Furthermore we compare the results to other stabilized methods. We find that the vertical velocity component is more sensitive to the choice of GLS stabilization parameter than horizontal velocity. Additionally, the accuracy of the vertical velocity component is especially important since errors in this component can cause ice surface instabilities and propagate into future ice volume predictions. If the element cell size is set to the minimum edge length and the stabilization parameter is allowed to vary non-linearly with viscosity, the GLS stabilization parameter found in literature is a good choice on simple domains. However, near ice margins the standard parameter choice may result in significant oscillations in the vertical component of the surface velocity. For these cases, other stabilization techniques, such as the interior penalty method, result in better accuracy and are less sensitive to the choice of the stabilization parameter. During this work we also discovered that the manufactured solutions often used to evaluate errors in glaciology are not reliable due to high artificial surface forces at singularities. We perform our numerical experiments in both \fenics{} and Elmer/Ice.%
}
\end{abstract}

%% main text
\section{Introduction}
Ice sheets and glaciers are important components of the climate system. Their melting is one of the main sources of sea level rise \citep{IPCC5} and in the past, major climatic events have been triggered by the dynamics of ice sheets \citep{Heinrich1988, AlleyMacAyeal1994}. Numerical modeling is a  key tool to understand both past and future evolution of ice sheets.

The dynamics of ice sheets can be described as a very viscous, incompressible, non-New\-to\-ni\-an free surface flow, driven by gravity. The velocity field and pressure are given by the solution of a non-linear (steady-state) Stokes system - the $\mathfrak{p}$-Stokes system. The position of the ice-atmosphere interface is computed by solving an additional convection equation where the velocity field enters as coefficients. 

Early ice sheet models used simple approximations of the governing equations discretized by the finite difference method \citep{Huybrechts1990,Greve1997,CalovMarsiat1998}. In these models, some stress components are neglected, so that they are inaccurate in e.g. the dynamical coastal regions but have the benefit of being computationally cheap. Since then, the complexity of the mathematical models as well as the numerical methods has increased dramatically. Today, many of the state of the art ice sheet models use the finite element method (FEM) to discretize the full, non-linear, $\mathfrak{p}$-Stokes equations \citep{Zhang2011,Larour2012,Petra2012,BrinkerhoffJohnson2013,Gagliardini2013}. Due to their computational expense, these models are limited to a few centuries if applied to the Greenland Ice Sheet, and an even shorter period if applied to the Antarctic Ice Sheet. For longer simulations on ice sheet scale, approximations are still needed.

Typically, standard FEM methods developed for other types of (usually Newtonian) fluids are imported into the field of computational ice dynamic, even though ice offers many specific characteristics such as a non-linear rheology and shallow domains. The accuracy, robustness and stability of the discretizations used in ice simulations are not well established. Furthermore, it is not clear how numerical errors resulting from solving the discretized $\mathfrak{p}$-Stokes system interplay with the simulations of the ice surface evolution and ultimately predictions of future ice volume. 

In this study we perform numerical experiments to assess the numerical errors, convergence, and robustness of one of the most common discretizations used for the $\mathfrak{p}$-Stokes system in ice sheet modeling: equal order linear elements with a Galerkin Least-Squares (GLS) stabilization \citep[e.g.][]{FrancaFrey1992}. We also investigate how the errors influence the accuracy of ice surface position calculations and how they depend on the GLS stabilization parameter. Finally, we test and compare with alternative equal order discretizations, namely a Pressure Penalty method, an Interior Penalty method \citep{BurmanHansbo2006}, a Pressure Global Projection Method \citep{CodinaBlasco1997} and the Local Projection Stabilization \citep{BeckerBraak2001}, some of which overcome issues found with the GLS stabilized method. The experiments are performed with two different finite element codes that have been used with success in glaciology: \fenics{} \citep{fenics:book, Alnaes2015}, which is the underlying framework of the ice sheet model VarGlaS \citep{BrinkerhoffJohnson2013}, and Elmer/Ice \citep{Gagliardini2013}, which is built on the Elmer software \citep{manual:Elmer}. \fenics{} uses tetrahedra (triangles in 2D), while Elmer/Ice commonly uses prismatic or quadrilateral elements (triangles or rectangles in 2D). Implementing the same problems in two different codes has helped to demonstrate problems related to the underlying equations and avoid any issues related to specific implementations. Once we established that the two codes behave similarly we make further experiments in \fenics{}. We also measure accuracy by two different measures, namely by using manufactured solutions \citep{Leng2013} and a reference solution. This is because we have observed some issues with the usage of the manufactured solutions, while a numerical reference solution is feasible only in 2D due to computational cost. By using both methods for some of our experiments we can differentiate between issues related to the manufactured problem and errors arising from the numerical method. It also allows us to discuss the usefulness of the manufactured solutions, which have become popular as a method of model validation in glaciology. 

The paper is structured as follows. The governing equations and the standard equal order GLS stabilized finite element method are described in Section 2, together with the FEM codes used. In Section 3 the accuracy and convergence of the equal order GLS stabilization is measured using both \fenics{} and Elmer/Ice using standard parameter settings. This is done both with manufactured and reference solutions for the ISMIP-HOM set-up. In Section 4, we study how the numerical errors measured in Section 3 propagate to the evolution of the free ice surface. The sensitivity of the accuracy to the to the GLS stabilization parameter is measured in Section 5, while Section 6 treats how the stability parameter depends on viscosity and finite element cell size. Finally, alternative stabilization techniques are tested in Section 7, with regards to accuracy and parameter sensitivity. Our main conclusions are presented in Section 8.

\section{Theory and Background}

\subsection{Mathematical Model}
\subsubsection{Governing Equations} 
\begin{figure}[th!]
\centering
\footnotesize{
  \def\svgwidth{1.0\textwidth}
  \input{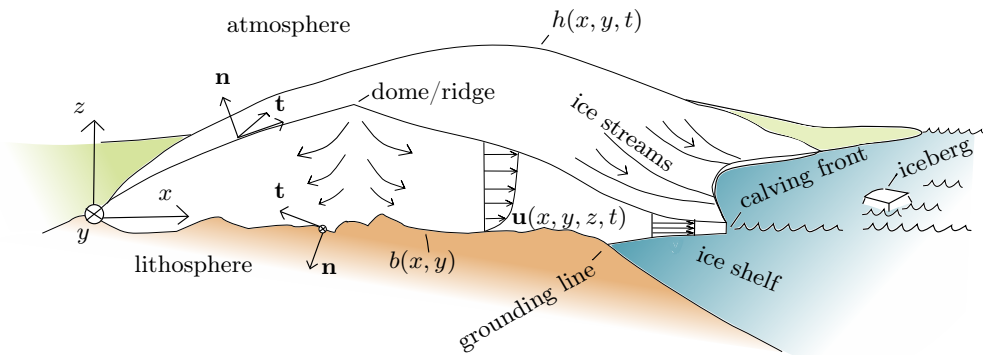}}
   \caption{A body of ice in a Cartesian coordinate system. The ice surface $s=s(x,y,t)$ evolves in time, while the base $b=b(x,y)$ is considered rigid. The surface normal vector, $\mathbf{n}$ is pointing outwards from the ice. In two dimensions there is one tangential vector $\mathbf{t}$, and in three dimensions two orthogonal vectors $\mathbf{t}_1$ and $\mathbf{t}_2$ span a tangential plane. Black arrows indicates velocity patterns, $\mathbf{u}(x, y, z, t)$.} 
   \label{fig:coords}
\end{figure}

We consider an ice sheet or glacier domain $\Omega$ in a Cartesian coordinate system $(x,y,z)$, see \cref{fig:coords}. Note that the sketch in \cref{fig:coords} is exaggerated in the vertical direction, and that in reality, ice sheets are very thin.
The ice velocity $\mathbf{u}=(u_x,u_y,u_z)$ and pressure $p$ are given by the solution to the $\mathfrak{p}$-Stokes equations
\begin{subequations}
\label{eq:vFS} 
\begin{alignat}{2}
\label{eq:momentum}
-\nabla p +\nabla \cdot \mathbf{S}  + \rho \mathbf{g} &= \mathbf{0},\\
\label{eq:mass} \nabla\cdot \mathbf{u} &= 0, 
\end{alignat}
\end{subequations}
where  $\rho$ is the ice density, $\rho\mathbf{g}$ the gravitational force, and $\mathbf{S}$ is the deviatoric stress tensor. The deviatoric stress tensor is related to the strain rate tensor $\mathbf{D}=\frac{1}{2}(\nabla \mathbf{u} + \nabla \mathbf{u}^T)$ through a power-law, 
\begin{subequations}
\label{eq:constitutive}
\begin{alignat}{2}
\label{eq:Glen} \mathbf{S(u)}&=2\eta(|\mathbf{D}|)\mathbf{D},\\
\label{eq:power} \eta &=\eta_0 |\mathbf{D}|^{\mathfrak{p}-2},
\end{alignat}
\end{subequations}
where $|\mathbf{D}|^2 = \mathbf{D}:\mathbf{D}$ is the second invariant of the strain rate tensor.
The power-law parameter $\mathfrak{p}$ indicates the non-linearity of the material and $\eta_0$ is a large temperature-dependent parameter. In this study, we consider the iso\-thermal case to focus on the effects resulting from the discretization of \cref{eq:vFS}. In glaciology, \cref{eq:power} is called Glen's flow law and $\mathfrak{p} = 1/n + 1$, where $n$ is the so-called Glen's parameter. We will apply the standard value $n=3$, so that $\mathfrak{p}=4/3$. Since $\mathfrak{p}<2$ ice is a shear thinning fluid and has infinite viscosity as the strain rate tensor approaches zero. 

Inserting \cref{eq:constitutive} into \cref{eq:momentum} yields
\begin{equation}
\label{eq:stokes} 
-\nabla p +\nabla\cdot\left(\eta(|\mathbf{D}|) (\nabla \mathbf{u} + (\nabla \mathbf{u})^T) \right)  + \rho \mathbf{g} = \mathbf{0}.
\end{equation}
The ice surface, $s$, deforms according to the velocity field, $\mathbf{u}|_{z=s}$, and its evolution given by free surface equation
\begin{equation}
\label{eq:freesurface}
{\partial_t s}+u_x {\partial_x s}+u_y {\partial_y s}-u_z=a_s\quad \text{for } z=s.
\end{equation}
The net accumulation of the ice sheet, $a_s$, depends on precipitation and surface air temperature. The free surface problem \cref{eq:freesurface} is solved only on the ice surface. 

%&
\subsubsection{Boundary conditions}
The boundary of the glacier domain, $\partial \Omega$, consists of the glacier bed, the surface, and the horizontal boundaries, $\partial \Omega =\Gamma_b\cup\Gamma_s\cup\Gamma_h$.  At the ice surface, $\Gamma_s$, and horizontal boundaries, $\Gamma_h$, wind stresses and atmospheric pressure are neglected so that the stress free condition holds.

At the ice-bedrock interface, $\Gamma_b$, the conditions depend on e.g. thermal and hydrological conditions. We consider frozen and grounded ice sheets, so that a no-slip condition applies. Together we have that 

\begin{subequations}
\label{eq:bcs}
\begin{alignat}{2}
\label{eq:bc_surface}
\eta(|\mathbf{D}|) (\nabla \mathbf{u} + (\nabla \mathbf{u})^T)\cdot \mathbf{n} - p\mathbf{n} &= \mathbf{0}, \quad \text{on } \Gamma_s \cup \Gamma_h, \\
\mathbf{u} &= \mathbf{0}, \quad \text{on } \Gamma_b.
\label{eq:bc_bed}
\end{alignat}
\end{subequations}

\subsection{Discretization Methods}

\subsubsection{Overall Solution Strategy}
The standard approach to solve for the velocity field, pressure and the evolution of the ice surface position is outlined in \cref{alg:AlgMovSurf}. The Picard iteration in Algorithm~1 is sometimes exchanged for a less robust but faster Newton method after a few iterations. 

\begin{algorithm}[h!]
\caption{General Solution Procedure. For each time step $k$, and non-linear iteration $n$, a linear system is solved. }
\label{alg:AlgMovSurf}
\begin{algorithmic}[1]
\State Set initial condition for velocity $\mathbf{u}^0_0$, pressure $p^0_0$, and ice surface $s^0$.
\For {each time step $k$}
\While {change > tol}
\State Compute viscosity $\eta_n^k=\eta(\mathbf{u}_{n}^k)$
\State Assemble Stokes model using $\eta_n^k$
\State Solve for velocity $\mathbf{u}_{n+1}^k$ and pressure $p_{n+1}^k$.
\State change = function of $\mathbf{u}^k_{n+1}-\mathbf{u}^k_{n}$
\State $n=n+1$
\EndWhile
\State Insert $\mathbf{u}_n^k$ into \cref{eq:freesurface}
\State Update the computational mesh according to the new $s$
\State $k=k+1$
\EndFor
\end{algorithmic}
\end{algorithm}

For zero strain rates, the linear system assembled in \cref{alg:AlgMovSurf} is ill-conditioned, and in case of a Newton iteration, the Jacobian matrix does not exist  \citep{Hirn2011}. The standard method to avoid this is to introduce a small parameter, the \textit{critical shear rate}, to the constitutive law \cref{eq:power}, so that infinite viscosity is avoided at zero strain.
 
\subsubsection{Finite Element Formulation}
\label{sec:EqualOrderStokes}
The weak formulation of the $\mathfrak{p}$-Stokes system in  \cref{eq:vFS} reads:\\
\noindent
\textit{Find $(\mathbf{u},p)  \in \mathbf{V} \times Q$ such that}
\begin{subequations}
\label{eq:weakform} 
\begin{alignat}{2}
A(\mathbf{u}, \mathbf{v}) + B(\mathbf{v}, p) &= F(\mathbf{v}) \quad \forall \mathbf{v} \in V,\\
B(\mathbf{u}, q) &= 0 \quad \forall q \in Q,
\end{alignat}
\end{subequations}
with
\begin{subequations}
\label{eq:weakform2} 
\begin{alignat}{4}
A(\mathbf{u}, \mathbf{v}) &= \int_\Omega \mathbf{S(u)}: \nabla \mathbf{v} d\Omega \\& \nonumber = \int_\Omega (\eta(|\mathbf{D}|) (\nabla \mathbf{u}) + \nabla (\mathbf{u})^T) : \nabla \mathbf{v} d\Omega, \\
B(\mathbf{v},p) &= \int_\Omega - p \nabla \cdot \mathbf{v} d\Omega, \\ 
F(\mathbf{v}) &= \int_\Omega \rho\mathbf{g} \cdot \mathbf{v} d\Omega,
\end{alignat}
\end{subequations}
where $\mathbf{V}:= \left\{\mathbf{v}\in [W^{1,\mathfrak{p}}(\Omega)]^d : \mathbf{v}|_{\Gamma_b} = \mathbf{0}  \right\}$ and $Q:= L^{\mathfrak{p}/(\mathfrak{p}-1)}(\Omega)$, $\mathbf{V}\times Q$ being either $x-y$ periodic or stress-free on $\Gamma_h$.  %We assume that the domain boundary is sufficiently smooth..
The associated discretized problem is:\\
\noindent
\textit{Find $(\mathbf{u}_h,p_h) \in \mathbf{V}_h \times Q_h$ such that}
\begin{subequations}
\begin{alignat}{2}
A(\mathbf{u}_h, \mathbf{v}_h)+B(\mathbf{v}_h,p_h)&= F(\mathbf{v}_h)  \quad \forall \mathbf{v}_h  \in \mathbf{V}_h \label{eq:discrete_momentum}\\
B(\mathbf{u}_h,q_h)&=0 \quad \forall q_h  \in  Q_h,\label{eq:discrete_mass}
\end{alignat}
\label{eq:discrete}%
\end{subequations}
or equivalently,\\
\noindent
\textit{Find $(\mathbf{u}_h,p_h) \in \mathbf{V}_h \times Q_h$ such that}
\begin{equation}
A(\mathbf{u}_h, \mathbf{v}_h)+B(\mathbf{v}_h,p_h)+B(\mathbf{u}_h,q_h) = F(\mathbf{v}_h)  \quad \forall (\mathbf{v}_h,q_h )  \in \mathbf{V}_h \times  Q_h,
\label{eq:discrete_compact}
\end{equation}
where $\mathbf{V}_h$ and $Q_h$ are finite element space restrictions of $\mathbf{V}$ and $Q$ respectively, defined on a partition $T_h$ of $\Omega$ with mesh size $h$.  These spaces have to fulfill the \emph{inf-sup} condition \citep{Babuska1973,Brezzi1974,Brezzi1991,Belenki2012}. If the inf-sup condition is violated, spurious pressure oscillations or locking of the velocity field may occur. Examples of choices that lead to stable discretizations are the so-called Taylor-Hood element \citep{taylorhood}, which consists of piece-wise quadratic polynomials for the velocity and piece-wise linear for the pressure, or the MINI element (residual free bubbles method), which enriches each component of the velocity space with an extra degree of freedom \citep{Arnold1984,Baiocchi1993}. Many ice sheet models tend to apply linear polynomial functions for both the velocity and pressure space \citep{Zwinger2007,Zhang2011,Tezaur2015} ($P1P1$ for tetrahedrons/ triangles or $Q1Q1$ for prisms/ quadrilaterals). Equal order elements are preferred due to their simple implementation and reduced computational cost, but unfortunately, these do not fulfill the inf-sup condition. To circumvent this issue, stabilization techniques are used. 

One of the most common stabilization techniques is the GLS method \citep{Hughes1986,FrancaFrey1992}, in which extra terms $S_{\mathtiny{GLS}}$ and $F_{\mathtiny{GLS}}$ are added to the discrete form as
% \textit{Find $(\mathbf{u}_h,p_h) \in \mathbf{V}_h \times Q_h$ such that}
%
\begin{equation}
\begin{aligned}
A(\mathbf{u}_h, \mathbf{v}_h) + &B(\mathbf{v}_h,p_h) \\
+ &B(\mathbf{u}_h,q_h)  +S_{\mathtiny{GLS}}((\mathbf{u}_h, p_h), (\mathbf{v}_h, q_h)) = \\ 
&F(\mathbf{v}_h) + F_{\mathtiny{GLS}}((\mathbf{v}_h, q_h)), \quad \forall (\mathbf{v}_h,q_h)  \in \mathbf{V}_h \times Q_h.
\label{eq:GLS}
\end{aligned}
\end{equation}
The GLS method is a consistent method in the sense that the extra terms vanish for the true strong solution. The stabilization terms $S_{\mathtiny{GLS}}$ and $F_{\mathtiny{GLS}}$ are defined as
\begin{subequations}
  \begin{alignat}{2}
    &S_{\mathtiny{GLS}}((\mathbf{u}_h, p_h), (\mathbf{v}_h, q_h)) = \nonumber\\
    &- \sum_{K\in T_h}\tau_{\mathtiny{GLS}} (-\nabla \cdot \mathbf{S(u)} + \nabla p_h,  -\nabla \cdot \mathbf{S(v)} + \nabla q_h) _{K}, \\
    &F_{\mathtiny{GLS}}((\mathbf{v}_h, q_h)) = - \sum_{K\in T_h}\tau_{\mathtiny{GLS}} (\rho\mathbf{g},
    - \nabla \cdot \mathbf{S(v)} + \nabla q_h) _{K},
  \end{alignat}
\end{subequations}
where $K$ denotes triangles/ quadrilaterals or tetra-/ hexahedrons in the partition $T_h$, and $(\cdot, \cdot)_K$ denotes the $L^2$ inner product in the element $K$. For $P1P1$ elements on triangles/ tetrahedrons second derivatives vanish, leading to $\nabla\cdot\mathbf{S}=\mathbf{0}$. For prismatic or quadrilateral elements a dependency on the gradient of the velocity in \cref{eq:GLS} remains. This contribution is however small unless elements are highly distorted. 

In \cite{FrancaFrey1992} the stabilization parameter is
\begin{equation}\label{eq:francatau}
\tau_{\mathtiny{GLS}} = \tau_0\frac{m_K h_K^2}{8\eta},
\end{equation}
where $m_K=1/3$ for linear interpolations. The size of the stabilization parameter thus depends on how the element measure, i.e. the cell size $h_K$, is defined (this will be discussed in \cref{sec:cell_size}). For non-Newtonian fluids, $\tau_{\mathtiny{GLS}}$ also varies with the viscosity $\eta$. The parameter $\tau_{0}$ is introduced in the present study to investigate the effects of under- and over-stabilization. As it does not occur in the original formulation in \cite{FrancaFrey1992} its default value is $\tau_{0}=1$.

For the \emph{linear} Stokes $\mathcal{O}(h^1)$ convergence in pressure and velocity \textit{gradient} is expected for linear element with GLS stabilization, measured in the $L^2$-norm, implying $\mathcal{O}(h^2)$ convergence in velocity if the solution to the problem is smooth enough \citep{ErnGuermond2004}. To our knowledge there are however no \textit{a-priori} estimates of the discretization error of the GLS method applied to the $\mathfrak{p}$-Stokes equations, but there are (optimal) estimates for equal order elements with local projection stabilization \cite{Hirn2012} and for inf-sup stable elements \cite{Belenki2012}. In \cite{Hirn2012} a $\mathcal{O}(h^1)$ convergence is suggested for the velocity \textit{gradient} measured in a $L^{\mathfrak{p}}$ norm, and in \cite{Hirn2012} and \cite{Belenki2012} a $\mathcal{O}(h^{2(\mathfrak{p}-1)/\mathfrak{p}})$ convergence in pressure, measured in a $L^{\mathfrak{p}/(\mathfrak{p}-1)}$ norm. 

\subsubsection{Implementation and mesh generation in \fenics{}}
\label{sec:fenics}
\fenics{} is a set of fully parallelized software components for solving partial differential equations by the finite element method.
The core is written in \CC{} and also has a Python-interface. Notably, one of the high-level features is the use of a (Python like) form language (UFL, \cite{UFL}) which closely resembles the mathematical language used for weak formulations of partial differential equations. \fenics{} supports a wide variety of finite elements defined on triangles or tetrahedrons.

In this study the $P1P1$ element was used. We used the built-in mesh generation to deform a  structured mesh on a cuboid (in 3D, consisting of tetrahedrons) or a  rectangular (2D, triangles)  to the desired surface and bed geometry. 

The non-linearity was solved by a manually programmed Newton method in which, for each iteration, the linear sub-problem was solved with a direct linear solver from the PETSc \citep{petsc-web-page} package. 

\subsubsection{Implementation and mesh generation in Elmer/Ice}\label{sec:Elmer}
Elmer/Ice is based on the multiphysics code Elmer \citep{Gagliardini2013}. It is written in Fortran 90, and has been used to perform large scale, parallelized ice sheet simulations \citep[e.g.][]{Seddik2012,GilletChaulet2012}.

\begin{figure}
    \centering
    \includegraphics[width=0.9\textwidth]{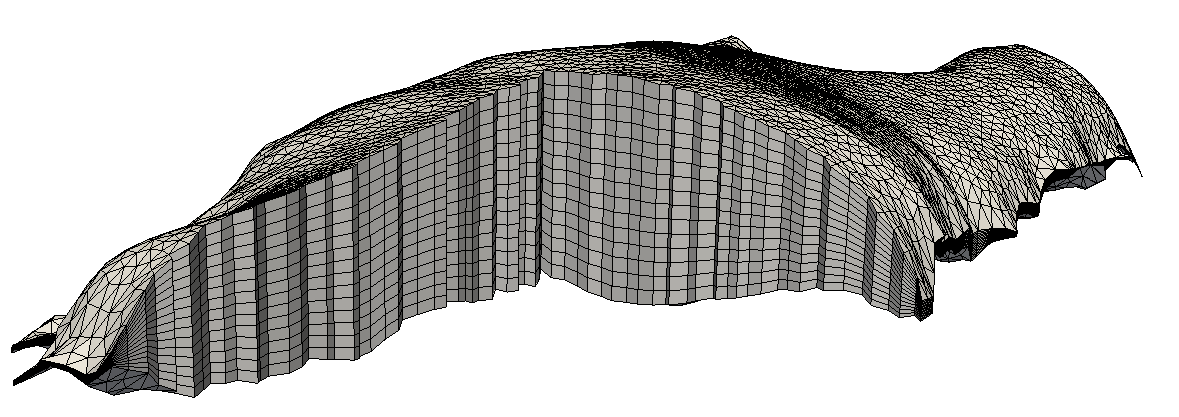} 
    \caption{Extruded mesh for the Greenland Ice Sheet}
    \label{fig:extrudedmesh}.
\end{figure}

So far, equal order linear elements are used for ice sheet simulations with Elmer/Ice. The standard approach to construct a 3D mesh is to first create a 2D footprint mesh consisting of triangles or quadrilaterals and extrude this in the vertical into a 3D mesh consisting of prismatic or hexagonal mesh, see \cref{fig:extrudedmesh}. In this study hexagonal meshes are  used in the Elmer/Ice simulations.
The number of vertical layers in an ice sheet simulation are usually 15--20. For an ice sheet that is a couple of kilometers thick the typical vertical edge size is thus about  \SI{200}{\metre} while the resolution in the horizontal plane can vary from \SI{50}{\kilo\metre} down to \SI{500}{\metre} if static mesh adaptation is used \citep{Seddik2012,GilletChaulet2012}. As the ice surface evolves, the extruded mesh allows for shifting nodes only in the vertical direction, which is computationally efficient.

Unlike \fenics{}, the weak forms and stabilization methods are already implemented, and only parameter choices are easily adjustable. Available stabilization techniques are GLS stabilization (as will be used in this study), or the MINI-element (residual free bubbles method)  \citep{Arnold1984,Baiocchi1993}. The residual free bubbles method is computationally costly due to the extra degrees of freedom.

The non-linearity in \cref{alg:AlgMovSurf} was solved by a Picard iteration, in which for each iteration, the linear sub-problem was solved with a direct linear solver from the UMFPACK package \citep{suitesparse-web-page}.
 
\section{Accuracy, Convergence, and Evaluating Measuring Techniques} \label{sec:convergence}
In this section we assess how the numerical errors or the finite element discretization of the $\mathfrak{p}$-Stokes equations decrease with the cell size, but also how large the errors are and how they are distributed. This is done for the default stabilization parameter ($\tau_0=1)$ on on the standard benchmark set-up ISMIP-HOM A and B  \citep{Pattyn2008}. 

\subsection{Experiment Set-Up}

\begin{figure}
    \centering
\includegraphics[width=0.75\textwidth]{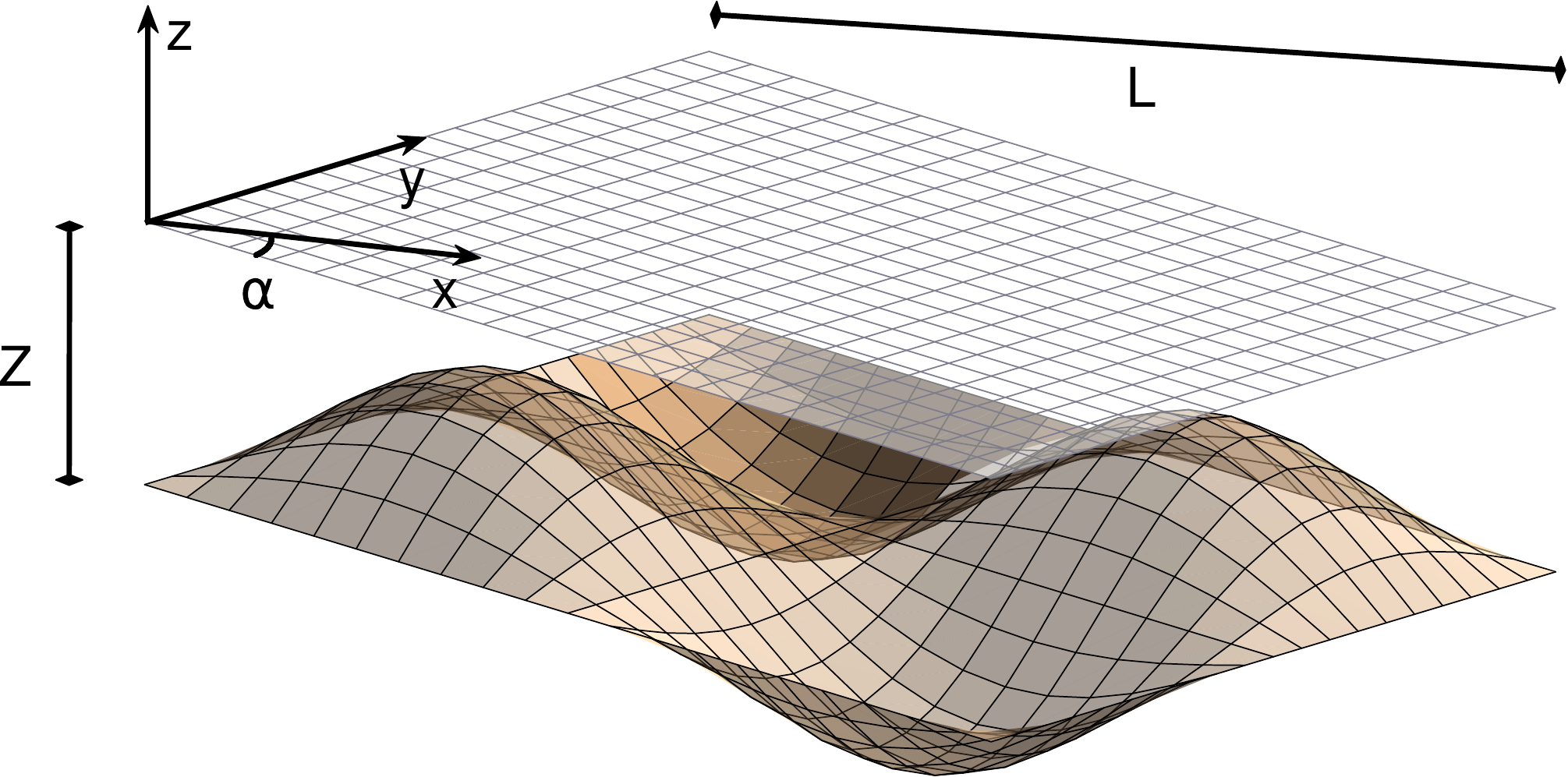} 
\caption{The ISMIP-HOM A domain. The ice is $Z= \SI{1}{\kilo\metre}$ thick on average and is flowing down a plane of inclination $\alpha=0.5^{\circ}$ and length $L=80$ km. The bed has sinusodial bumps of amplitude $Z/2=500$ m.}
\label{fig:ismiphom}
\end{figure}

The ISMIP-HOM A experiment  describes ice flow over an infinite, inclined sinusoidal bed (see \cref{fig:ismiphom}) with no slip conditions at the base. The bed, $b$, and surface, $s$, are
\begin{equation}
\begin{aligned}
  s(x,y) &= -\tan(\alpha) x  \\
  b(x,y) &= s(x,y) - Z + \frac{Z}{2}\left[\sin\left(\frac{2 \pi x}{L}\right) \sin\left(\frac{2 \pi y}{L}\right)\right],
 \label{eq:ISMIP}
\end{aligned}
\end{equation}
where $Z= \SI{1}{\kilo\metre}$ is the typical thickness of the ice, $L = \SI{80}{\kilo\metre}$ is the horizontal ($x$ and $y$) extension of the domain and $\alpha=0.5^\circ$ is the inclination of the surface. For the ISMIP-HOM B experiment the $y$-dependency is excluded. The ISMIP-HOM benchmark is described in more detail in \cite{Pattyn2008}.

Measuring accuracy and convergence is non-trivial on a non-linear problem with singularities. No analytical solution exists for the ISMIP-HOM problem, and obtaining a numerical reference solution for a three dimensional problem is computationally demanding. Therefore, \cite{SargentFastook2010} and \cite{Leng2013} developed a manufactured solution which provides an analytical reference solution to a problem with altered surface boundary conditions and body force. These solutions have been used to measure convergence rate and accuracy for the ISMIP-HOM benchmark experiment A in e.g. \cite{Leng2013}, \cite{Gagliardini2013} and \citet{BrinkerhoffJohnson2013} (slightly altered).  

As the manufactured solutions introduce artificial surface and body forces, there is a risk that the accuracy observed for the manufactured problem does not apply for the original problem. To our knowledge, the quality of the manufactured solutions of \cite{Leng2013} has not been studied. Therefore we also use a numerical reference solution for ISMIP-HOM B experiment, the two dimensional version of the ISMIP-HOM A experiment. This reference solution is obtained by solving the problem on a very fine mesh. It would of course be optimal to have a numerical reference solution for the three-dimensional ISMIP-HOM A experiment, but this was too computationally demanding. To generate the manufactured solution and artificial boundary conditions and forcing a code modified from the supplementary material of \citet{Leng2013} was used in both \fenics{} and Elmer/Ice. 

We employ $20$ different meshes, the reference solution excluded. These meshes include five different resolutions in the horizontal direction, $N_X=10,$ $20, 40,$ $80, 160$ and four different resolutions in the vertical direction, $N_Z=5, 10,$ $20, 40$, where $N_X$ and $N_Z$ denotes the number of elements in respective direction. Throughout this paper the mesh with $N_X=20$ and  $N_Z=20$ will be regarded as the resolution closest to those used in practical ice sheet simulations, i.e. a horizontal resolution of  $4$ km and a vertical resolution of about  $500$ meters. The numerical reference solution is computed for $N_X=1280$ and $N_Z=640$ with \fenics{}, whereas a coarser resolution of $320\times80$ is used in Elmer/Ice. 
The simulations were performed with $h_K$ defined as the \emph{minimum edge length}.

We choose to follow the approach in earlier studies \citep[e.g.][]{Leng2013,Gagliardini2013}, and present the errors in the $L^2$-norm instead of the for the $\mathfrak{p}$-Stokes problem more natural spaces  $L^{\mathfrak{p}/(\mathfrak{p}-1)}$ and $W^{1,\mathfrak{p}}$.
\subsection{Results}
\subsubsection{Convergence}

\begin{figure}[h!]
  \centering
    \stackinset{l}{0.5cm}{b}{-0.1cm}{a)}{%
      \includegraphics[width=0.46\textwidth]{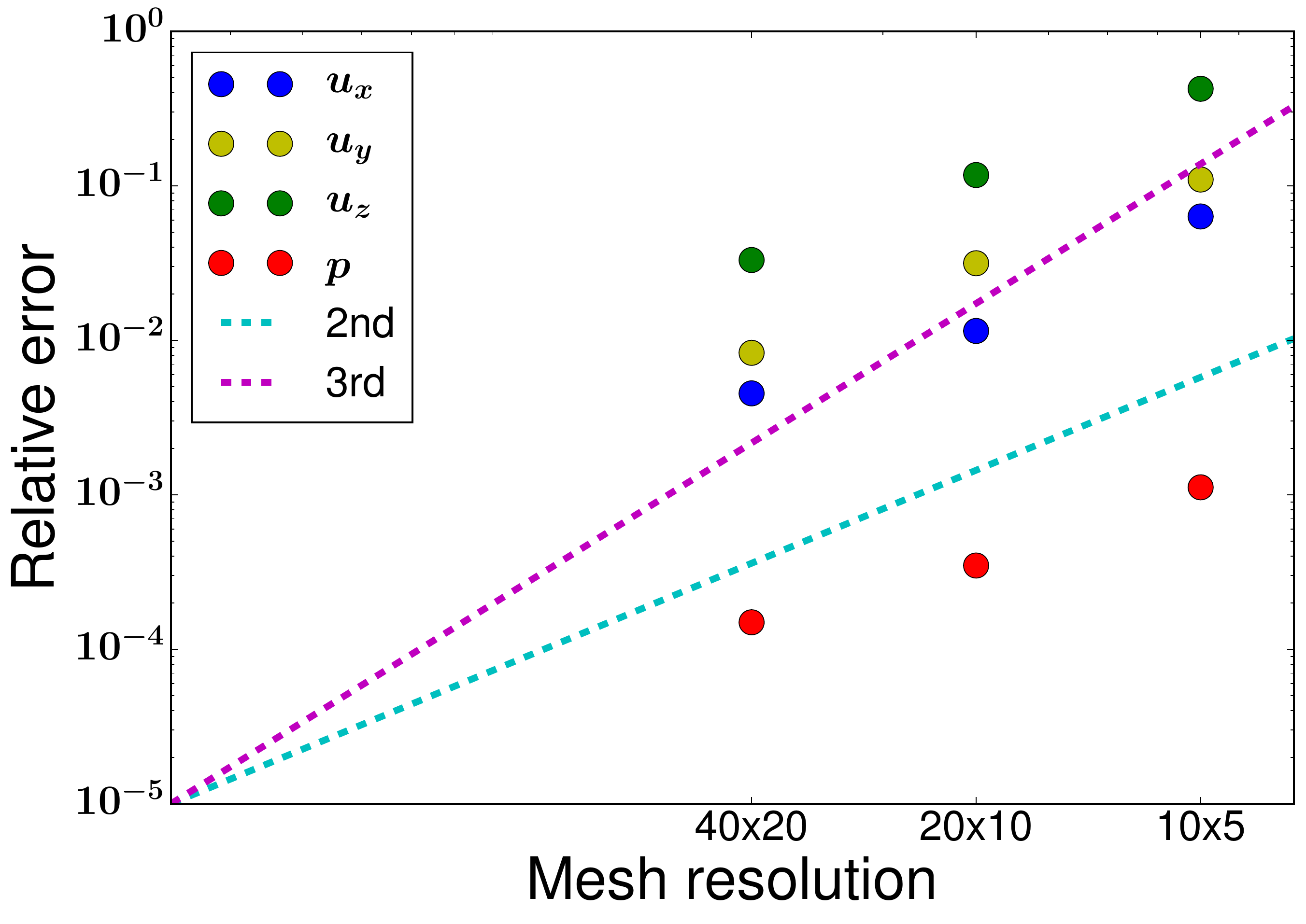}}
    \stackinset{l}{0.5cm}{b}{-0.1cm}{b)}{%
      \includegraphics[width=0.46\textwidth]{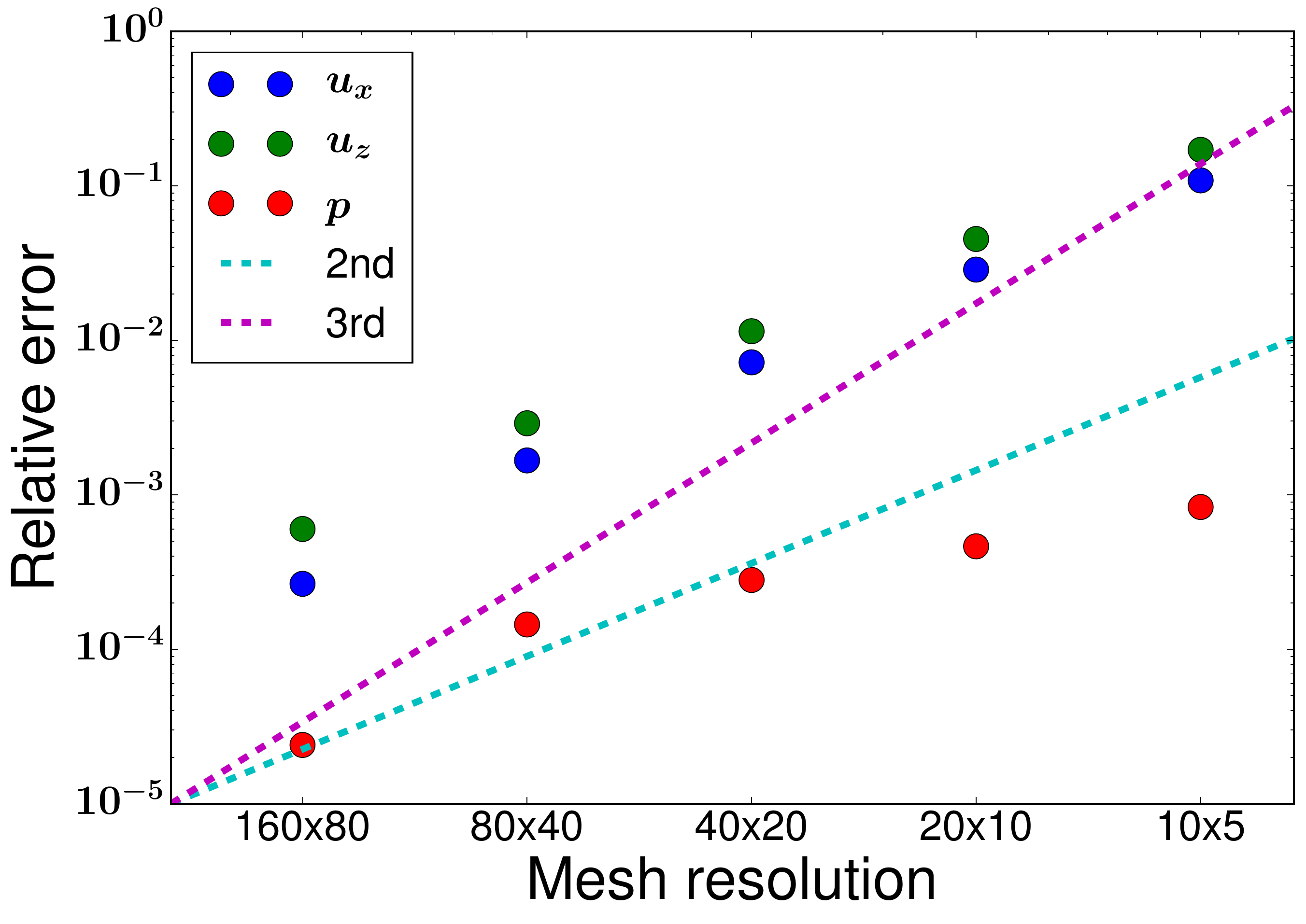}}\\
    \vspace{0.5em}
    \stackinset{l}{0.5cm}{b}{-0.1cm}{c)}{%
      \includegraphics[width=0.46\textwidth]{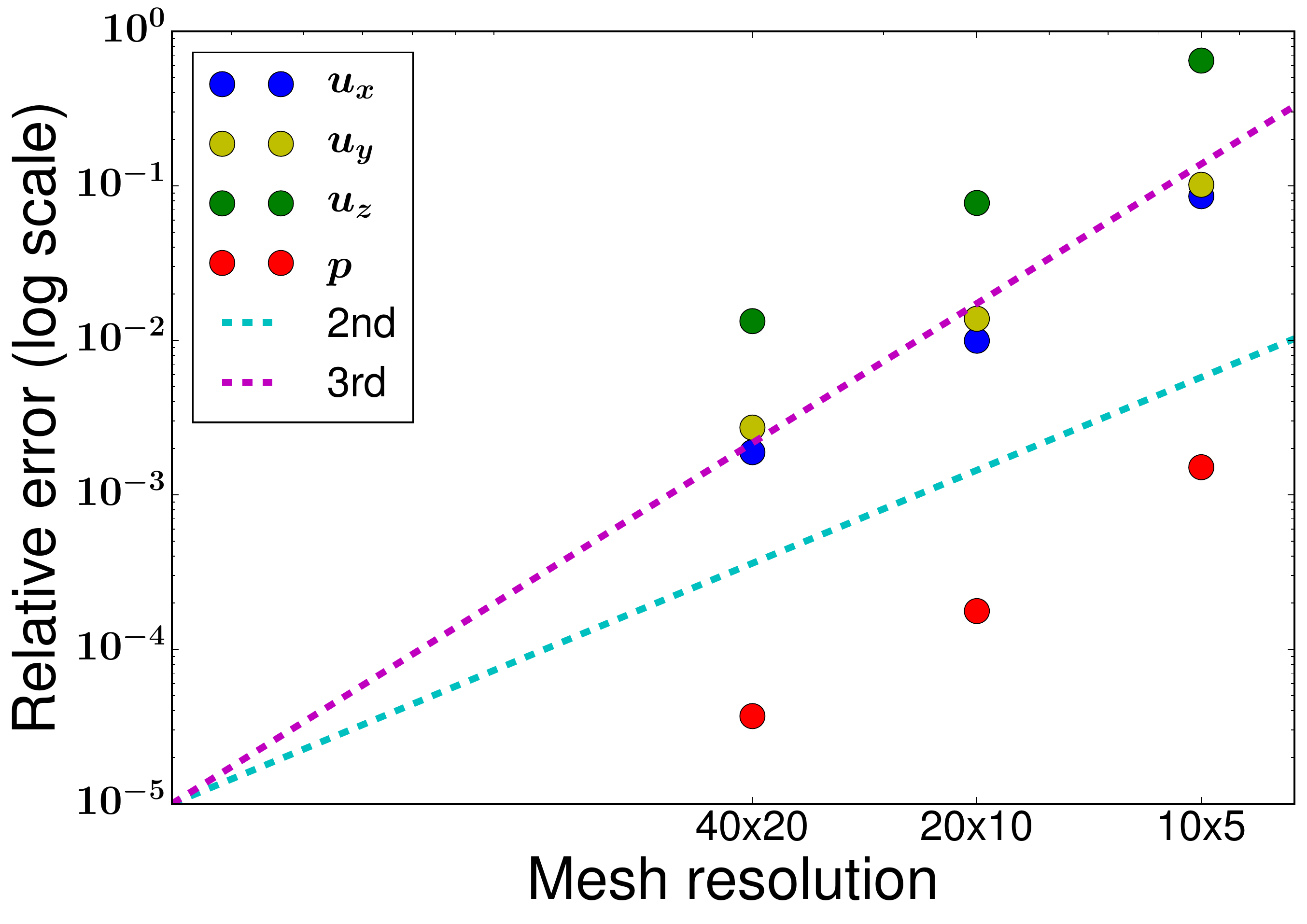}}
    \stackinset{l}{0.5cm}{b}{-0.1cm}{d)}{%
      \includegraphics[width=0.46\textwidth]{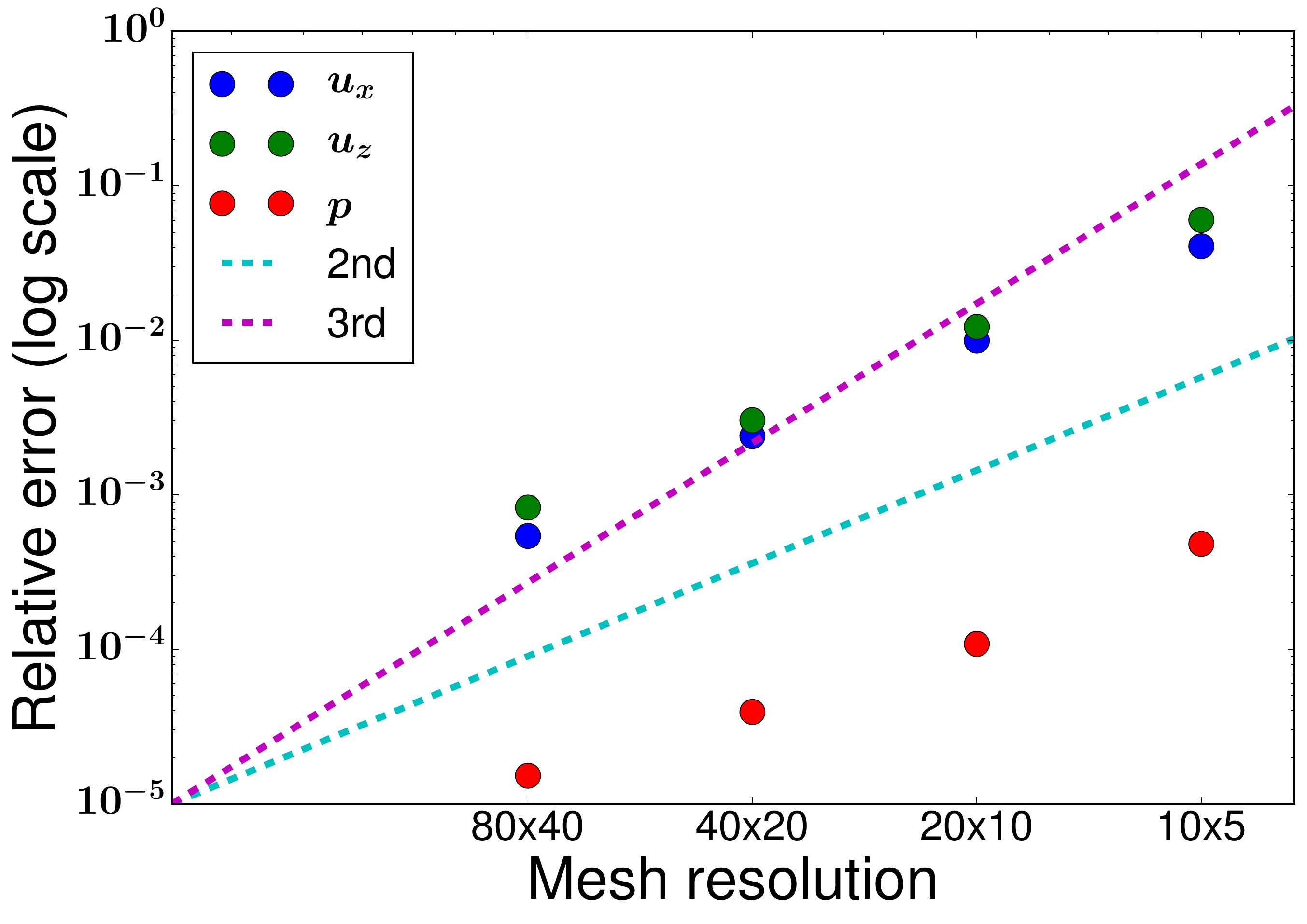}}\\
\caption{Convergence of velocity components and pressure for equal order \mbox{(bi-)linear} elements with standard GLS stabilization, computed by \fenics{} and Elmer/Ice. The accuracy was measured in comparison to both a manufactured solution (panel a,c) and for the original problem using a numerical reference solution (panel b,d). Dashed lines indicate second (cyan) and third (magenta) order convergence rates. The panels show the accuracy compared to a a) manufactured solution using \fenics{}, b) numerical reference solution using \fenics{}, c) manufactured solution using Elmer/Ice, d) numerical reference solutions using Elmer/Ice.}
\label{fig:convergence}
\end{figure}

The relative error in velocity components and pressure for varying grid resolutions is shown in \cref{fig:convergence} for \fenics{} and Elmer/Ice, using both a manufactured solution and a numerical reference solution.

Both \fenics{} and Elmer/Ice show second order convergence in the velocity when compared to a numerical reference solution. In Elmer the convergence rate is also second order for the pressure. This is higher than expected even for the linear Stokes problem. Such super-approximation can occur for smooth enough solutions \citep{ErnGuermond2004} and was also observed for the $\mathfrak{p}$-Stokes problem in \cite{Hirn2011}. When compared to the manufactured solution, the rate of convergence in \fenics{} is two, while in Elmer/Ice it increased to three. The unexpected  third order convergence was obtained also in \cite{Gagliardini2013} for the manufactured problem, not only with linear elements with GLS stabilization, but also with residual free bubbles stabilization and higher order elements. When comparing Elmer/Ice to the manufactured solutions, the convergence rate is two under horizontal refinement, and almost four under vertical refinement. Refining in both the horizontal and vertical direction combines to the third order convergence. The higher convergence order under vertical refinement is seen also in \fenics{}, where it is three instead of Elmer/Ice's four, but the combined horizontal and vertical refinement results in the expected second order convergence rate. For a linear viscosity law, the convergence rate is two even with Elmer/Ice on the manufactured solutions, and the and numerical errors are overall lower.

A probable explanation to the unexpected results is that for the manufactured problem, the artificial surface stresses and body force added are functions of the (calculated) viscosity and may therefore approach infinity. This is of course not viable in the discrete case and an upper limit is set for the viscosity. However, even with this limit, the artificial stresses become very high and the convergence to the solution is poor. To obtain convergence, we follow \citet{Leng2013} and \citet{Gagliardini2013} and set the artificial surface stress to zero at these manufactured singular points. The high artificial stresses can affect the problem to a degree where it neither is representative for ice sheet scenarios nor a problem for which it is possible to get a representative convergence rate. The errors and convergence is sensitive to implementation details and it is likely that the singularity in the manufactured problem is the cause of the unexpected third order convergence of Elmer/Ice. Note that the above discussion does not exclude the possibility that some of these effects may be due to that the manufactured solutions are for the 3D ISMIP-HOM A problem and the numerical reference solution is for the 2D ISMIP-HOM B problem. To be certain, we did also perform the simulations using the manufactured simulations in 2D, with parameters from \citet{SargentFastook2010}, which qualitatively gave the same results. 

\subsubsection{Error Distribution}
\begin{figure}[ht!]
  \centering
    \stackinset{l}{0.5cm}{b}{-0.1cm}{a)}{%
      \includegraphics[width=0.46\textwidth]{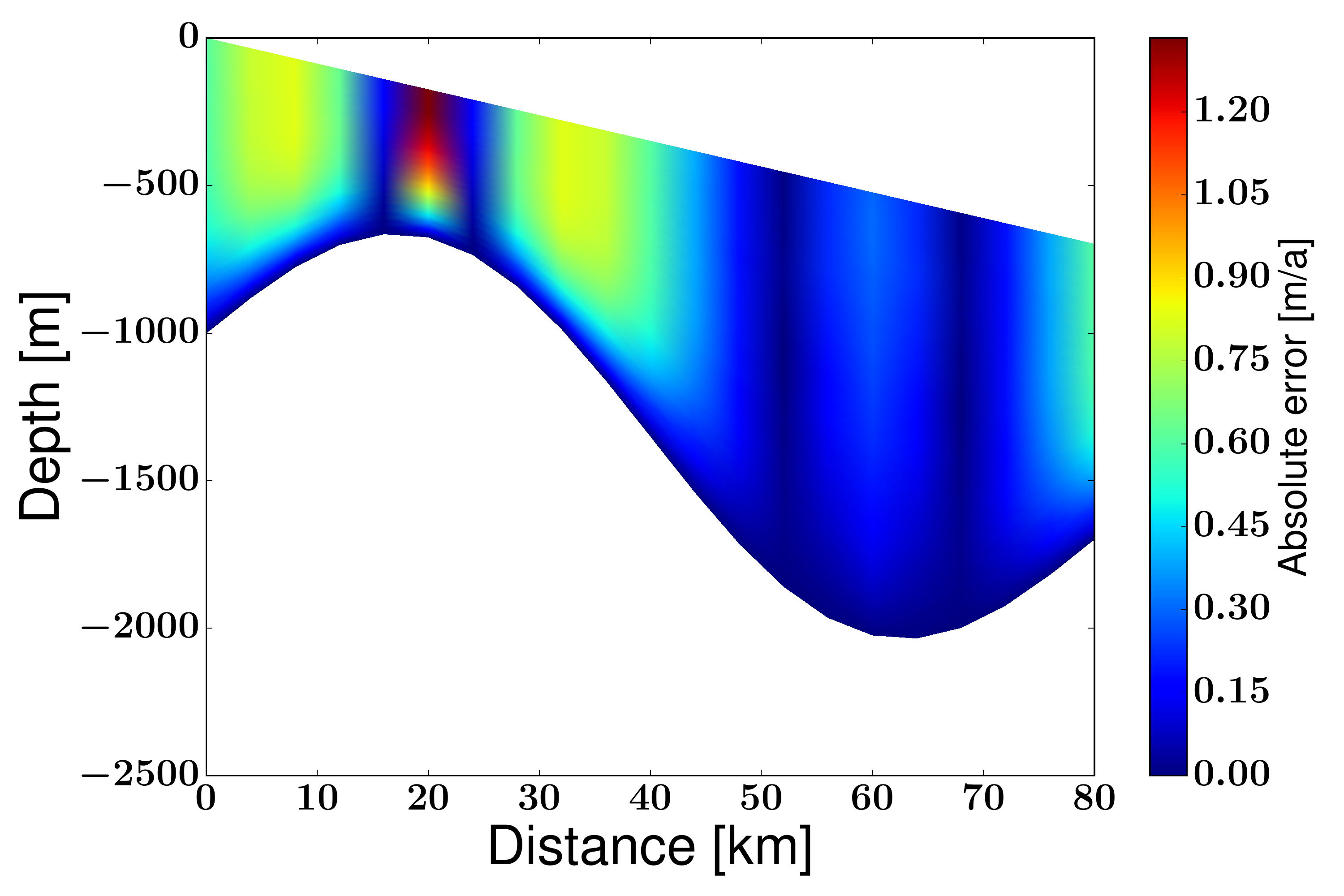}}
    \stackinset{l}{0.5cm}{b}{-0.1cm}{b)}{%
      \includegraphics[width=0.46\textwidth]{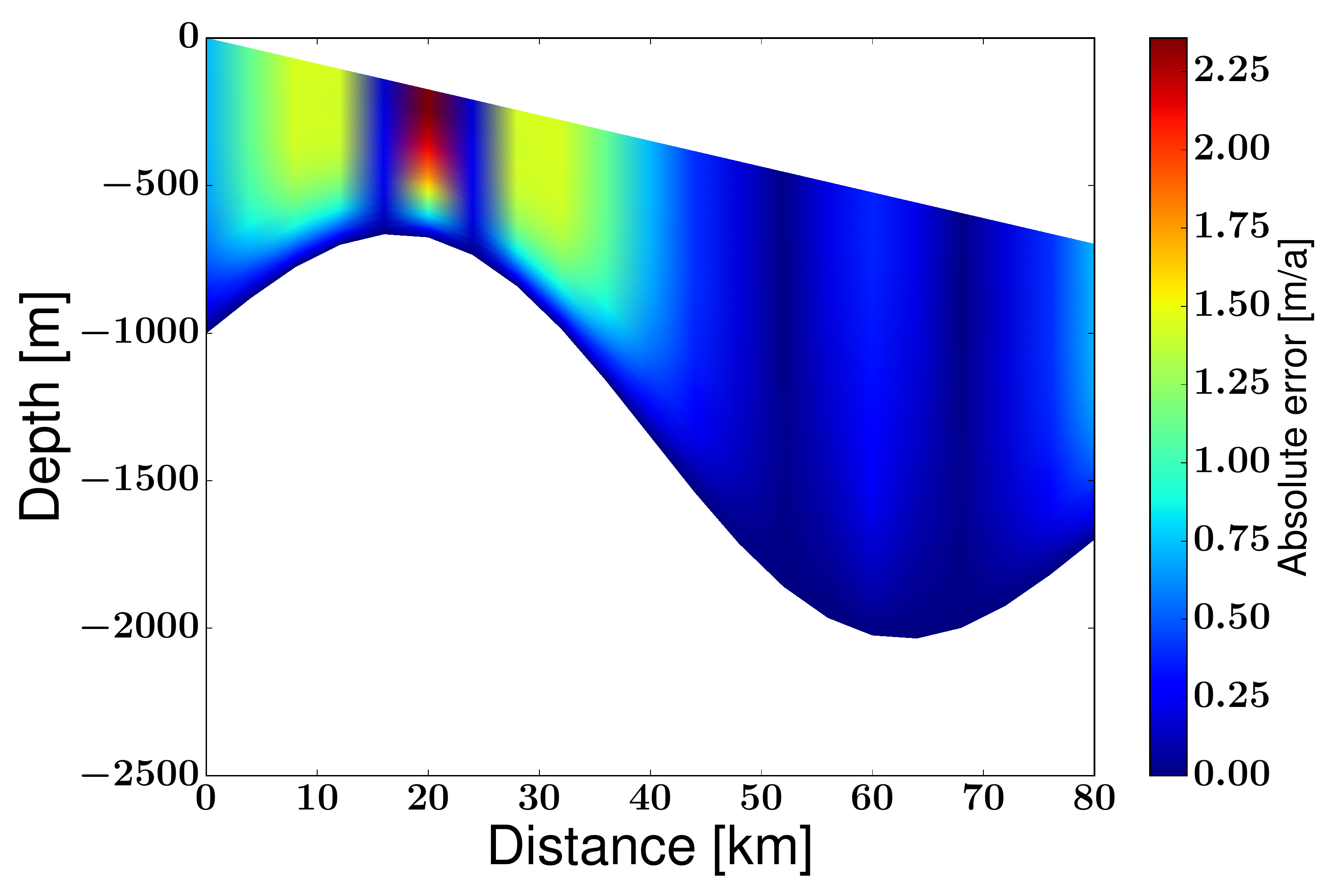}}\\
    \vspace{0.5em}
    \stackinset{l}{0.5cm}{b}{-0.1cm}{c)}{%
      \includegraphics[width=0.46\textwidth]{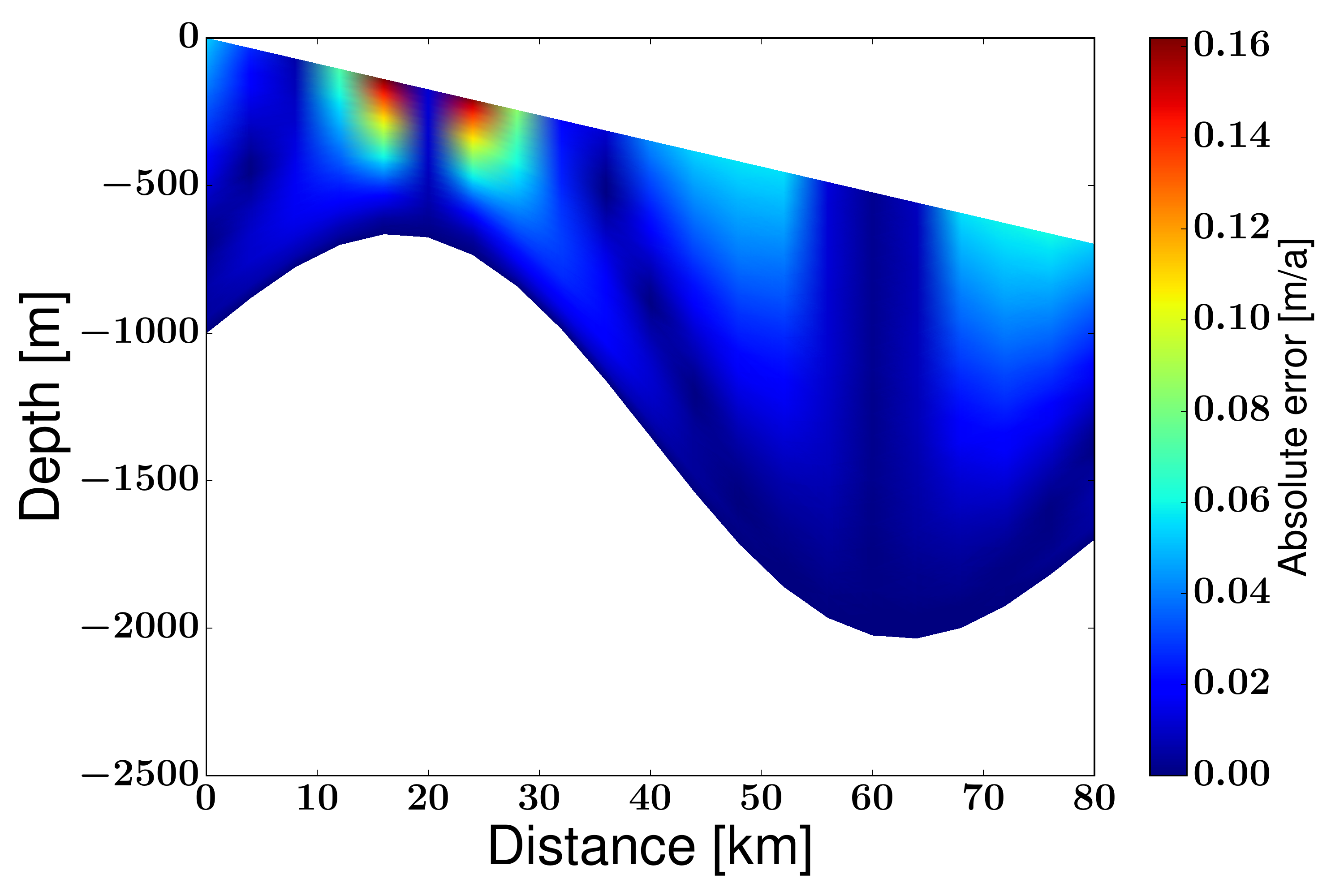}}
    \stackinset{l}{0.5cm}{b}{-0.1cm}{d)}{%
      \includegraphics[width=0.46\textwidth]{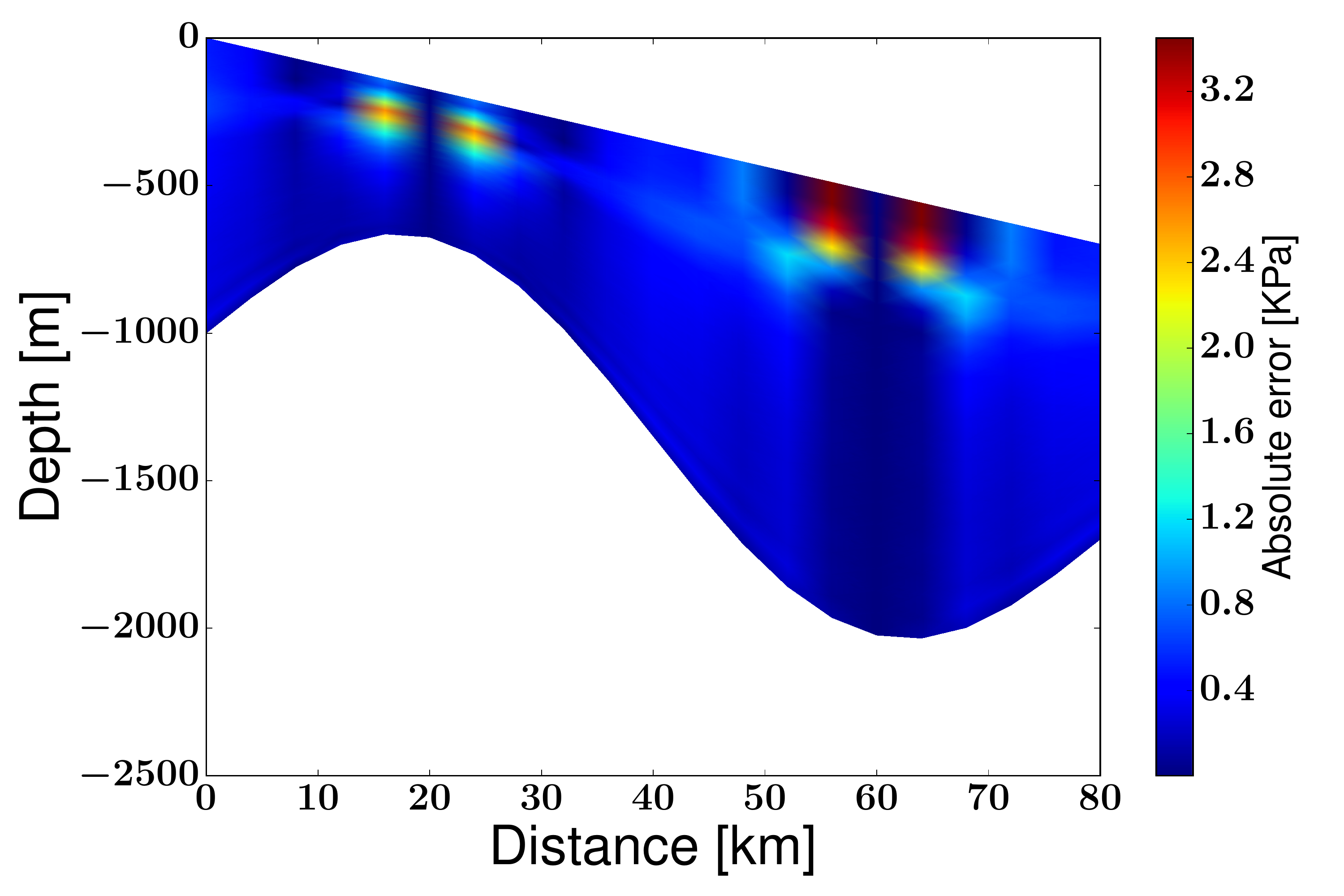}}\\
    \caption{Absolute error of Elmer/Ice in comparison for the manufactured problem on a $20\times 20\times 20$ grid, shown along the cross-section $y=L/4$. The panels show the error of a) horizontal velocity $u_x$, b) horizontal velocity $u_y$, c) vertical velocity $u_z$, d) pressure $p$.}
    \label{fig:manufactureddistribution}
\end{figure}

\begin{figure}[ht!]
  \centering
    \stackinset{l}{0.5cm}{b}{-0.1cm}{a)}{%
      \includegraphics[width=0.46\textwidth]{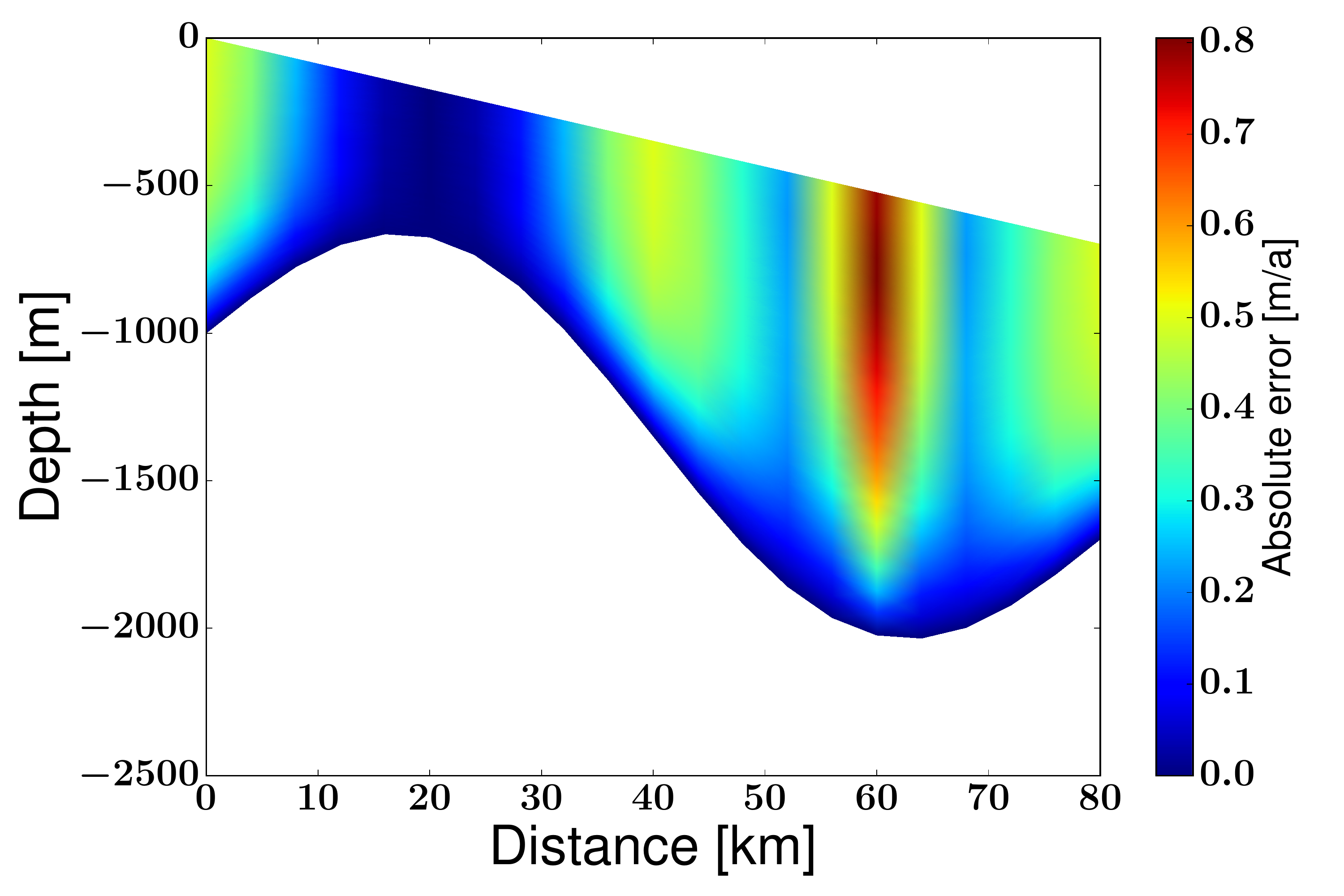}}
    \stackinset{l}{0.5cm}{b}{-0.1cm}{b)}{%
      \includegraphics[width=0.46\textwidth]{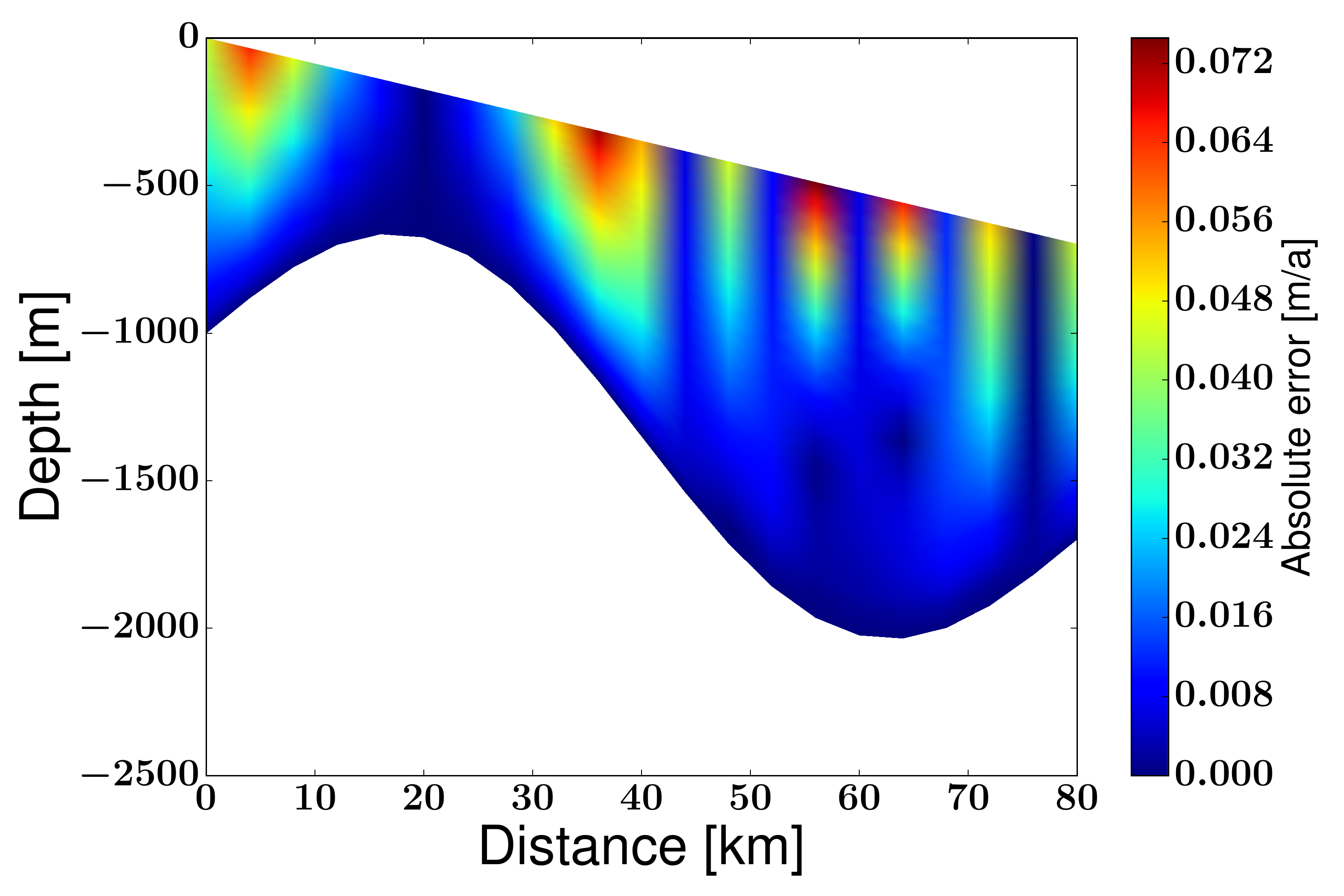}}\\
    \vspace{0.5em}
    \stackinset{l}{0.5cm}{b}{-0.1cm}{c)}{%
      \includegraphics[width=0.46\textwidth]{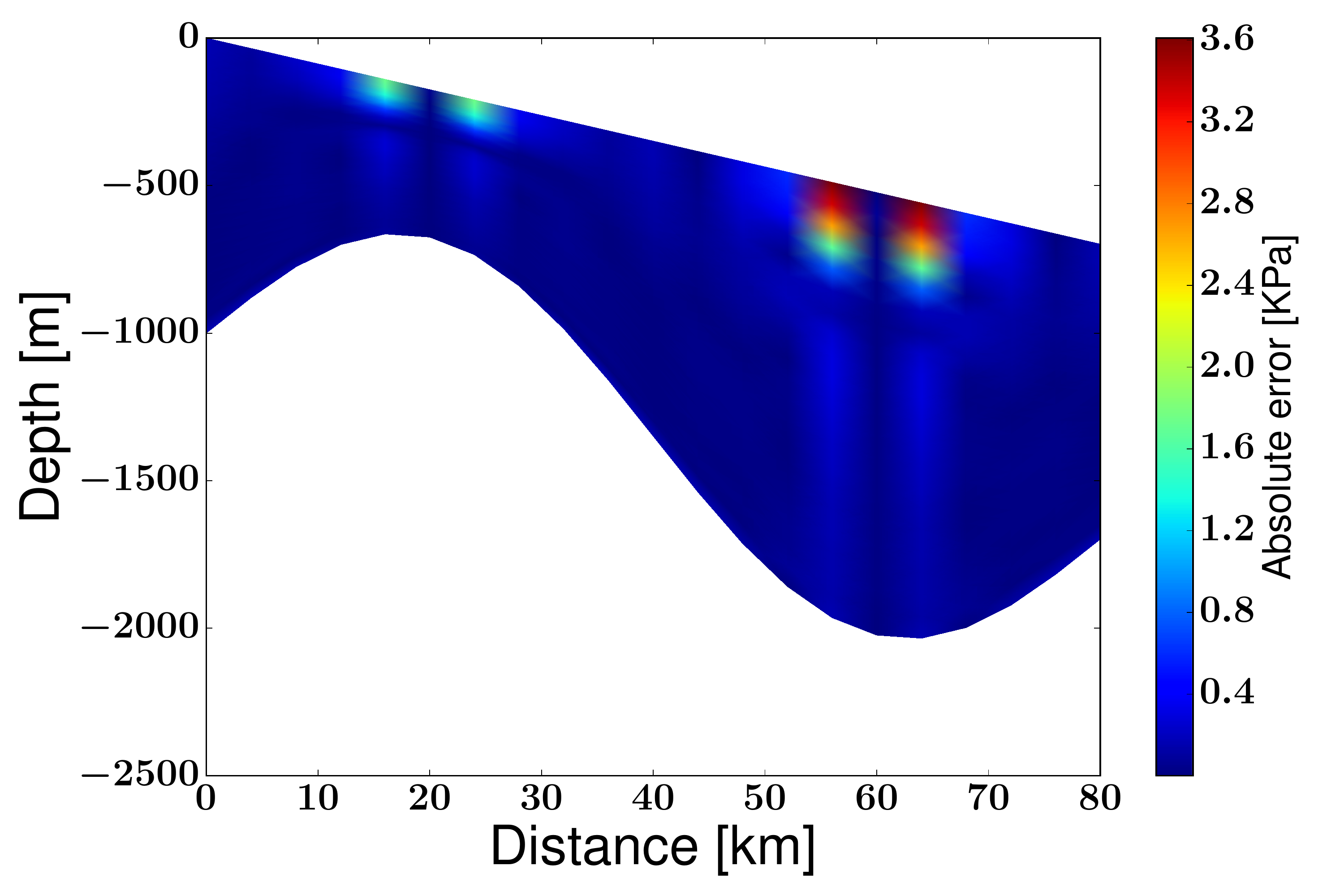}}\\
    \caption{Absolute error of Elmer/Ice in comparison to the numerical reference solution on the original case, on a $20\times 20$ grid. The panels show the error of a) horizontal velocity $u_x$, b) vertical velocity $u_z$, c) pressure $p$.} 
    \label{fig:numericaldistribution}
\end{figure}

The absolute errors for the velocity and pressure fields on the manufactured problem are shown in \cref{fig:manufactureddistribution} (along cross-section $y=L/4$). The errors relative to the numerical reference  are shown in \cref{fig:numericaldistribution}. Since the results are similar for \fenics{} and Elmer/Ice in these simulations, we here only show the Elmer/Ice output. The errors at $x=L/4$ and $x=3L/4$ are due to the non-linear viscosity, and for a linear case only the high errors at $x=0$, $x=L/2$, and $x=L$ remains. These errors are expected to increase if the wave-length of the bedrock-variation decreases.
Both the magnitude and distribution is very different for the manufactured case in \cref{fig:manufactureddistribution} compared to the original case in \cref{fig:numericaldistribution}, especially so at $x=L/4$ and $x=3L/4$ where singularities occur and the viscosity is infinite in the continuous case. Another difference between using a numerical reference solution and manufactured solution is that for the latter, the vertical velocity error is significantly higher than the horizontal velocity error, both in \fenics{} and Elmer/Ice . This effect is not as prominent when the numerical reference is used. 

Since neither convergence rates, nor error magnitude and distribution is well represented using the manufactured solutions, we will in the remaining experiments of this paper study the original problem using the numerical reference solution,  Only in \cref{sec:ismipgls}  will we use the manufactured problem again to ensure that the stability term in the numerical reference solution does not bias certain results. 

\section{Propagation of Errors to the Free Surface Equation}
\label{sec:freesurf}

In this section we investigate how the errors in the velocity propagate to the solution of the ice surface position, which they do since the velocity components at the ice surface enter as coefficients in the free surface evolution equation \cref{eq:freesurface}. As can be seen from \cref{fig:numericaldistribution}, these errors are highest near the surface which directly affects the free surface.

The expected local error (between two consecutive time steps) arising from errors in the velocity for an Euler forward time-stepping scheme is
\begin{equation}
\begin{aligned}
s^{t+\Delta t}&=s^{t}+\Delta t \big(-{\partial_x s}(u_x+\epsilon_x) +u_z+\epsilon_z+a_s\big)\\
&=s^{t}+\Delta t \big(-{\partial_x s}u_x +u_z+a_s\big)+\Delta t \big(-{\partial_x s}\epsilon_x +\epsilon_z\big),
\label{eq:freesurfaceerror}
\end{aligned}
\end{equation}
where $\epsilon_x$ is the numerical error in the horizontal surface velocity $u_x|_{z=s}$ and $\epsilon_z$ is the numerical error in the horizontal surface velocity $u_z|_{z=s}$. The initial surface slope in the ISMIP-HOM B (and A) experiment is $0.5^\circ$, so that $\frac{\partial s}{\partial x}=-0.0087$. The error term in \cref{eq:freesurfaceerror} is thus $\Delta t (0.0087\epsilon_x+\epsilon_z)$. Considering that $\epsilon_x \leq 0.8$ and $\epsilon_z \leq 0.07$ in \cref{fig:numericaldistribution} and using a time step of $\Delta t = 0.1$ years, the magnitude of the error is expected to be $0.1 \cdot 0.0087 \cdot 0.8+0.1 \cdot 0.07 \approx 0.007$. The error in the horizontal surface velocity is significantly lower than $0.08$ in all places except for at the singularity above the through, so overall we expect that the errors in the vertical surface velocity will influence the accuracy of the free surface position more than the errors in the horizontal surface velocity.

\subsection{Experiment Set-Up}
To verify the above discussion and identify the errors propagating from the errors in the velocity field as separate from the errors arising from the discretization of \cref{eq:freesurface} , we run two transient simulations, called Simulation I and Simulation II, spanning 10 years. Both simulations employs a forward Euler time-step\-ping scheme, with a fairly small time step of $\Delta t = 0.1$ years to ensure stability. In Simulation I, the fine grid of $320\times80$ is used both for computing the velocity and pressure, and for solving the free surface evolution equation \cref{eq:freesurface}. In Simulation II, the fine grid is used only for solving the free surface evolution equation, and the grid $20\times 20$ is used for computing the velocity field. As mentioned previously, the $20\times 20$ resolution is equal to, or better than, a resolution (\SI{4}{\kilo\metre} cell size in the horizontal with $20$ vertical layers) of an ice sheet simulation. By comparing the result of the two simulations, we can investigate the effect that the numerical errors in the velocity have on the free surface position. 

We also repeated the experiments with a backward Euler time-stepping scheme with almost identical results. All the experiments are performed with Elmer/Ice as there is already an existing implementation of the free surface evolution.

\subsection{Results}
\begin{figure}[ht!]
  \centering
    \stackinset{l}{0.5cm}{b}{-0.1cm}{a)}{%
      \includegraphics[width=0.46\textwidth]{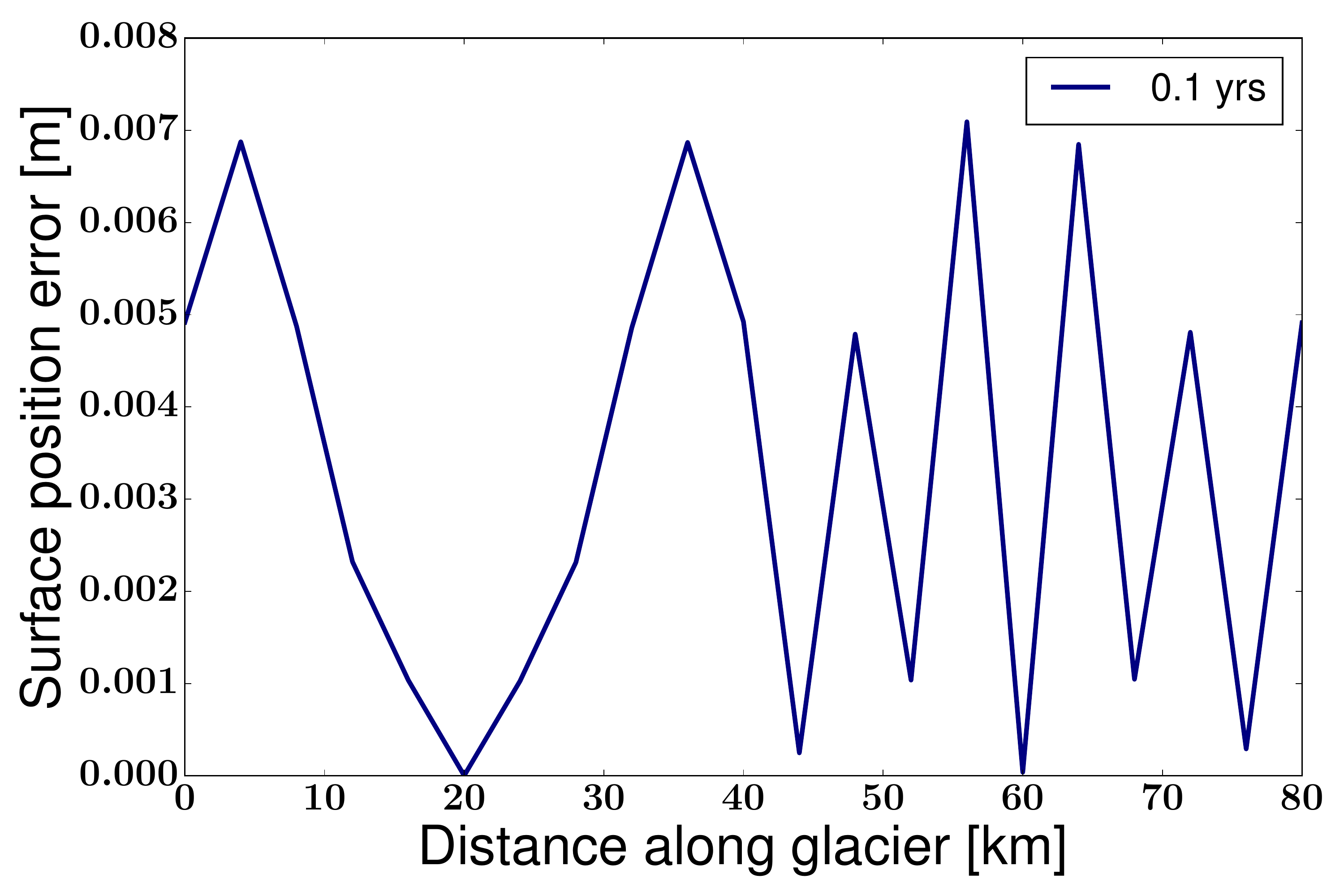}}
    \stackinset{l}{0.5cm}{b}{-0.1cm}{b)}{%
      \includegraphics[width=0.46\textwidth]{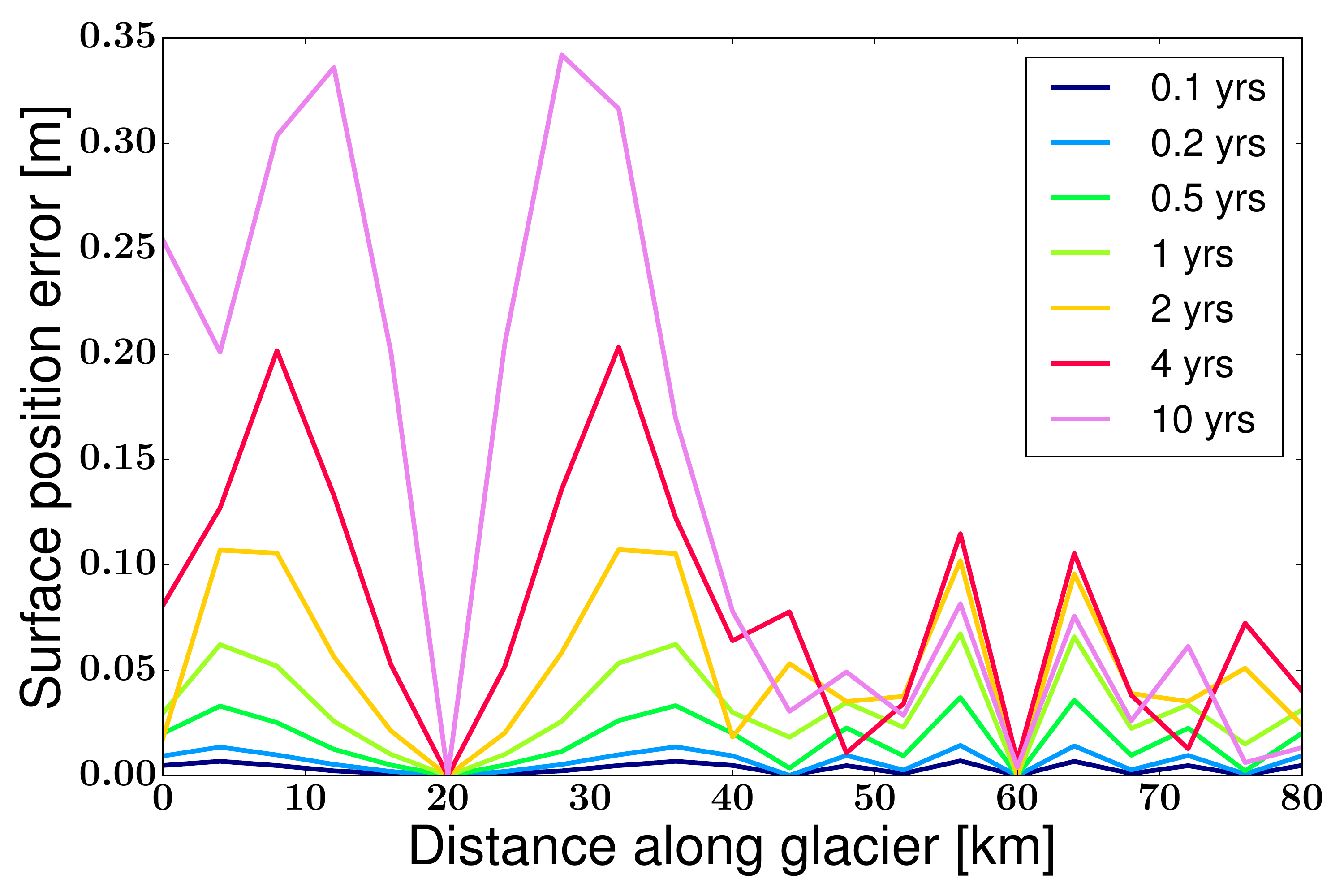}}\\
    \caption{Errors in the ice surface position due to velocity errors. Panel a) shows the errors propagating from velocity inaccuracies after one time step while panel b) shows the accumulation of propagated errors over \SI{10}{\yr}.} 
    \label{fig:freesurfacepropogation}
\end{figure}

The difference between the free surface position in Simulation I and Simulation II is shown \cref{fig:freesurfacepropogation}. 
After one time step, the magnitude of the errors is indeed $0.007$, and the pattern of the errors in the free surface after one time step (\cref{fig:freesurfacepropogation}a) corresponds to the pattern of the errors in the vertical velocity in \cref{fig:numericaldistribution}.  

The local errors accumulate throughout the simulation, especially over the bump, see \cref{fig:freesurfacepropogation}b. After one year the error amounts to about $\SI{0.1}{\metre}$, which is on the order of magnitude of a typical net surface accumulation/ ablation \citep{Bindenschadler2013}. After ten years it is about $\SI{0.3}{\metre}$. The importance of the vertical velocity errors compared to the horizontal velocity errors is large for small surface slopes. For the ISMIP-HOM A and B experiment the surface gradient is $0.0087$. In the Greenland SeaRISE data set \citep{data:searise,Bindenschadler2013}, it is never larger than $0.08$, and often much smaller.

Clearly, it is important to control the errors in the vertical velocity, in order to accurately predict ice surface position and ultimately the ice sheet volume. Despite this, the vertical velocity error is however rarely discussed, as the absolute vertical velocity error is lower than the absolute error in the horizontal velocity, and since the horizontal velocity is often of more interest to glaciologists than the vertical velocity. As we will see in the following section, an unfortunate choice of the GLS stabilization parameter has a larger impact on the vertical velocity than on the horizontal components. Especially considering the accuracy of the ice surface, it is important to pay attention to the choice of stabilization parameters. 

\section{Sensitivity to the GLS Stabilization Parameter}
\label{sec:tau}

If the stability parameter $\tau_{\mathtiny{GLS}}$ in \cref{eq:GLS} is not chosen carefully, significant errors will be introduced to the solution. For some cases GLS stabilization also introduces artificial boundary conditions \citep{BraackLube2009}. In this section we perform experiments to assess whether the value in \cite{FrancaFrey1992} (i.e. $\tau_0=1$ in \cref{eq:francatau}), which was given for a Newtonian fluid on a simple domains and isotropic meshes, is a good choice for ice sheet simulations. To investigate the issues of over- and under-stabilization and artificial boundary conditions, and how sensitive these effects are to the stability parameter, we here perform experiments on three different experiment set-ups where we vary $\tau_0$. The three setups are: 
\begin{itemize}
\item The ISMIP-HOM domain
\item An outlet glacier simulation
\item Channel flow
\end{itemize} 
In all simulations in this section,  $h_K$ was defined as the \emph{minimum edge} of the cell and the non-linear viscosity $\eta$ is used in the GLS stabilization parameter. A justification of these choices follows in \cref{sec:parameterchoice}.

\subsection{The ISMIP-HOM domain}\label{sec:ismipgls}
\subsubsection{Experiment Set-Up}
The domain and boundary conditions are set-up as in the previous ISMIP-HOM experiments. We run repeated simulations on the $20\times 20$ mesh, varying between $\tau_0 = 10^{-4}$ and $\tau_0 = 10^{4}$, and measuring the accuracy of the solutions.  Despite our findings in \cref{sec:convergence} putting the manufactured problem in question, we here measure the errors both using a numerical reference solution, and on the manufactured problem. This is because the numerical reference solution will contain a stabilization term and we use the manufactured solutions to ensure that this term does not influence our results. For the reference solution we set $\tau_0=1$. We perform the experiments with both \fenics{} and Elmer/Ice to ensure that the behavior of the stabilized problem is independent of implementation details. 

\subsubsection{Results}
\begin{figure}[ht!]
  \centering
    \stackinset{l}{0.5cm}{b}{-0.1cm}{a)}{%
      \includegraphics[width=0.46\textwidth]{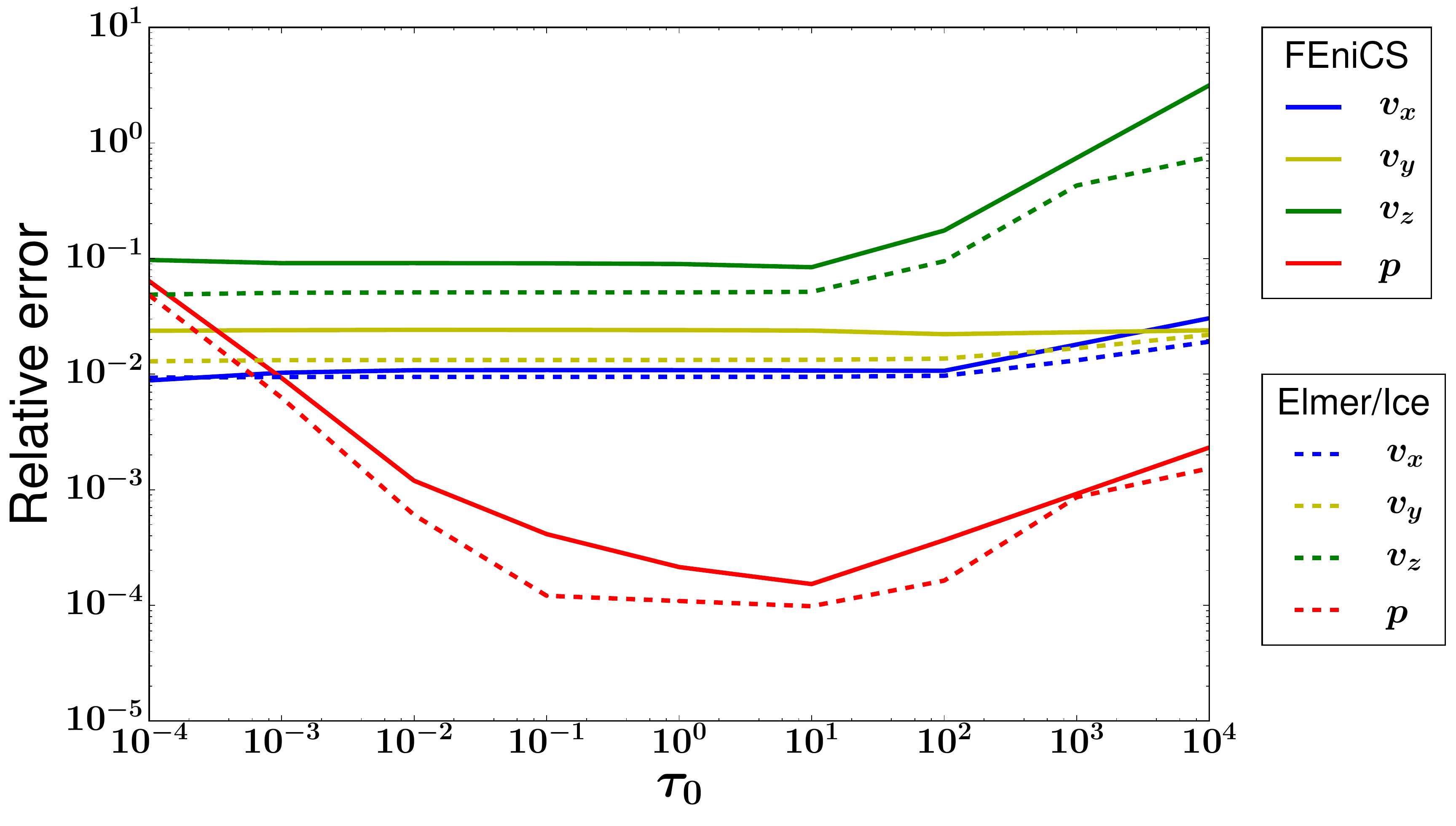}}
    \stackinset{l}{0.5cm}{b}{-0.1cm}{b)}{%
      \includegraphics[width=0.46\textwidth]{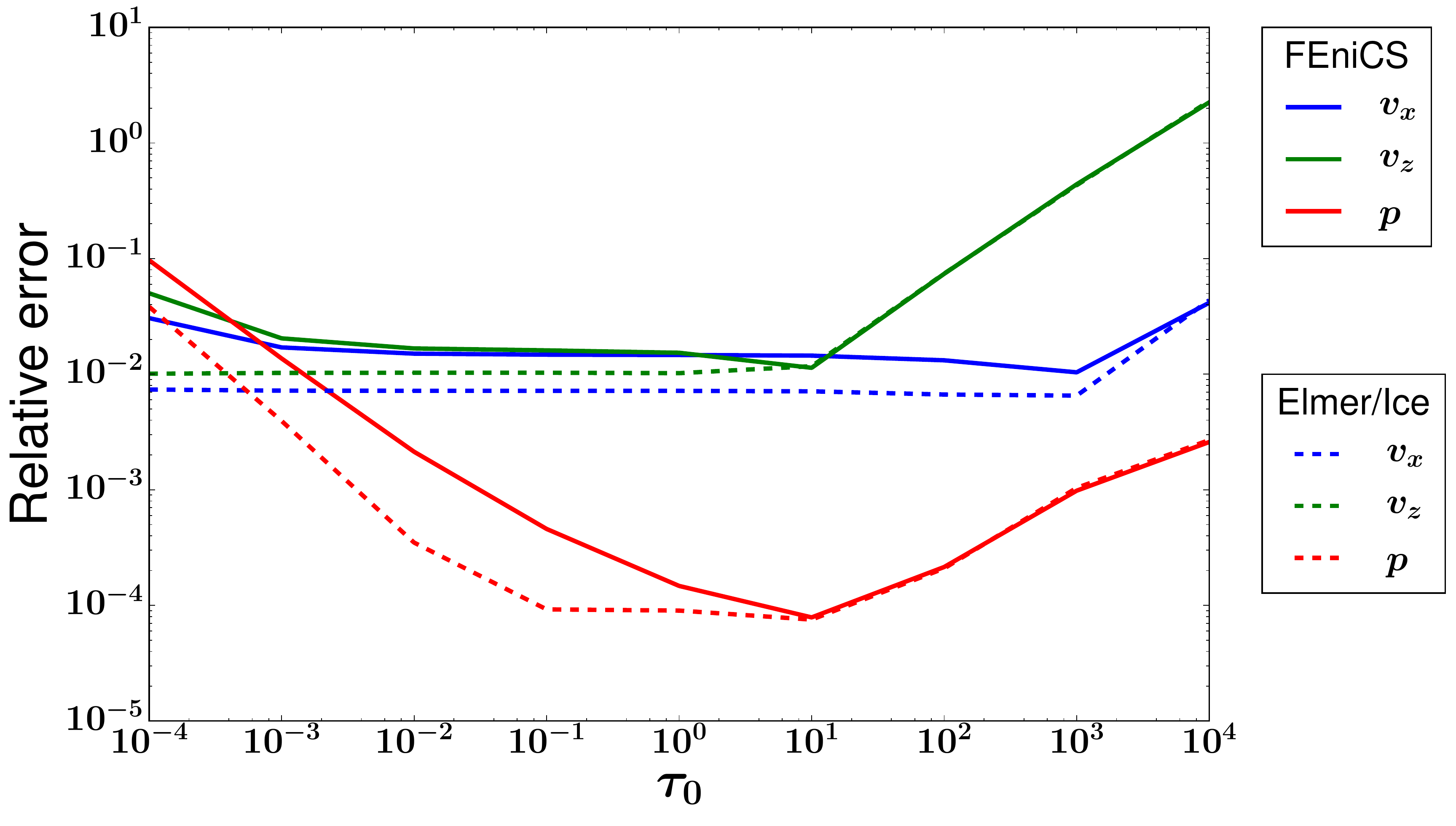}}\\
    \caption{The $L^2$-norm of the relative errors compared to a a) manufactured solution (3D) and b) numerical reference solution (2D) in velocity components and pressure for a GLS stabilization. The errors are computed with \fenics{} (solid lines) and Elmer/Ice (dashed lines) on a $20\times 20\times 20$ (for 2D $20\times20$) grid for varying $\tau_0$.
\label{fig:fenics_tausensitivity}}
\end{figure}
\Cref{fig:fenics_tausensitivity} shows how the $L^2$-norm of the errors in the velocity components and pressure vary with $\tau_0$. For an under-stabilized problem, the error in the pressure is high due to spurious oscillations, while for a too large $\tau_0$, the stability terms alters the original problem significantly. The default parameter $\tau_0=1$ is a good choice while $\tau_0=10$ renders a slightly lower pressure error in \fenics{}. The overall behavior is however very similar in \fenics{} and Elmer/Ice, and therefore, all experiments from now on will be performed only in \fenics{} for simplicity.

\begin{figure}[ht!]
\begin{center}
{\includegraphics[width=0.75\textwidth]{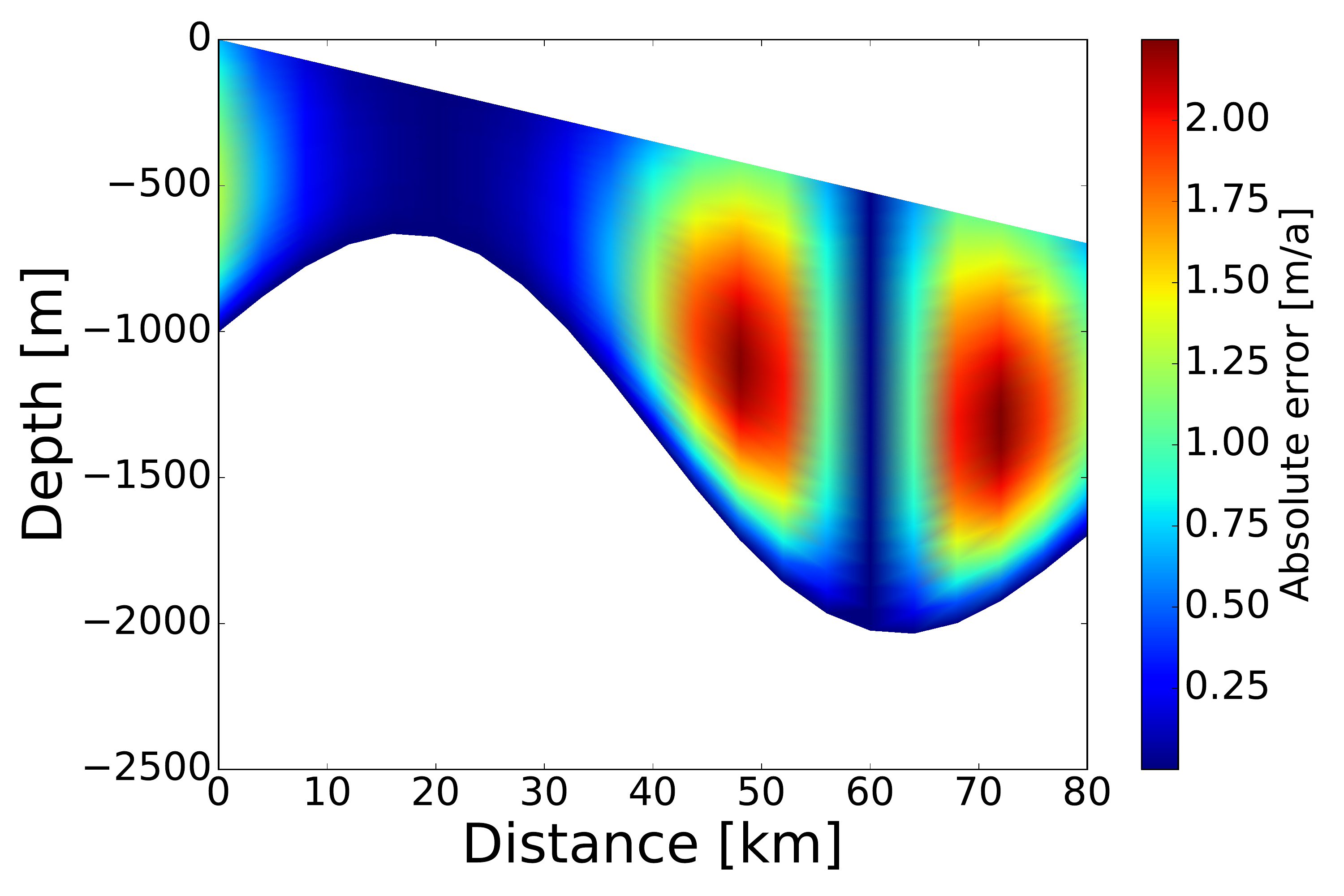}}
\end{center}
\caption{Absolute error in the vertical velocity using $\tau_0=\num[retain-unity-mantissa = false]{1e3}$. An over-stabilized problem introduces smooth errors in the vertical velocity component. The errors are computed by comparing to a numerical reference solution on the ISMIP-HOM B domain \label{fig:smoothvz} in \fenics{}. Results from Elmer/Ice are very similar. }
\end{figure}

As seen in \cref{fig:fenics_tausensitivity}, an over-stabilized problem introduces errors mainly in the vertical velocity and pressure. This is because of the shallowness of ice sheets and the direction of gravitational force. The errors introduced in the vertical component by over-stabilization are smooth and influence the surface, see \cref{fig:smoothvz}. These errors therefore risk to pass unnoticed in a complex ice sheet simulation. Since the free surface position (and ultimately ice volume prediction) is sensitive to errors in the vertical velocity, the sensitivity of the vertical velocity to the stabilization parameter is unfortunate. 

By considering the Shallow Ice Approximation (SIA) we may gain some further understanding of why over-stabilization mainly influence the vertical velocity component. Even though the SIA is a  significant simplification of the problem, it is quite accurate for the ISMIP-HOM domain with $L=\SI{80}{\kilo\metre}$ \citep{Pattyn2008, Ahlkrona13}. According to the SIA, the $\mathfrak{p}$-Stokes system of \cref{eq:stokes} (in two dimensions) can, together with the incompressibility constraint, be reduced to
\begin{subequations}  \label{eq:SIA}
\begin{alignat}{3}  \label{eq:SIA1} 
-\frac{\partial p}{\partial x}+\frac{\partial}{\partial z}\left(\eta \frac{\partial u_x}{\partial x}\right)&=0\\ \label{eq:SIA2} 
   -\frac{\partial p}{\partial z}&=\rho g\\  \label{eq:SIA3}    
 \frac{\partial u_x}{\partial x}+\frac{\partial u_z}{\partial z}&=0 
\end{alignat}
\end{subequations}
In the SIA equations, the horizontal velocity $u_x$ is thus solely determined by the horizontal momentum balance of \cref{eq:SIA1}, the pressure is given by the vertical momentum balance of \cref{eq:SIA2}, and the vertical velocity $u_z$ is given by the balance of mass in \cref{eq:SIA3}. 
For linear elements, the stabilization term in \cref{eq:GLS} mostly (exclusively, on a tetrahedron mesh) affects the balance of mass, and as the horizontal velocity is strongly restricted by the horizontal momentum balance, the vertical velocity will balance any alterations introduced by the stabilization parameter. For thicker domains, where the SIA does not hold, the sensitivity to $\tau_0$ is more equally distributed among the velocity components. Note that the non-linearity of the viscosity is not important here, and also for a linear viscosity the errors associated with over-stabilization are larger in the vertical component. 

\subsection{An Outlet Glacier} \label{sec:vialov}
In this section we study a more realistic set-up than the ISMIP-HOM problem. In glaciology, a common problem is to  model a single outlet glacier which is part of a larger ice sheet. Inlet velocity boundary conditions are typically given by a coarse mesh simulation of the whole ice sheet using the Stokes equation, or a cheaper method like the shallow ice approximation \citep{Greve1995} or the Blatter-Pattyn \citep{Blatter95,Pattyn2003} formulation of the system.  This set-up introduces two separate complications that we study: 1) A more complicated geometry with a glacier front and 2) a more involved lateral boundary conditions. 
\subsubsection{Experiment Set-Up}
We model the front of a two-dimensional glacier specified by a Vialov type surface profile \citep{vialov1958} underlain by a bumpy bed. The surface, $s$, and bed, $b$, of the entire ice sheet are defined by
\begin{equation}
\begin{aligned}
  s(x) &= H\left(1 - \left(\frac{|x|}{L}\right)^{\frac{n+1}{n}} \right)^{\frac{n}{2n+2}} + 200,\\
  b(x)&=  150\cos\left(\frac{10\pi |x|}{L}\right)\left(1-\cos\left(\frac{\pi x}{2L}\right)\right)\\ &+250 \cos\left(\frac{\pi x}{2L}\right) + 25\cos\left(\frac{80 \pi |x|}{L}\right),
\label{eq:bedvialov}
\end{aligned}
\end{equation}
where the dome is at $x=0$, the radius of the ice sheet is $L=\SI{300}{\kilo\metre}$, the maximum height is $H=\SI{2000}{\metre}$ and as before, $n=\frac{1}{\mathfrak{p}-1}=3$. We focus on the region stretching from $x=\SI{275}{\kilo\metre}$ to the glacier front at $x=\SI{300}{\kilo\metre}$.

To produce an inflow boundary condition at $x=\SI{275}{\kilo\metre}$ we first simulate the entire ice sheet simulation on a coarse mesh. This coarse mesh is an extruded $160\times20$ mesh which is equidistant in the horizontal direction. The bedrock undulations with the highest frequency (i.e. the last term in \cref{eq:bedvialov}) cannot be resolved properly on the coarse mesh and is therefore excluded when generating the inflow conditions. 

The mesh used for the front simulation is a horizontally refined mesh with horizontal resolution of $\SIrange{100}{850}{\metre}$ (see \cref{fig:stab_ratio}b). This is done to resemble a practical simulations, as it is common to apply horizontal plane mesh refinement near glacier fronts. Horizontal mesh refinement will be discussed further in \cref{sec:cell_size}.  At the glacier front, we have specified a natural, stress free condition.
 
Bearing the findings of \cref{sec:freesurf} and \cref{sec:tau} in mind, we focus our attention to the surface velocity and how this changes with the type of stabilization method and size of stabilization parameter (Note that although the surface gradient is steep at the very edge of the glacier it is still only $0.05$ at  $x=\SI{297}{\kilo\metre}$ and   $0.09$ at $x=\SI{299}{\kilo\metre}$). More specifically, we measure changes in the vertical velocities when compared to an unstabilized velocity field.  While this approach does not allow us to measure exact errors, it does show the sensitivity to the stabilization parameter. The unstabilized velocity field is non-oscillatory.

\subsubsection{Results}

\begin{figure}[ht!]
\begin{center}
{\label{fig:Vialov_GLS_hmin}\includegraphics[width=0.75\textwidth]{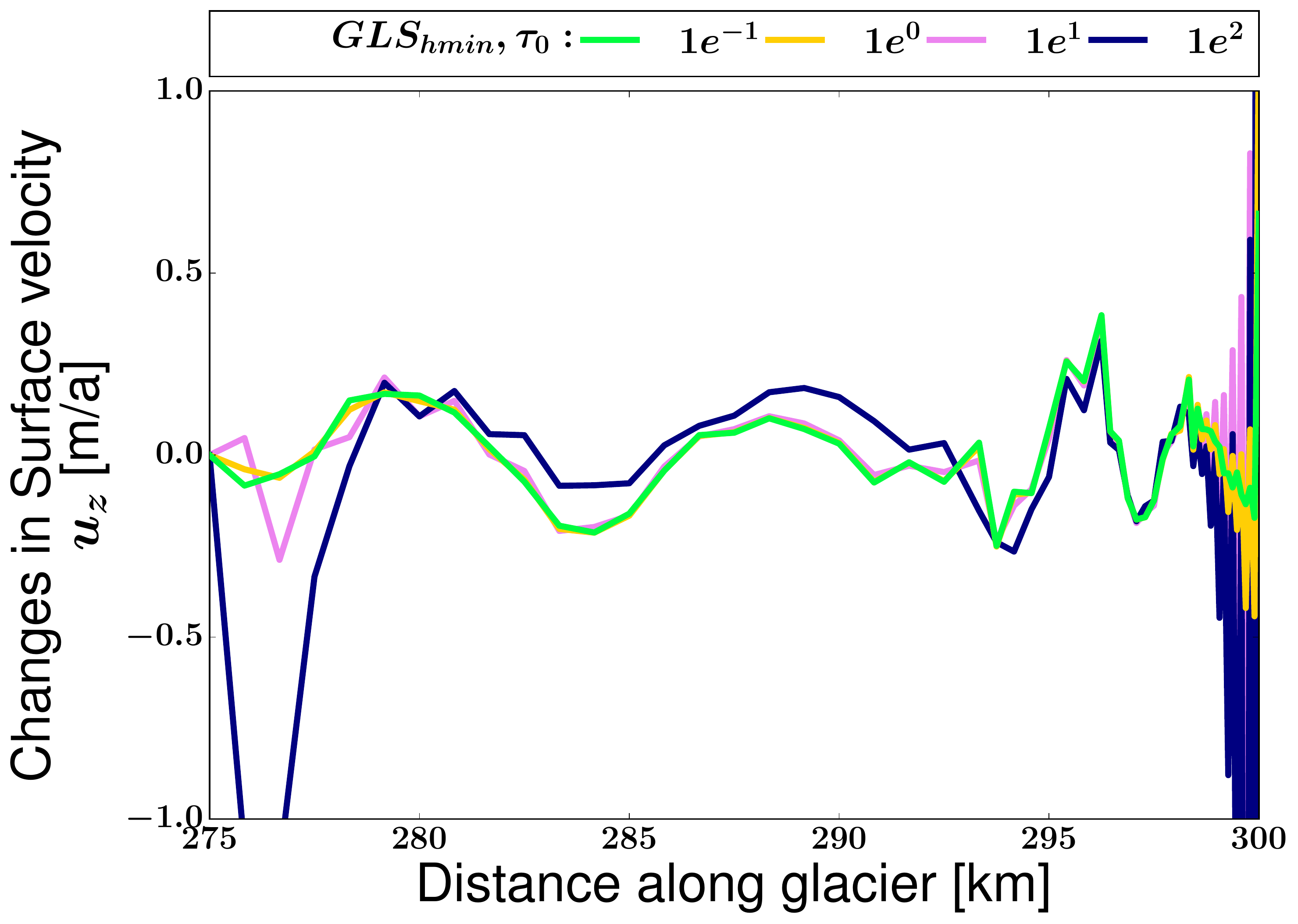}} \hspace{0.2cm} \\
%\
\end{center}
\caption{Changes in vertical velocity for GLS for different values of $\tau_0$, compared to an unstabilized velocity solution. The values of $\tau_0$ have been chose such that any smaller values produced visible oscillations in the pressure field.}
\label{fig:Vialov_stabilization_hmin}
\end{figure}

The result is presented in \cref{fig:Vialov_stabilization_hmin}. Firstly, we note that setting an inlet boundary conditions makes the pressure more prone to oscillations, specifically near the inlet. The smallest $\tau_0=0.1$ shown is the minimal value that resulted in an acceptable, oscillation-free, pressure field. Secondly, any value above $\tau_0=0.1$ introduces oscillations in the velocity at the front of the glacier. This effect does not seem to be related to the inlet boundary condition, and occurs also for simulations of the entire ice sheet. Because of such over-stabilization of the vertical velocity, the default GLS parameter as given in \citet{FrancaFrey1992} (obtained for $\tau_0=1$) does not seem to be an appropriate choice. The two effects together narrows the range of possible stability parameters.
The changes in the horizontal surface velocity were of the same absolute magnitude.

\subsection{Poiseuille Flow - Artificial Boundary Conditions}
In the previous examples, interior effects of under and over-stabilization were considered. In this section we focus on artifacts at the boundary due to \textit{artificial boundary conditions}. It is a known problem of GLS stabilization that it introduces the artificial boundary condition 
\begin{equation}
  \int_{\partial \Omega} \tau_{\mathtiny{GLS}}(\partial_n p_h - \rho\mathbf{g}\cdot \mathbf{n}, \nabla q_h) \approx 0,
\label{eq:artificial_bcs}
\end{equation}
where $\partial_n$ denotes the normal derivative.  \citet{Droux1994} and \citet{BeckerBraak2001} showed that such artificial boundary conditions can have  significant effect on the numerical solution of the pressure,  using simulations of Newtonian so-called Poiseuille flow as an example.

\subsubsection{Experiment Set-Up}
To show the influence of artificial boundary conditions, we perform two-dimensional Poiseuille flow experiments with ice. Poiseuille flow is a pressure driven flow in a channel. The simulations are performed in the horizontal plane so that the body force is zero ($\rho\mathbf{g}\cdot \mathbf{n}=0$). The analytical solution is given by
\begin{align*}
(u_x,u_y) &= \left(\frac{2^{\frac{1-n}{2}}}{n+1} \Big(\frac{p_x}{\eta_0}\Big)^n ( R^{n+1} - |R - y|^{n+1} ), 0.0\right) \\
p &= p_x (L - x),
\end{align*}
where $n=3$ and $\eta_0$ are the parameters described in \cref{eq:constitutive}, $L=\SI{0.5}{\kilo\metre}$ is the horizontal extension of the domain, $R = \SI{0.5}{\kilo\metre}$ is the radius of the channel, and $p_x=\SI{200}{\pascal\per\metre}$ is the $x$-directional gradient of the pressure. We prescribed the analytical solution as an inflow velocity at $x=0$ and no-slip boundary conditions at $y=0$ and $y=2R$. As a outflow condition we prescribed a Neumann condition with the additional requirement that the vertical velocity be zero. Since the pressure is not unique, we set a point Dirichlet condition at $(x,y) = (L, 0)$ where the pressure equals zero. The simulations were preformed on a $20\times 20$ grid, with $h_K$ set as diameter of the cell (maximum cell edge length for triangles). However, note that in this horizontal plane simulation the mesh is isotropic so that $h_{\mathtiny{min}} \sim h_{\mathtiny{max}}$.

\subsubsection{Results}
\begin{figure}[ht!]
  \centering
    \stackinset{l}{0.5cm}{b}{-0.1cm}{a)}{%
      \includegraphics[width=0.46\textwidth]{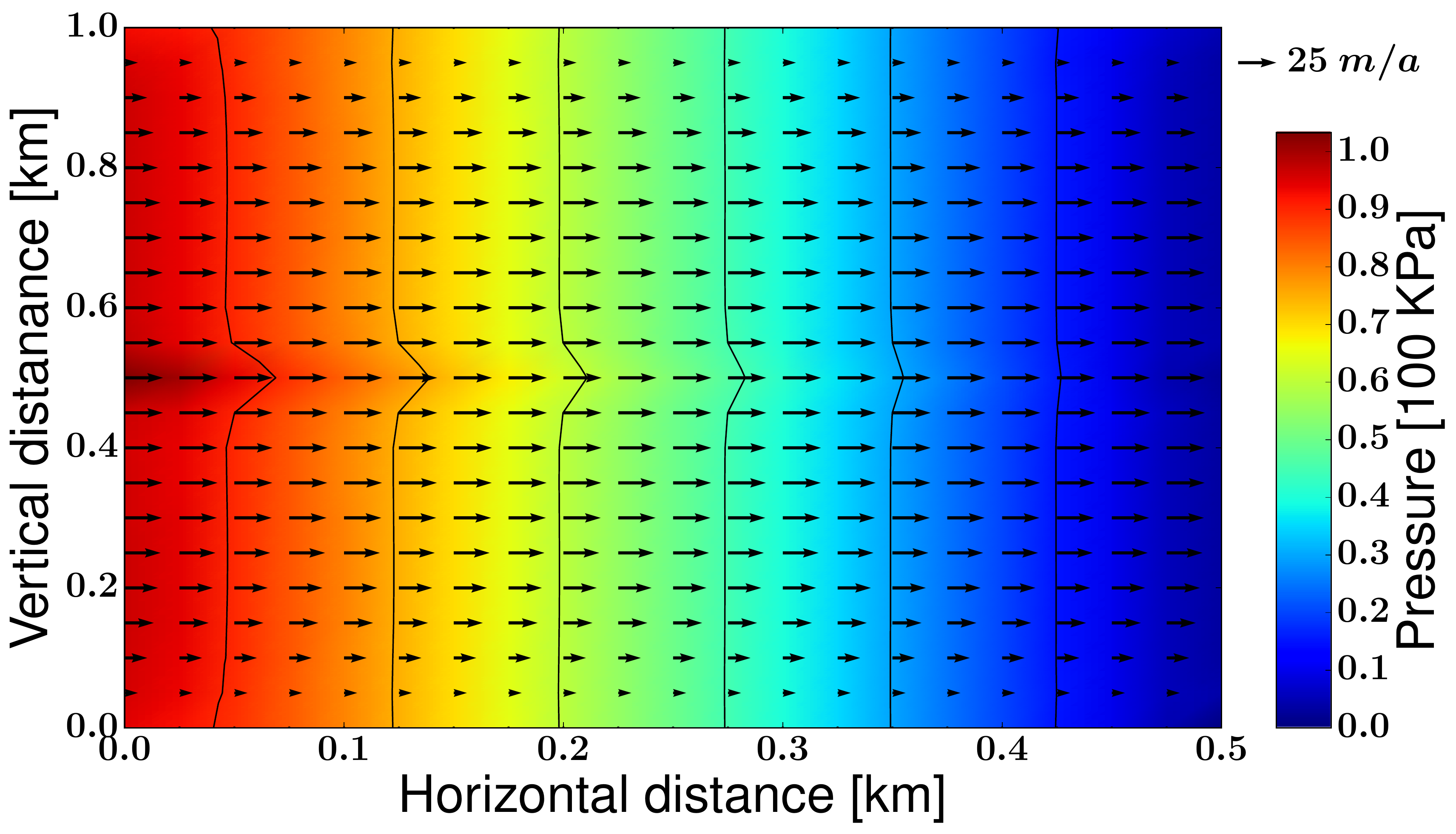}}
    \stackinset{l}{0.5cm}{b}{-0.1cm}{b)}{%
      \includegraphics[width=0.46\textwidth]{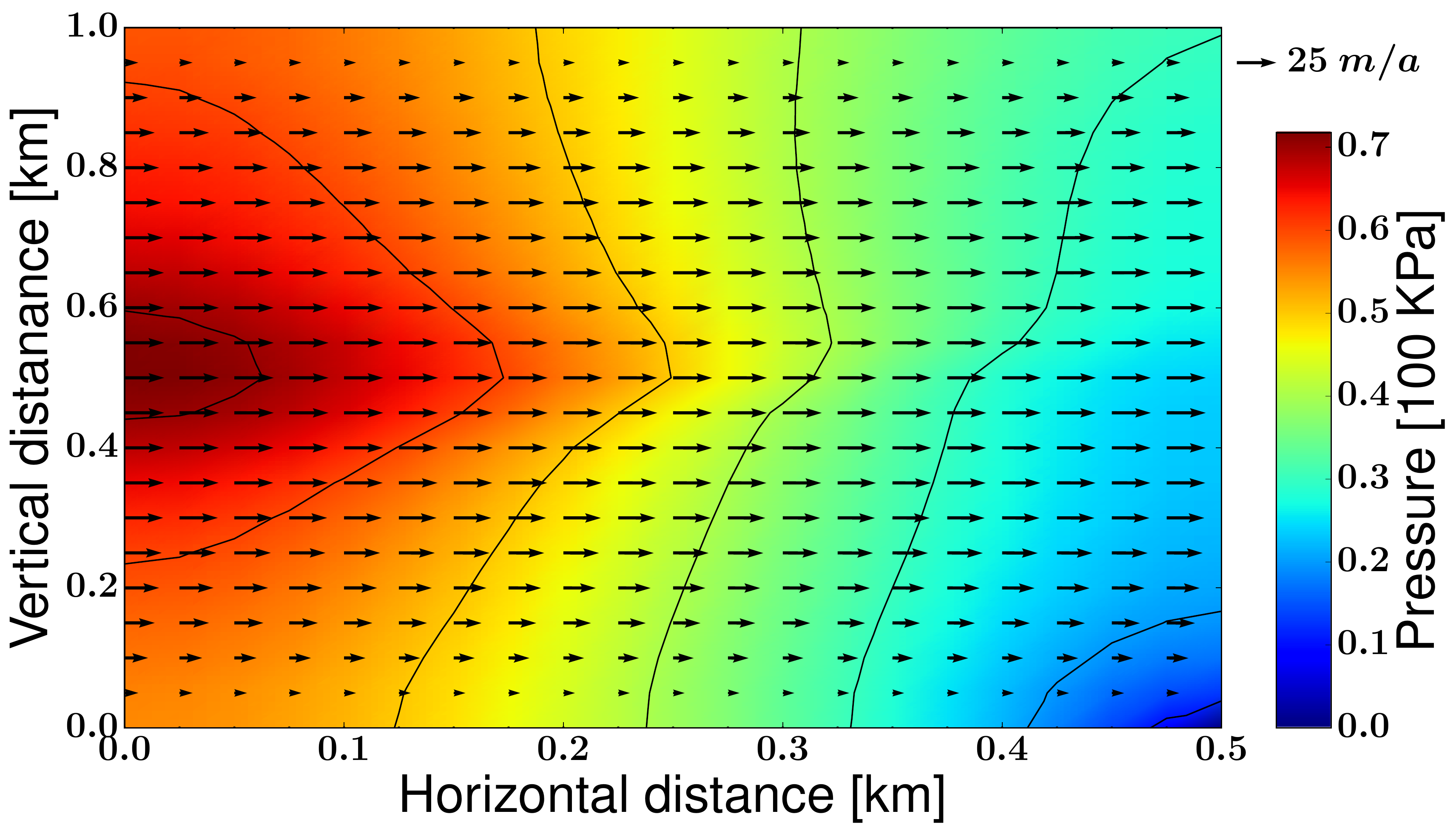}}\\
    \vspace{0.5em}
    \stackinset{l}{0.5cm}{b}{-0.1cm}{c)}{%
      \includegraphics[width=0.46\textwidth]{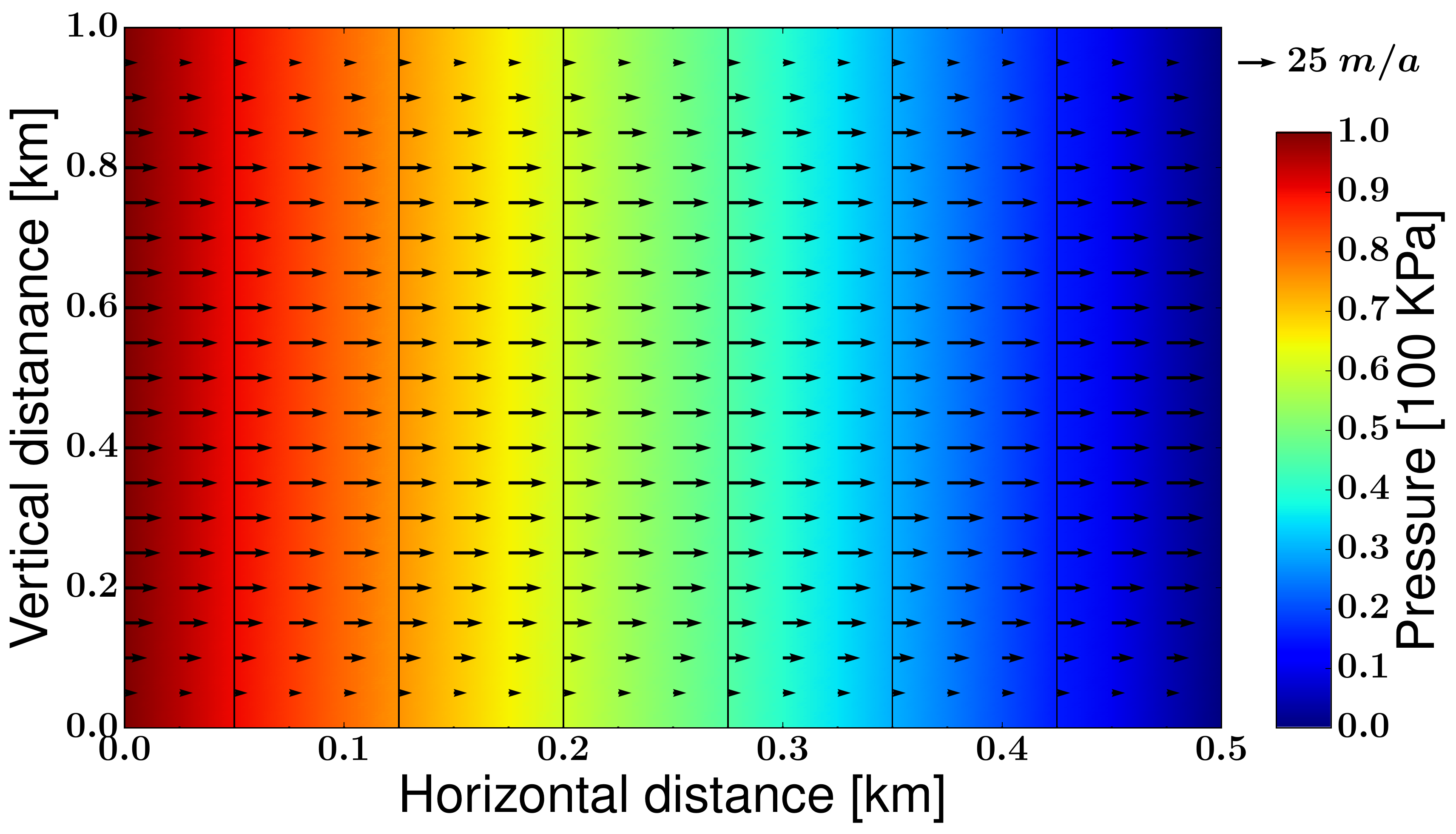}}\\
    \caption{Poiseuille flow stabilized with the GLS method (a,b) and the analytical solution (c). Stabilization parameter is $\tau_0=1$ (a) or $\tau_0=100$ (b), $h_K$ is taken to be maximum cell edge or facet. Velocity is depicted as \emph{black arrows} and pressure as \emph{color} with \emph{black lines} as pressure contours.}
    \label{fig:poiseuille_GLS}
\end{figure}

The results are shown in \cref{fig:poiseuille_GLS}. We included the analytical solution as a reference.

The pressure contours in (\cref{fig:poiseuille_GLS}a) bends towards the edges, which is the typical effect of artificial boundary conditions. Hence, the choice of $\tau_0=1$ is large enough to stabilize the pressure, but too large to avoid a significant effect of the artificial boundary condition.

Another interesting feature of \cref{fig:poiseuille_GLS}a is the spike of the pressure contours observed along $y=\SI{500}{\metre}$ in \cref{fig:poiseuille_GLS}a due to a singularity in the viscosity. The problem is not easily amendable by varying the critical shear rate that regularizes the viscosity. The spikes decrease as $\tau_0$ increases, but it is not possible to increase the stability parameter enough to visibly eliminate the pressure spike without over-stabilizing the solution grossly, see \cref{fig:poiseuille_GLS}b. The problem becomes less severe with mesh refinement, but for practical meshes the problem remains. The simulation was also performed with the inf-sup stable $P2P1$ element, which indicated that the pressure was affected by the singularity in a similar way to the GLS method, although to a lesser extent.

Poiseuille flow does remind of a horizontal plane simulation of a glacier in a confined bay with low basal friction. Tentative experiments indicate that the pressure field can be affected in such glacier simulations. However, further studies are need to determine whether this is due to setting inflow boundary conditions of due to over-stabilization.  Note that in a  three dimensional simulation $\rho\mathbf{g}\cdot \mathbf{n}\neq 0$. In fact, for a three dimensional simulation the artificial boundary condition will enforce a hydrostatic pressure near boundaries. 

\section{Stability Parameter Choices}\label{sec:parameterchoice}

The size of the stability parameter $\tau_{\mathtiny{GLS}}$ of  \cref{eq:francatau} obviously depends on the definition of $h_K$ and viscosity parameter. In the previous chapters, we defined $h_K$ as the minimum edge length (apart from the Poiseuille case) and let the viscosity vary with the velocity, without further discussion. We saw that with these choices, the stabilization parameter given in \citet{FrancaFrey1992} (i.e. $\tau_0=1$) is often, but not always, appropriate. In this chapter we perform experiments to justify these choices and illustrate the impact of alternative approaches.

\subsection{The Cell Size $h_K$}
\label{sec:cell_size}
While standard stabilization techniques were originally developed for isotropic meshes, a typical mesh as described in \cref{sec:Elmer} can render elements with an aspect ratio as high as $200$. The aspect ratio can affect the accuracy and stability of the numerical solution \citep{Shewchuk2002}. 
Given the flat elements, it is important to consider the definition of $h_K$ in $\tau_{\mathtiny{GLS}}$ in \cref{eq:GLS}.

If the cell size is as the \emph{minimum edge length} (denoted as $h_{\mathtiny{min}}$), the stabilization parameter will scale with the topography of the domain. However, since most meshes used in glaciology are of extruded type, refinements are commonly applied at the footprint level (horizontal refinement) \citep{Seddik2012,GilletChaulet2012,Larour2012} according to some measure such as velocity changes. Such a refinement will have little effect on the size of $h_{\mathtiny{min}}$ while the aspect ratio of the cells may vary substantially.

Alternatively, one can set $h_K$ as the diameter of the cell (denoted as $h_{\mathtiny{max}}$), i.e. as the length of the \emph{maximum edge}  for a triangle/ tetrahedron or the largest diagonal of the cell for a  rectangle/ hexahedron. This measure is commonly used in FEM (see e.g \cite{Boffi2013}) and allows for a potential horizontal refinement to affect the stabilization parameter.  However, this definition may cause problems in glaciology, since for an extruded mesh with anisotropic elements such as in \cref{fig:extrudedmesh}, the diameter does not account well for the variations in topography. Since the dominating effect of the stabilization is in the vertical, it could lead to over-stabilization for elements with high aspect ratio on a highly varying topography.

Another possibility is to use methods developed specifically for anisotropic meshes. In \cite{Blasco2008} a GLS stabilization for high aspect ratios is developed. The technique is based on using he Jacobian matrix, $\mathbf{J}_K$, of the affine transformation from the reference element $\hat{K}$ to the current element $K$. The modified stabilization for our case then becomes
\begin{equation}
\label{eq:blasco}
  \begin{aligned}
    &S_{\mathtiny{J}}((\mathbf{u}_h, p_h), (\mathbf{v}_h, q_h)) = \\
    &-\sum_{K\in T_h}\frac{m_K}{8\eta} (\mathbf{J}_K\cdot (-\nabla\cdot\mathbf{S(u)} + \nabla p_h), \mathbf{J}_K\cdot (-\nabla\cdot\mathbf{S(v)} + \nabla q_h))_{K},\\
    &F_{\mathtiny{J}}((\mathbf{v}_h, q_h)) = \\
    &-\sum_{K\in T_h} \frac{m_K}{8\eta} (\mathbf{J}_K\cdot\rho\mathbf{g},
    \mathbf{J}_K\cdot(\nabla\cdot\mathbf{S(v)}) + \nabla q_h) _{K}.
  \end{aligned}
\end{equation}
We now make numerical experiments on the ISMIP-HOM problem and the outlet glacier of \cref{sec:vialov} to investigate the effect of using 1) a standard GLS formulation $h_K=h_{max}$, 2) a standard GLS formulation $h_K=h_{min}$, and 3) the anisotropic stabilization of \cref{eq:blasco}.

\subsubsection{The ISMIP-HOM domain - Horizontal Mesh Refinement}\
\paragraph{Experiment Set-Up}
We simulate the ISMIP-HOM B domain on five different meshes with different horizontal resolution, $20\times20$, $40\times20$, $80\times20$, $160\times20$, and $320\times20$ and the three different approaches to the anisotropic meshes. For each mesh and method, we also vary $\tau_0$ to show how the appropriateness of the default parameter $\tau_0=1$ varies.  

For the experiments in this section we used \fenics{} and the numerical reference solution on the ISMIP-HOM B problem. 

\begin{figure}[ht!]
  \centering
    \stackinset{l}{0.5cm}{b}{-0.1cm}{a)}{%
      \includegraphics[width=0.46\textwidth]{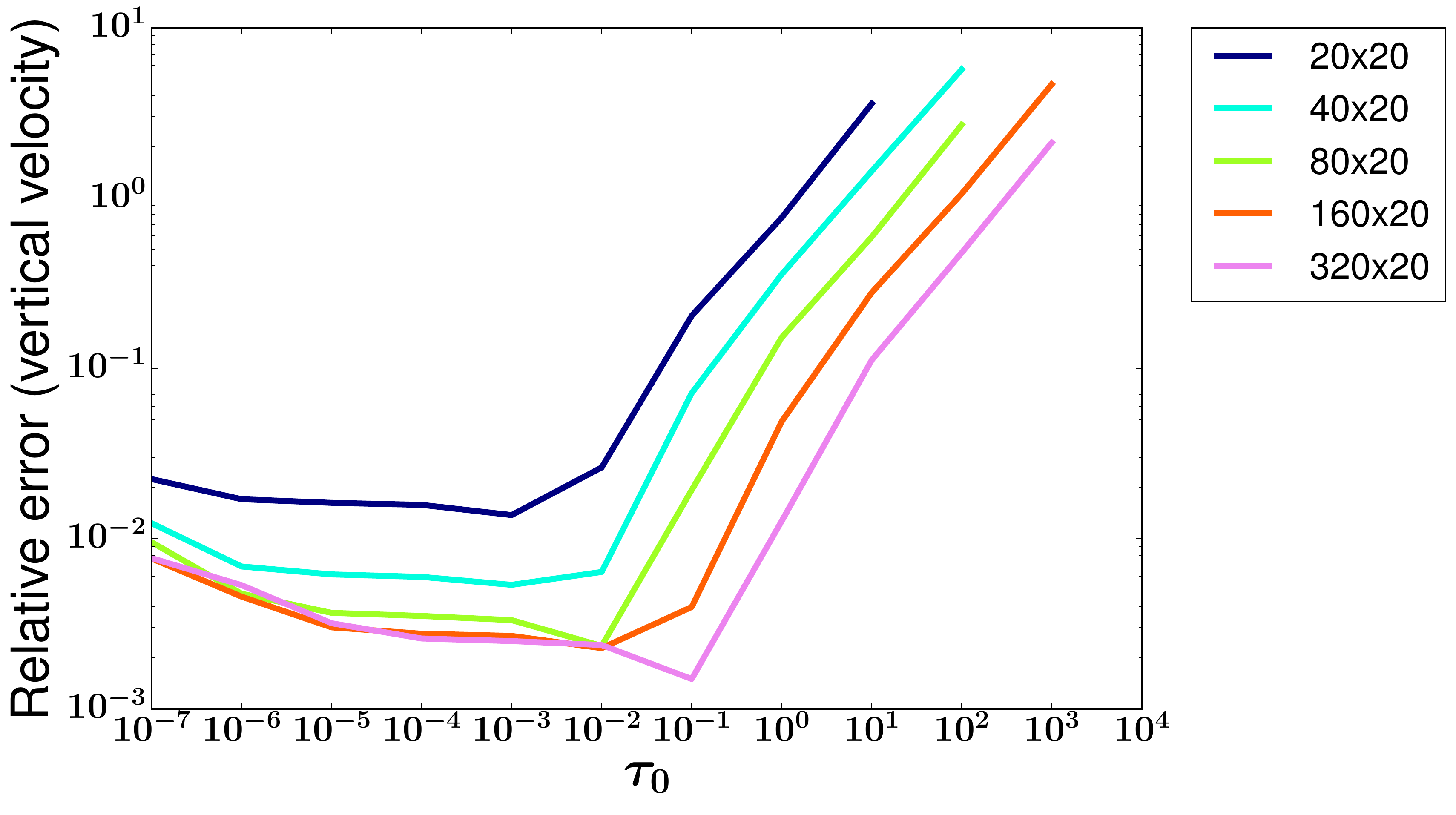}}
    \stackinset{l}{0.5cm}{b}{-0.1cm}{b)}{%
      \includegraphics[width=0.46\textwidth]{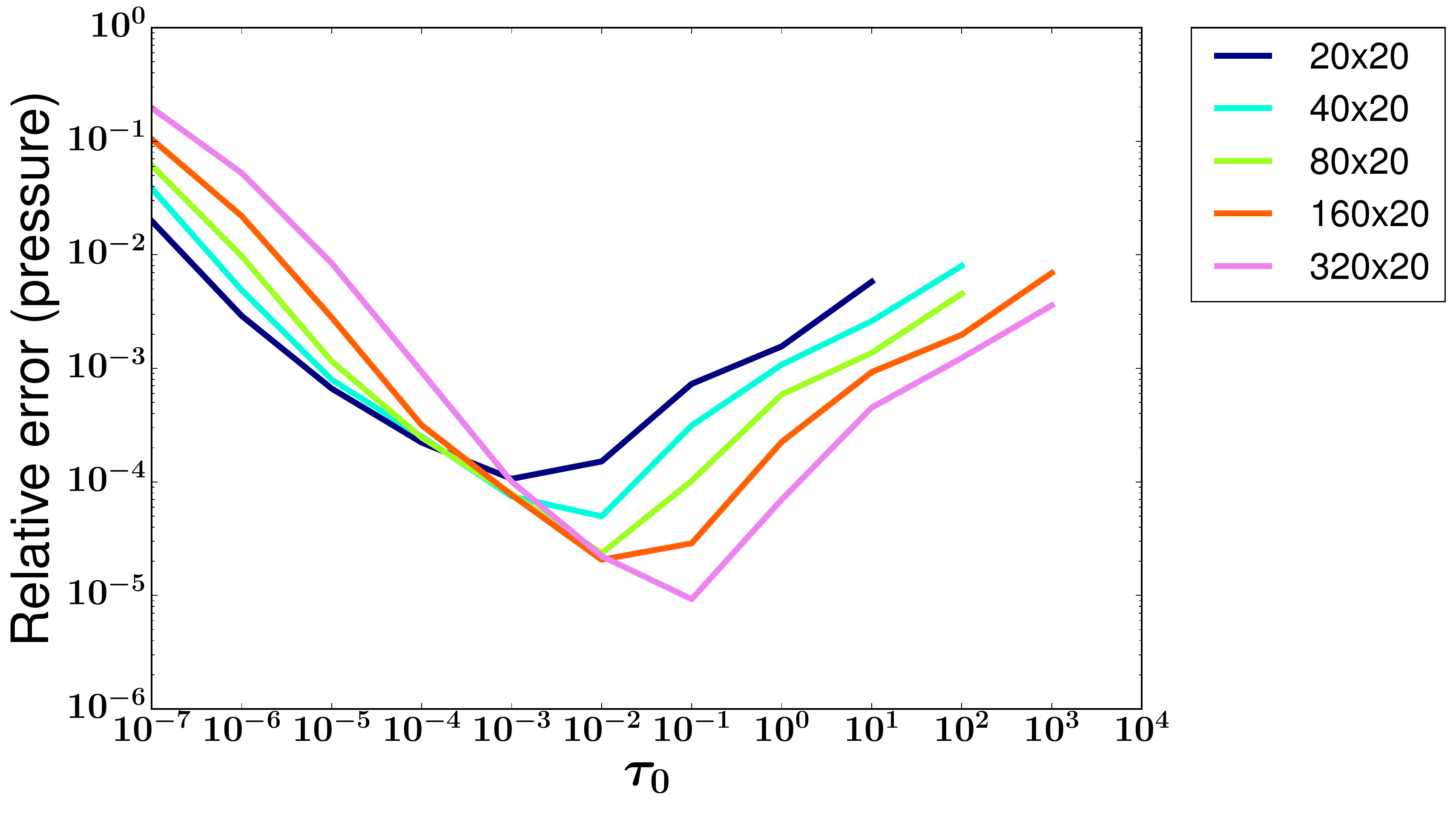}}\\
    \vspace{0.5em}
    \stackinset{l}{0.5cm}{b}{-0.1cm}{c)}{%
      \includegraphics[width=0.46\textwidth]{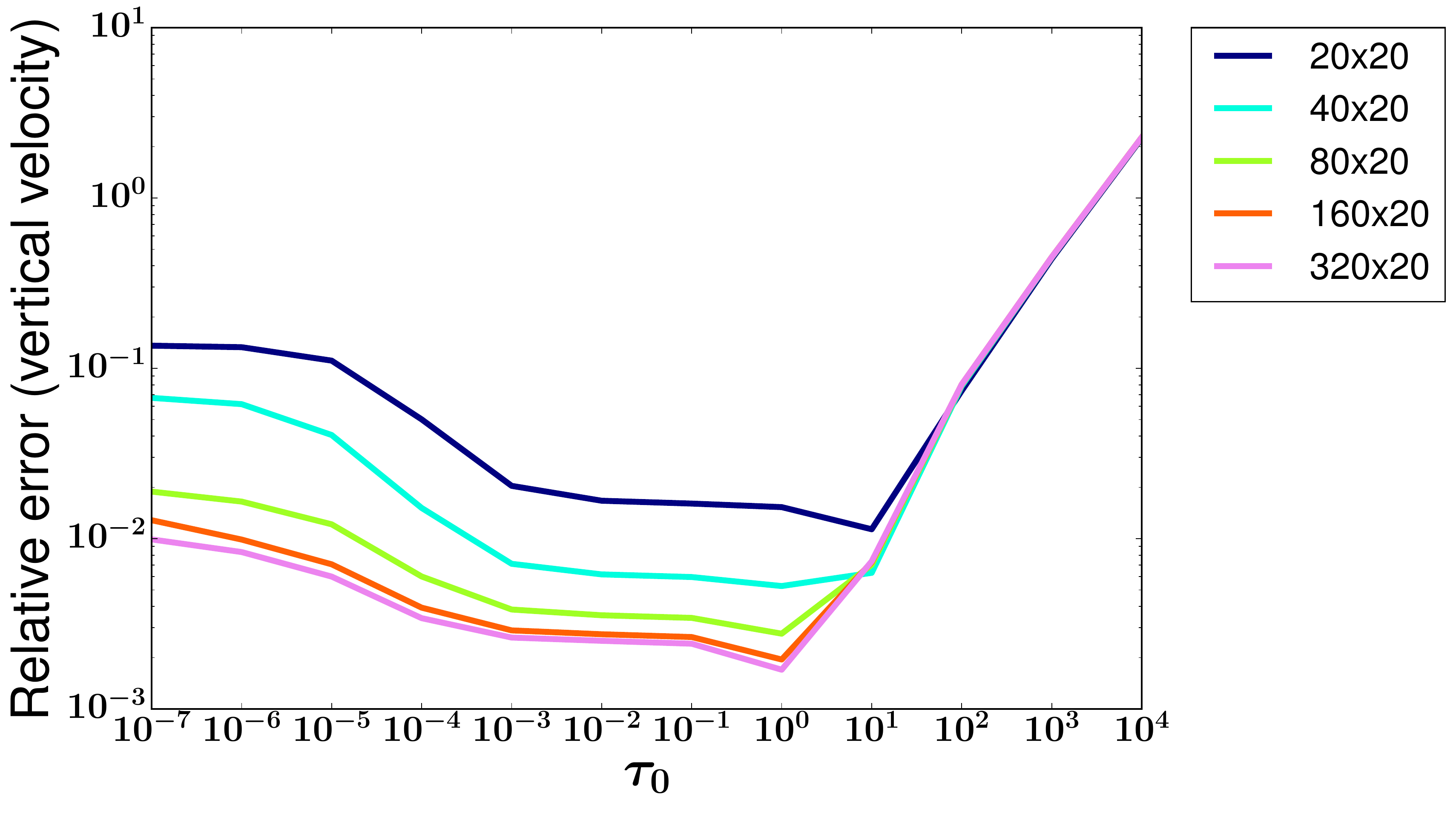}}
    \stackinset{l}{0.5cm}{b}{-0.1cm}{d)}{%
      \includegraphics[width=0.46\textwidth]{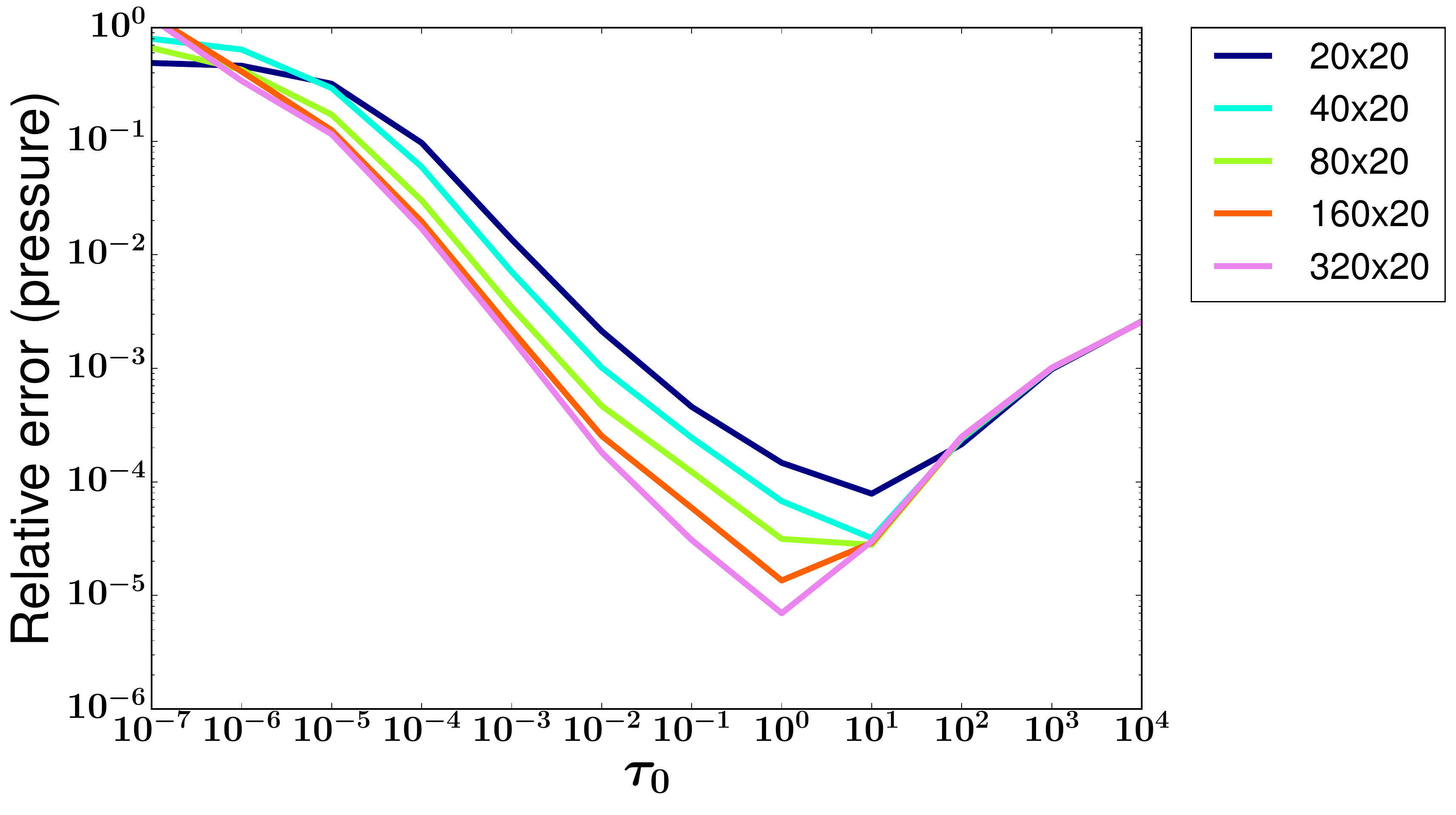}}\\
    \vspace{0.5em}
    \stackinset{l}{0.5cm}{b}{-0.1cm}{e)}{%
      \includegraphics[width=0.46\textwidth]{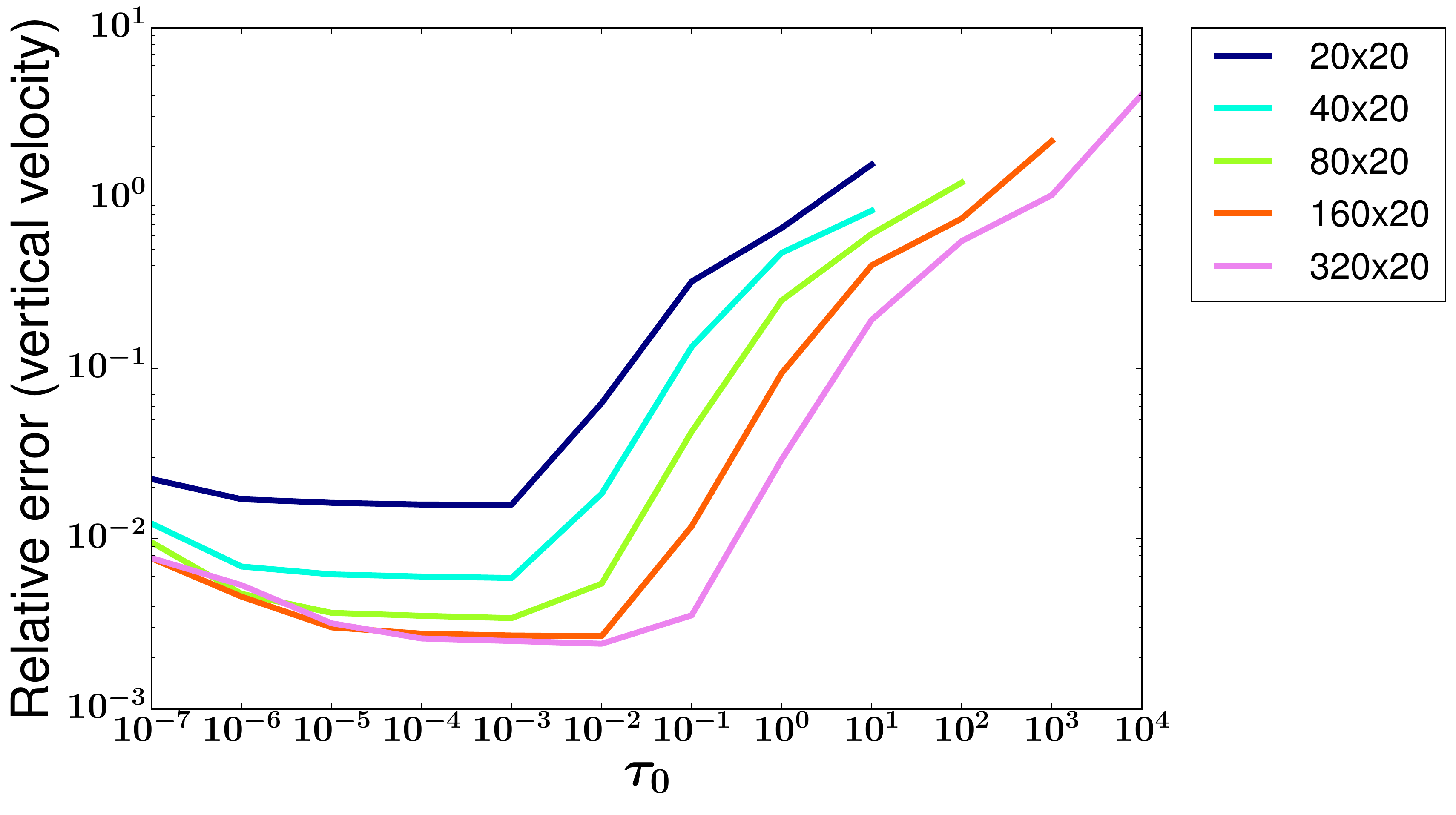}}
    \stackinset{l}{0.5cm}{b}{-0.1cm}{f)}{%
      \includegraphics[width=0.46\textwidth]{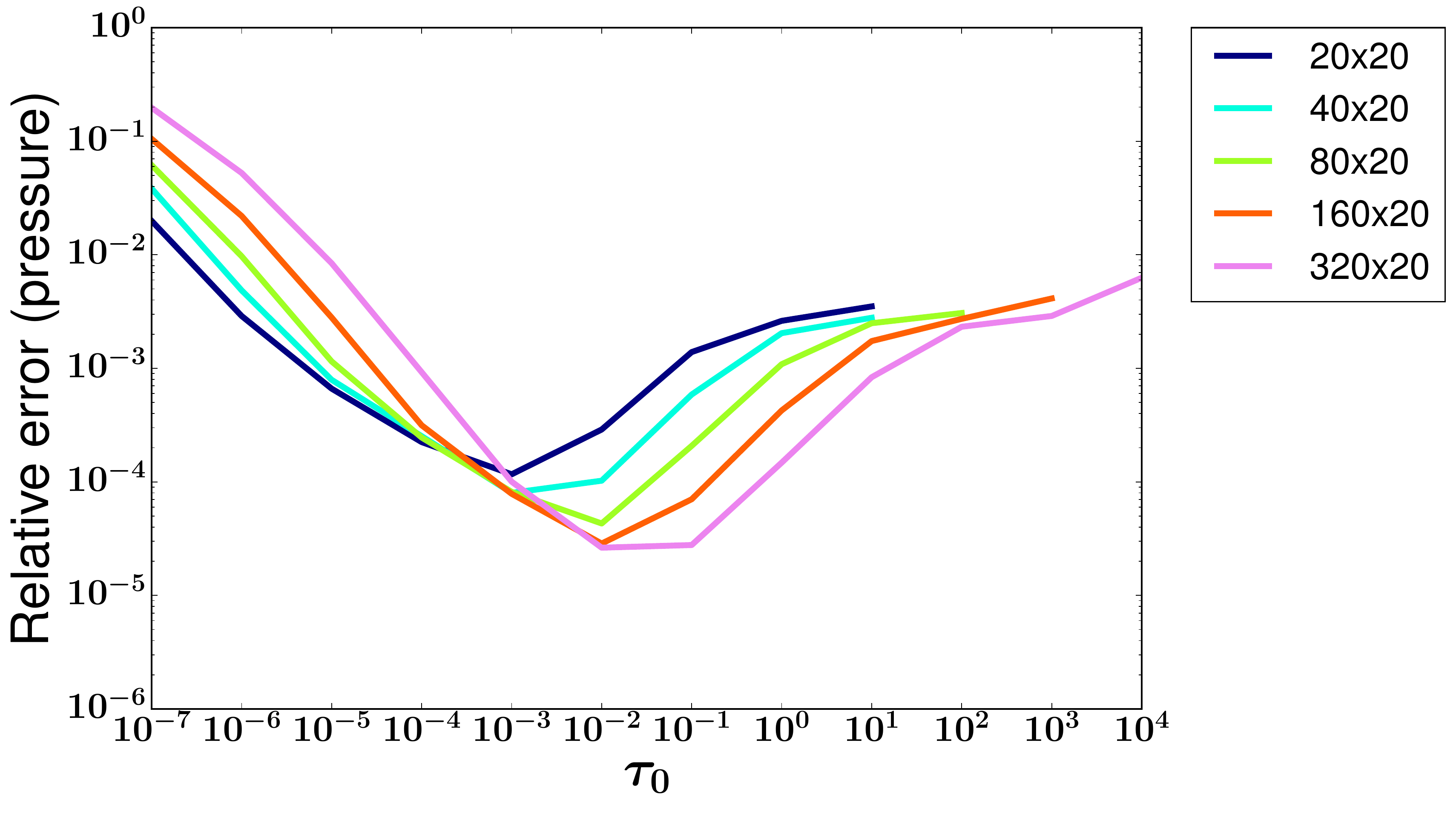}}\\
    \caption{Dependency of relative errors of vertical velocity and pressure on $\tau_{\mathtiny{GLS}}$ and the horizontal refinement of the ISMIP-HOM A domain using isotropic stabilization with $h_K=h_{\mathtiny{max}}$, $h_K=h_{\mathtiny{min}}$, or using an anisotropic formulation. The panels show vertical velocity (left) and pressure (right) for $h_K=h_{\mathtiny{max}}$ (a, b) for $h_K=h_{\mathtiny{min}}$ (c, d) and the anisotropic formulation (e, f).}
    \label{fig:horizontal_refinement}
\end{figure}

\paragraph{Results}
The results are shown in \cref{fig:horizontal_refinement}.  As we saw in previous sections, the default value of $\tau_{\mathtiny{GLS}}$ ($\tau_0=1)$ is appropriate on the ISMIP-HOM domain, and even optimal on lower aspect ratios. The value of $h_{\mathtiny{max}}$ is much larger than $h_{\mathtiny{min}}$, and the stability parameter needs to be decreased if $h_K=h_{\mathtiny{max}}$, in order to balance the larger element measure. In this case it is not clear how to choose an appropriate $\tau_0$ when no reference solution is available.

For $h_K=h_{\mathtiny{max}}$, a horizontal refinement of the mesh (\emph{decrease} in $h_{\mathtiny{max}}$) leads to an \emph{increase} in the optimal values of $\tau_0$, since the quadratic decrease in the size of $\tau_{\mathtiny{GLS}}$ with refinement is too rapid (\cref{fig:horizontal_refinement}a and \cref{fig:horizontal_refinement}b). Defining the cell size as $h_{\mathtiny{min}}$ leads to an \emph{decrease} in the optimal $\tau_0$ (\cref{fig:horizontal_refinement}c and \cref{fig:horizontal_refinement}d), since the size of  $h_{\mathtiny{min}}$ and thereby  $\tau_{\mathtiny{GLS}}$ is not affected by horizontal refinement. 

Clearly, to achieve best possible results, the stability parameter has to be tuned according to the mesh, regardless of cell size definition. This obviously makes it hard to find an optimal stability parameter on a mesh with varying cell size. More specifically, for $h_K=h_{\mathtiny{max}}$ this could affect the pressure variable by not achieving the increase of accuracy as the mesh is refined, while for $h_K=h_{\mathtiny{min}}$ it could result in slightly over-stabilizing the velocity and pressure in finer cells. It should be noted that the effects discussed above might be more severe for a more complex topography than the ISMIP-HOM experiment on a $80$ km long domain. 

The isotropic GLS stabilization clearly poses some problems on anisotropic meshes. However, as can be seen in \cref{fig:horizontal_refinement}e and \cref{fig:horizontal_refinement}f, the anisotropic GLS stabilization does not seem to be of any noteworthy improvement compared to the regular GLS method and due to this we choose to not discuss it further. We performed the same simulation using a linear flow law with similar results. 

\subsubsection{The Outlet Glacier Experiment}
\paragraph{Experiment Set-Up}
To investigate the impact of cell size definition for a more realistic scenario we here repeat the outlet-glacier simulation of \cref{sec:vialov} with $h_K=h_{max}$ instead of $h_K=h_{min}$. 
\begin{figure}[ht!]
\begin{center}
{\label{fig:Vialov_GLS_hmax}\includegraphics[width=0.75\textwidth]{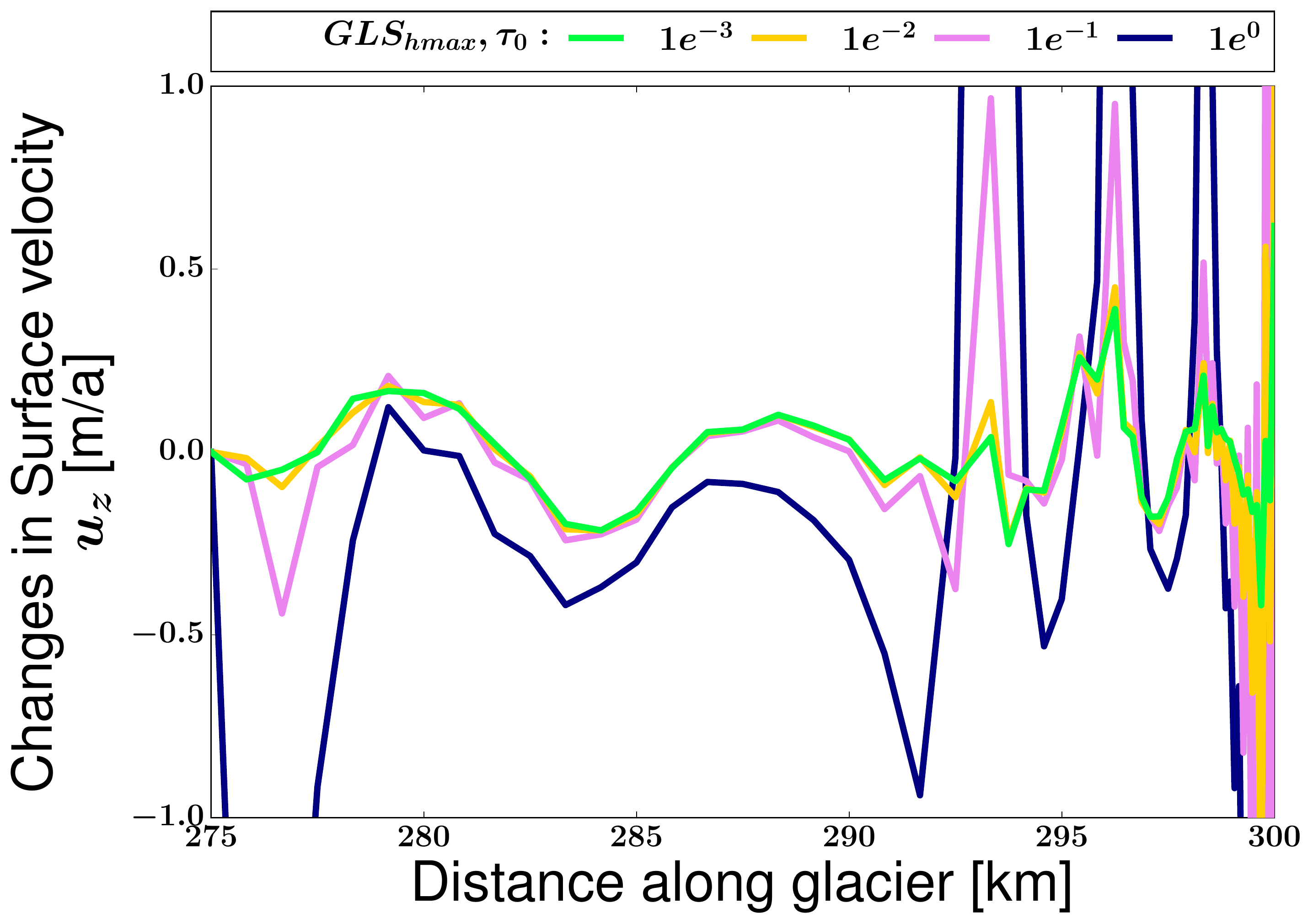}} \hspace{0.2cm}
\
\end{center}
\caption{Changes in vertical velocity for GLS for different values of $\tau_0$ with $h_K=h_{max}$, compared to an unstabilized velocity solution. The smallest value of $\tau_0$ shown is the smallest $\tau_0$ that ensured a stable pressure field \label{fig:Vialov_stabilization_hmax}} 
\end{figure}

\paragraph{Results}
The result is presented in \cref{fig:Vialov_stabilization_hmax}. As expected, the feasible values of $\tau_0$ are lower than for $h_K=h_{\mathtiny{min}}$ (compare to to $h_K=h_{\mathtiny{min}}$ in \cref{fig:Vialov_stabilization_hmin}). For the best choice of the stabilization parameter ($\tau_0 = \SIrange{e-4}{e-3}{}$), the accuracy at the front is comparable to the minimum edge length case, but increasing $\tau_0$ leads to over-stabilization more rapidly than for $h_K=h_{\mathtiny{min}}$.  Furthermore, it seems that the maximum edge length, even for moderate values of $\tau_0$, introduces larger changes than $h_{\mathtiny{min}}$ in the vertical velocity inland, far from the front and inlet.

\subsection{The Viscosity $\eta$}
The GLS stabilization in \citet{FrancaFrey1992} was developed for Newtonian fluids, where  $\eta$ is constant. In Elmer/Ice, the non-Newtonian nature of ice is accounted for by allowing the viscosity to vary also in the stabilization parameter. An alternative approach is to simply use a constant, approximation of the ice viscosity, $\eta_{lin}$, in the stabilization parameter. This approach has been used by \cite{BrinkerhoffJohnson2013}, where the authors reported that this choice resulted in better numerical stability. \cite{Hirn2011} however found that the convergence of the iterative solvers was better when the stability parameter varied with the viscosity.

\begin{figure}[ht!]
  \centering
    \stackinset{l}{0.5cm}{b}{-0.1cm}{a)}{%
      \includegraphics[width=0.46\textwidth]{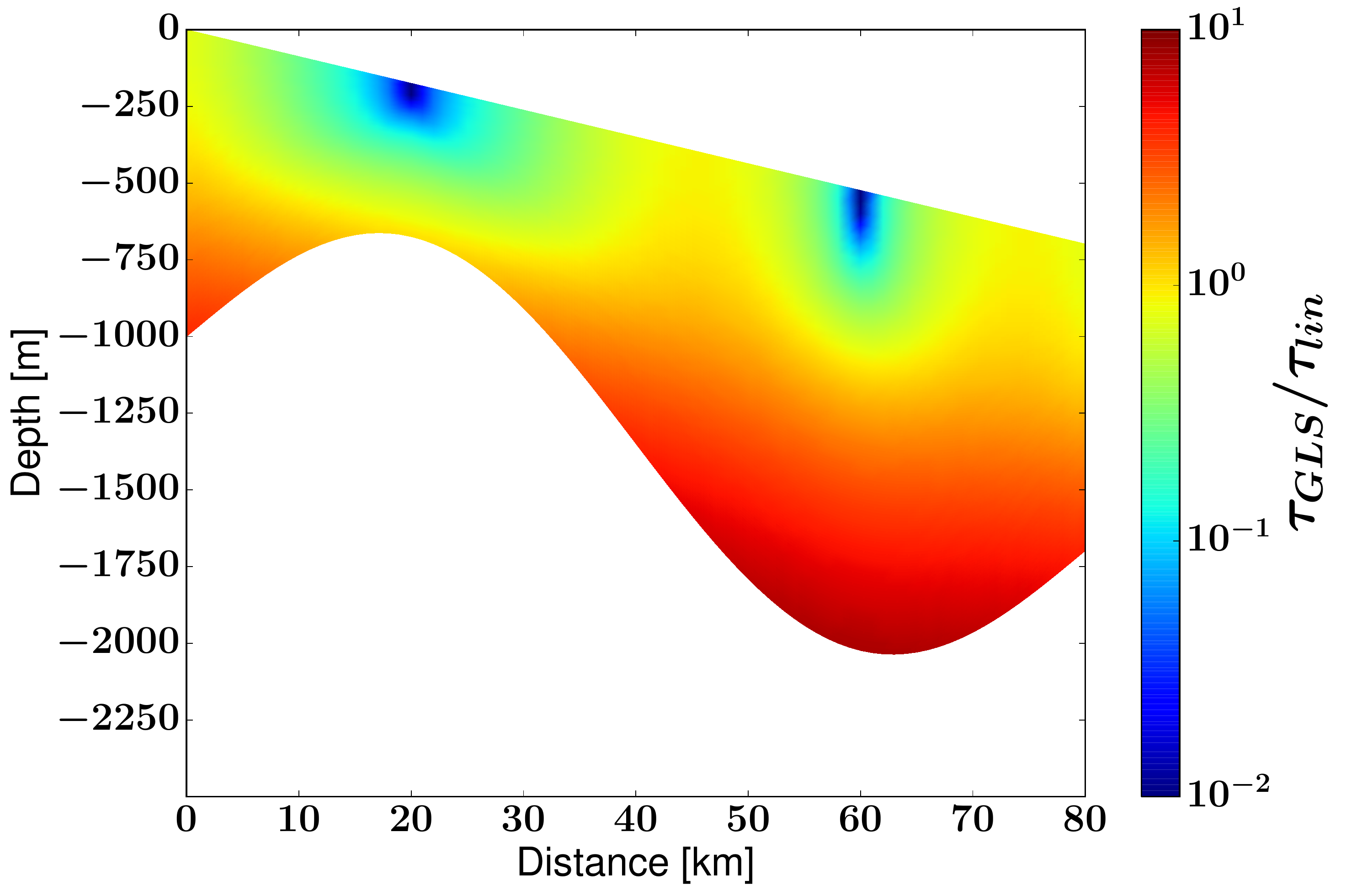}}
    \stackinset{l}{0.5cm}{b}{-0.1cm}{b)}{%
      \includegraphics[width=0.46\textwidth]{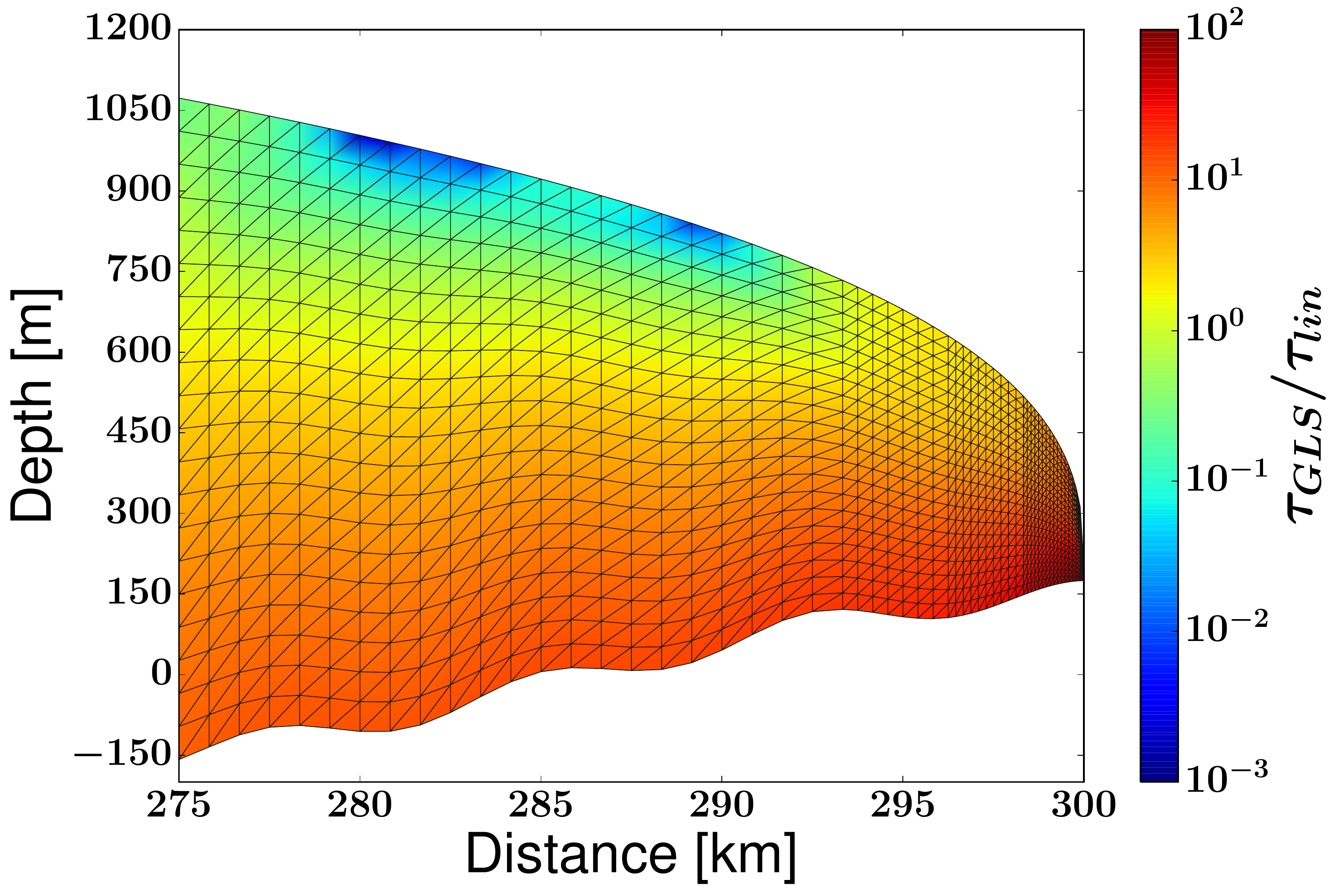}}\\
    \caption{Ratios of the stabilization parameters $\tau_{\mathtiny{GLS}}/\tau_{\mathtiny{lin}}$ for the domains used in this study.}
    \label{fig:stab_ratio}
\end{figure}

We denote the constant viscosity stabilization parameter by $\tau_{lin}$, using the common choice of $\eta_{lin} = \SI[retain-unity-mantissa = false]{e14}{\pascal\yr}$. \Cref{fig:stab_ratio} shows the ratio of $\tau_{\mathtiny{GLS}}/\tau_{lin}$ for the ISMIP-HOM and outlet glacier domain. The magnitude of $\tau_{\mathtiny{GLS}}$ is an order of magnitude higher at the bed, but perhaps more importantly, $\tau_{\mathtiny{lin}}$ is two (ISMIP-HOM) or three (outlet glacier) orders of magnitude higher than $\tau_{\mathtiny{GLS}}$ at points close to the surface where deformation is low and viscosity is high. This may lead to over-stabilization and thereby high errors in the vertical velocity. Since it occurs at the surface this will effect the accuracy of the free surface position. Further comparison studies between the two methods are need but out of the scope of this work.

\section{Alternative Stabilization Techniques} 
\label{sec:alternative_stab}
In this section we compare GLS to other types of stabilization methods. We will mainly focus on how the pressure and and in particular vertical surface velocities change with the size of the stabilization parameter. The implemented methods include a Pressure Penalty method (PP), an Interior Penalty (IP), Pressure-Global-Projection (PGP), and a linear Local-Pressure-Stabilization (LPS). After a brief description of these alternative stabilization methods, we repeat the stabilization parameter sensitivity experiment 
from \cref{sec:tau}. Based on the performance of each method, we select the IP and PGP methods to redo the outlet glacier and finally examine how the IP method performs in the non-linear Poiseuille flow.

The stabilization methods in this section have all been implemented in \fenics{}, which allows for easy implementation of new forms. The elements are triangles (for 2D) or tetrahedrons (for 3D). Since second order derivatives are identically zero on such elements we have omitted them in the descriptions below. In all of the below implemented stabilization methods, we use the constant viscosity stabilization parameter $\tau_{\mathtiny{lin}} = \tau_0\frac{m_K h_K^2}{8\eta_{lin}}$, using either $h_K = h_{min}$ or $h_K = h_{max}$.

\subsection{Description}

\subsubsection{Pressure Penalty method}
This minimalistic stabilization method is penalizing pressure gradients by adding the stabilization term
\begin{equation}
  \label{eq:pp}
  -\sum_{K\in T_h}\tau_{\mathtiny{lin}} (\nabla p_h , \nabla q_h )_{K}.
\end{equation}
The method is very easy to implement and has been used for glaciological applications in \citep{Zhang2011}. 

\subsubsection{Interior penalty method}
In \citep{BurmanHansbo2006} the authors penalized pressure gradients on the interior edges of the domain to stabilize the linear Stokes equations. The stabilization term here is
\begin{equation}
  \label{eq:ip}
  -\sum_{K\in T_h}\tau_{\mathtiny{IP}} \int_{\partial K} h_{K}^{s+1} [n\cdot\nabla p_h][n\cdot\nabla q_h] ds,
\end{equation}
where $[\cdot]$ indicates the jump over an interior edge $\partial K$, and we have defined $\tau_{\mathtiny{IP}} = \tau_0\frac{m_K}{8\eta_{lin}}$ have a user parameter of comparable size to the other stabilization methods.
   
The parameter $s$ depends on how the viscosity relates to a characteristic mesh size $h$ and should be $s=2$ if $\eta\geq h $ and $s=1$ otherwise. To simplify we have  chosen to set $s=2$ everywhere. Due to the large ice viscosity this should be appropriate in most areas and was true when comparing the viscosity of the solved problems to their mesh size.

In a study of Poiseuille flow for a Newtonian fluid, \citet{BurmanHansbo2006} show that this method does not introduce artificial boundary conditions as strongly as the GLS method does.

\subsubsection{Pressure-Global-Projection method}
In \citep{CodinaBlasco1997} the gradient of the pressure is projected onto the velocity space, resulting in an augmented system:
\begin{equation}
\label{eq:PGP} 
\begin{aligned}
&A(\mathbf{u}_h, \mathbf{v}_h) + B(\mathbf{v}_h, p_h) = F(\mathbf{v}_h) \quad \forall \mathbf{v}_h \in V_h,\\
&B(\mathbf{u}_h, q_h) + \tau_{\mathtiny{lin}} (\nabla p_h - \pmb{\xi}_h, \nabla q_h) = 0 \quad \forall q_h \in Q_h,\\
&(\nabla p_h, \pmb{\eta}_h) - (\nabla \pmb{\xi}_h, \pmb{\eta}_h) = 0 \quad \forall \pmb{\eta}_h\in \mathring{V}_h,
\end{aligned}
\end{equation}
where $\mathring{V}_h$ is the same space as the velocity space but without boundary conditions. The method introduces an additional variable, $\pmb{\xi}$, and therefore increases the computational cost.

\subsubsection{Local Projection Stabilization method}

Related to the pressure projection method, is a method by \citet{BeckerBraak2001} where the gradient of the pressure is projected onto a space of piece-wise constants defined on a coarser mesh. In the coarse mesh every element $M$ is a patch consisting of elements $K$ from the original mesh. The stabilization term is
\begin{equation}
\tau_{lin} (\nabla p_h - \overline{\nabla p_h}, \nabla q_h) = 0 \quad \forall q_h \in Q_h
\label{eq:lps}
\end{equation}
where $\overline{\nabla p_h}$ is the onto the coarse mesh projected pressure gradient.
Compared to the GLS, the LPS reduces the coupling between pressure and velocity but the stencil for the pressure variable increases due to the evaluation of the projected gradient on the coarser mesh.

In this study, each $M$ is subdivided into 16 triangles with similar properties as the parent patch. To avoid the larger stencil for the pressure, we use the pressure solution from the previous Newton iteration to solve for the projected pressure gradient. 

One of the benefits of the LPS method is that there exists a well-studied adaptation to the $\mathfrak{p}$-Stokes equations \citep{Hirn2012}, which is advantageous for three dimensional simulations or low regularity data. The method has not been implemented in \fenics{} or Elmer/Ice for the current study, but is recognized as a potential option to investigate in future studies.

\subsection{Sensitivity to Stability Parameter on the ISMIP-HOM domain}
\label{sec:alternative_stabilization_accuracy} 

\begin{figure}[ht!]
  \centering
    \stackinset{l}{0.5cm}{b}{-0.1cm}{a)}{%
      \includegraphics[width=0.46\textwidth]{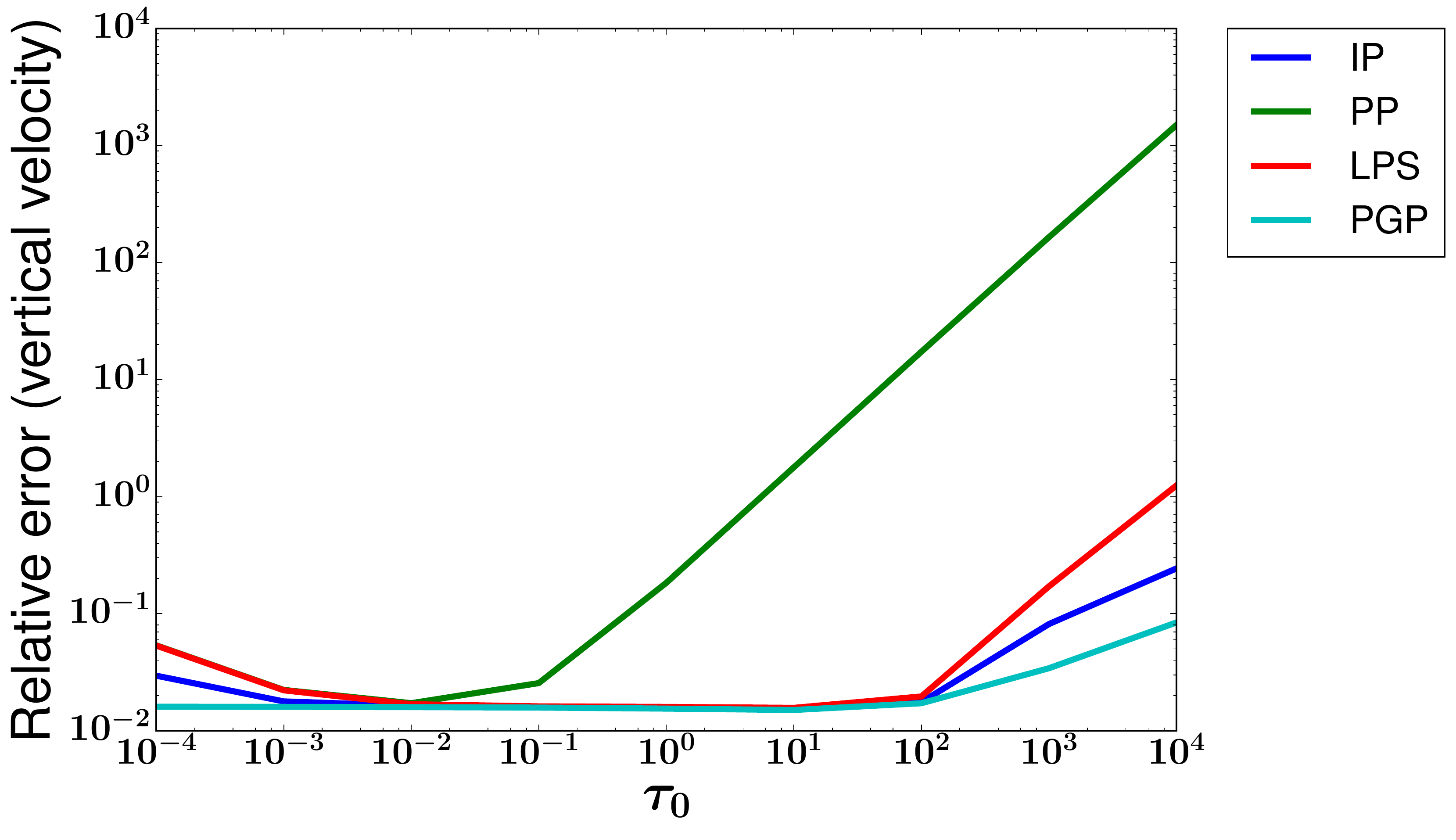}}
    \stackinset{l}{0.5cm}{b}{-0.1cm}{b)}{%
      \includegraphics[width=0.46\textwidth]{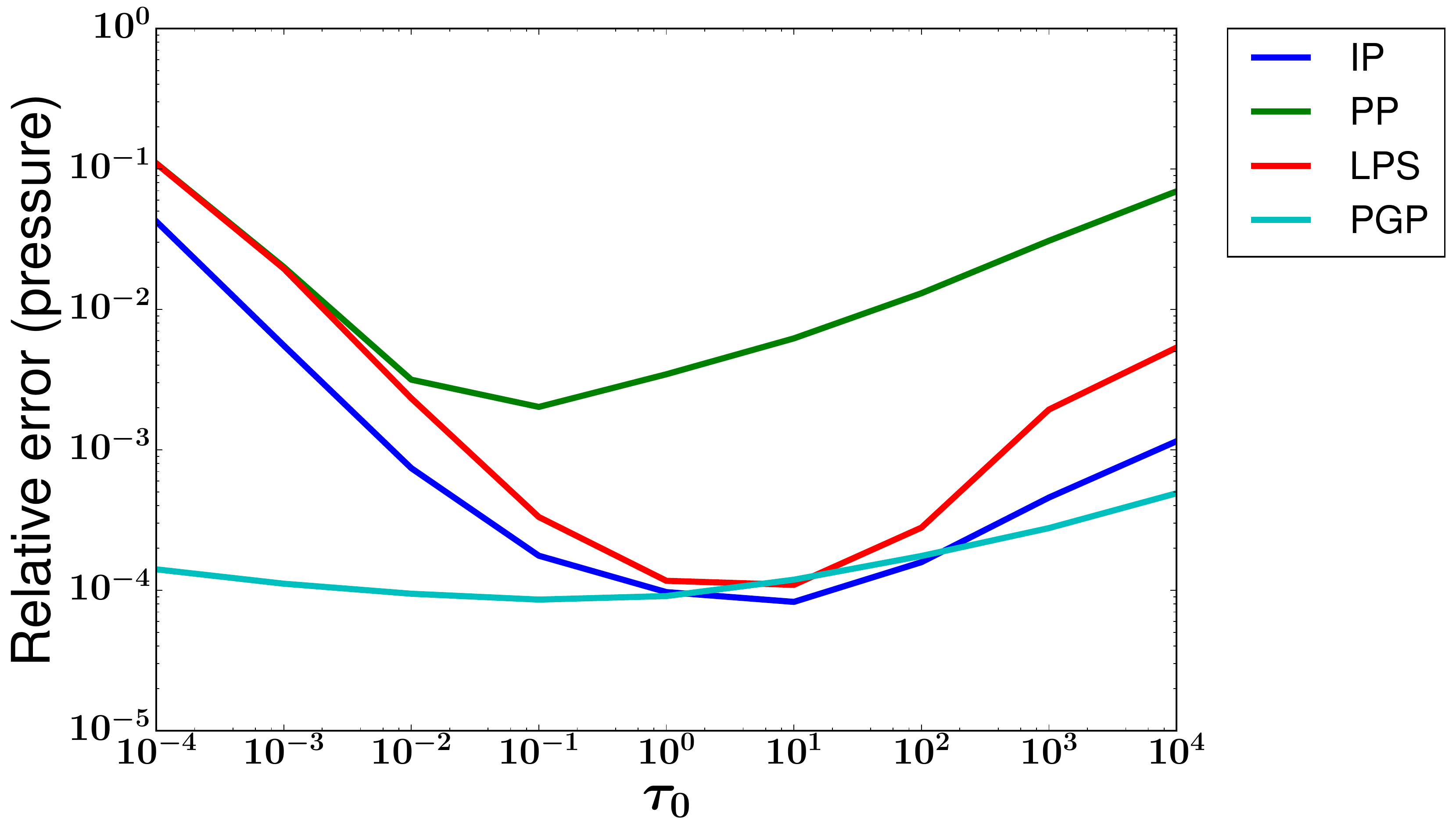}}\\
    \caption{The accuracy of alternative stabilization methods, and their sensitivity to the parameter $\tau_0$. The $L^2$-norm of the relative errors in vertical velocity and pressure are computed using a numerical reference solution.}
    \label{fig:fenicsismiptausensitivity}
\end{figure}

For the ISMIP-HOM domain we use the same type of mesh resolution and cell size definition as in \cref{sec:ismipgls}, i.e $20\times 20$ and $h_K = h_{min}$. As can be seen in \cref{fig:fenicsismiptausensitivity}, the relative errors for the LPS method behave almost identically to that of GLS (compare \cref{fig:fenics_tausensitivity}b). The PP method shows a much narrower span for both the pressure and velocity, and results in high pressure errors for all $\tau_0$. Both the IP and PGP methods have a wide span and a low relative error for the vertical velocity, even for high values of $\tau_0 = \SIrange{e3}{e4}{}$. PGP has the lowest relative errors for both pressure and velocity over the largest range of $\tau_0$.  The PGP and IP methods both performed very well under even more extreme conditions, with $\tau_0=\num[retain-unity-mantissa = false]{e4}$ giving relative errors in velocity of $\num[retain-unity-mantissa = false]{e-1}$.

We also performed the horizontal mesh refinement experiment with all of the above stabilization methods. The results were qualitatively similar to the results obtained with GLS and both the PGP and IP methods performed well over a wide span of $\tau_0$.

\subsection{The Outlet Glacier}
\label{sec:vialov_alternative}

\begin{figure}[ht!]
  \centering
    \stackinset{l}{0.5cm}{b}{-0.1cm}{a)}{%
      \includegraphics[width=0.46\textwidth]{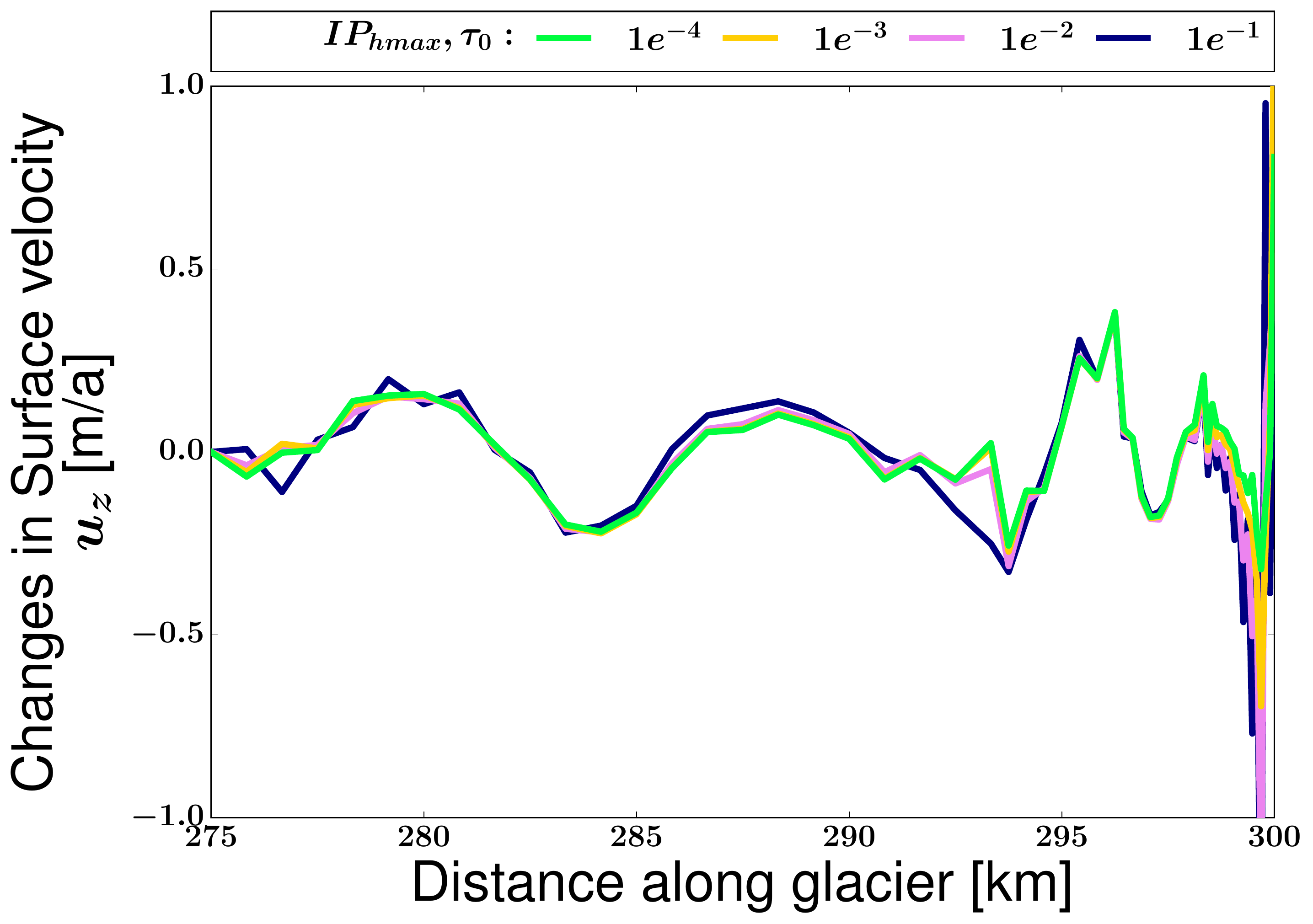}}
    \stackinset{l}{0.5cm}{b}{-0.1cm}{b)}{%
      \includegraphics[width=0.46\textwidth]{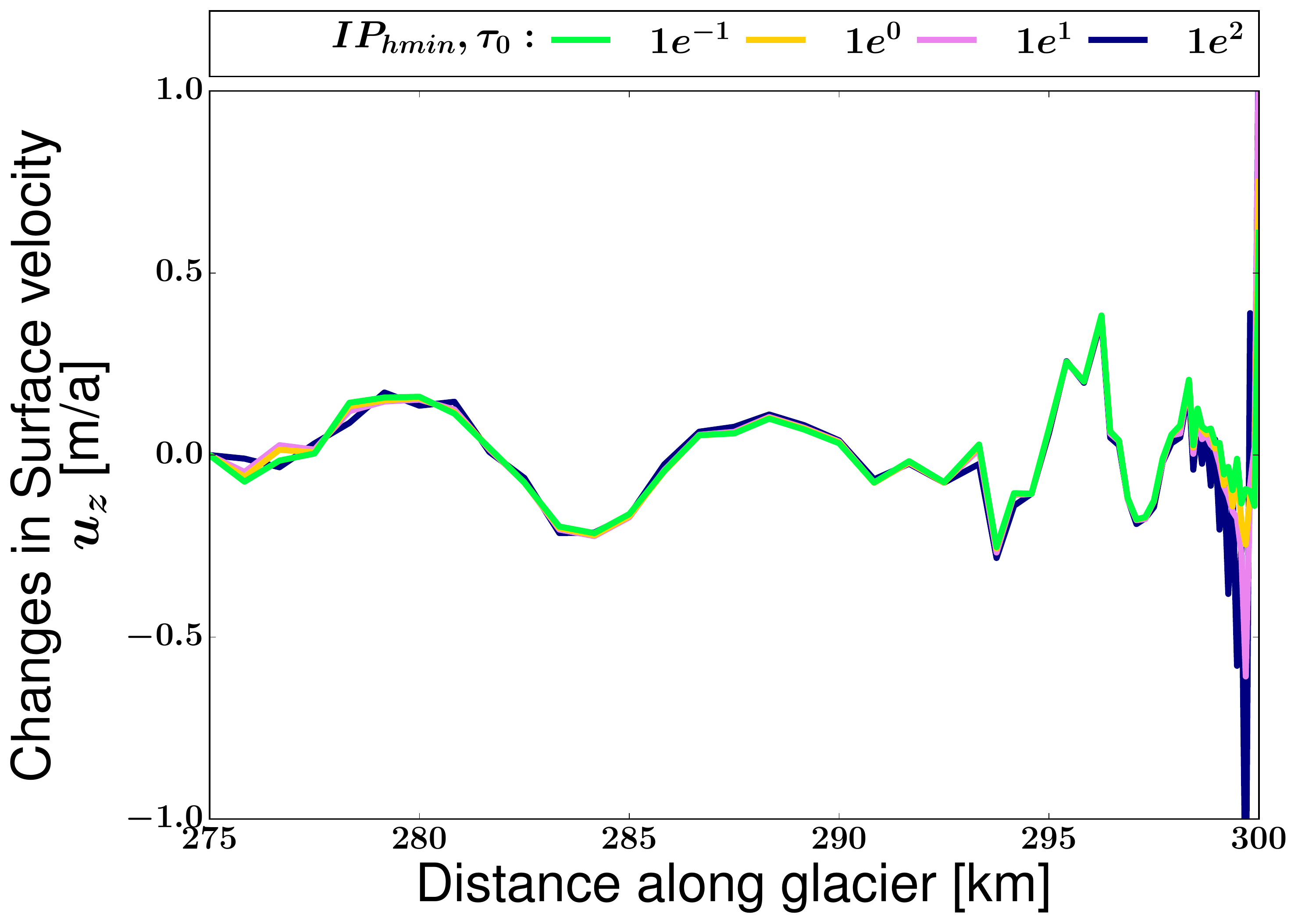}}\\
    \vspace{0.5em}
    \stackinset{l}{0.5cm}{b}{-0.1cm}{c)}{%
      \includegraphics[width=0.46\textwidth]{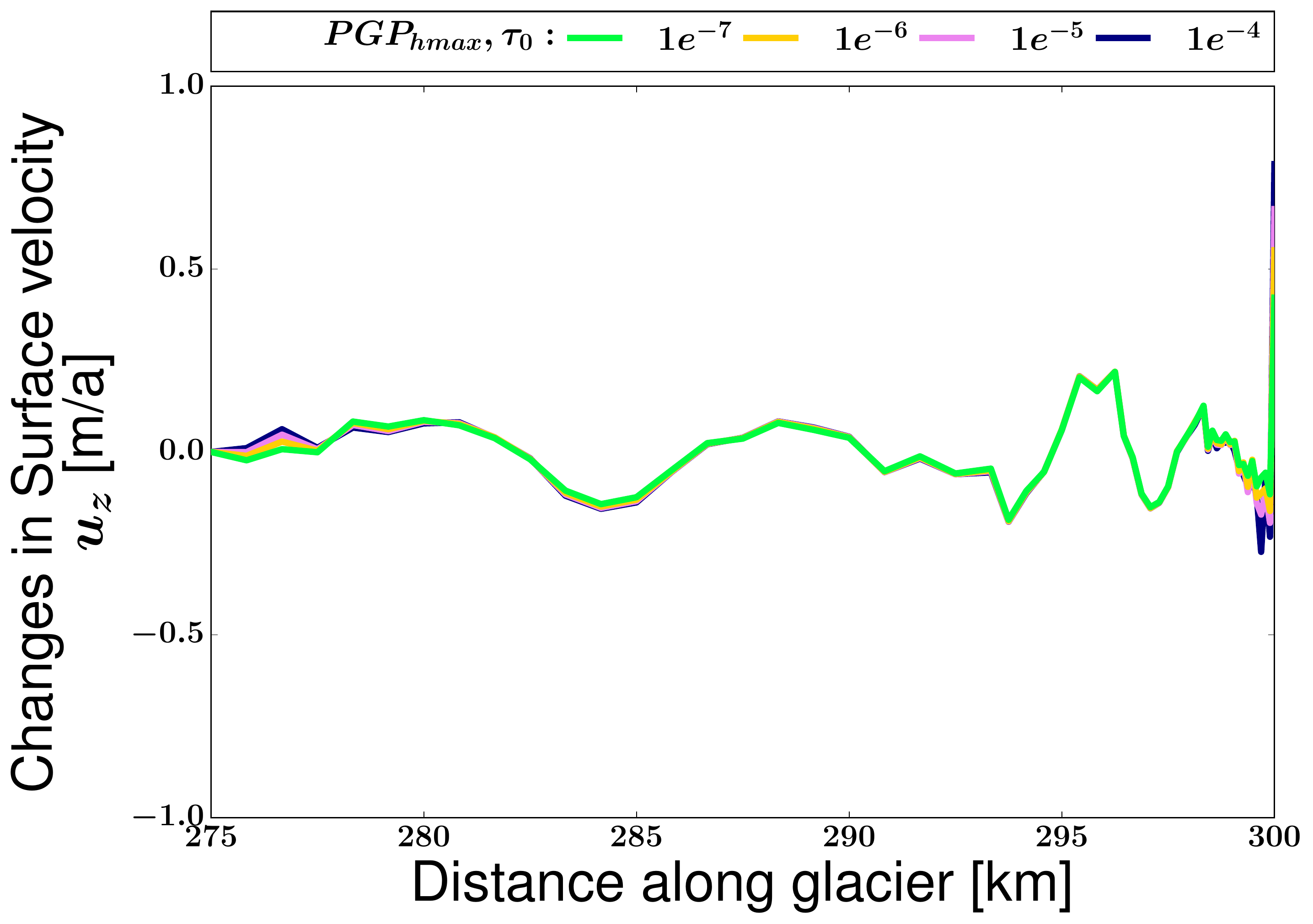}}
    \stackinset{l}{0.5cm}{b}{-0.1cm}{d)}{%
      \includegraphics[width=0.46\textwidth]{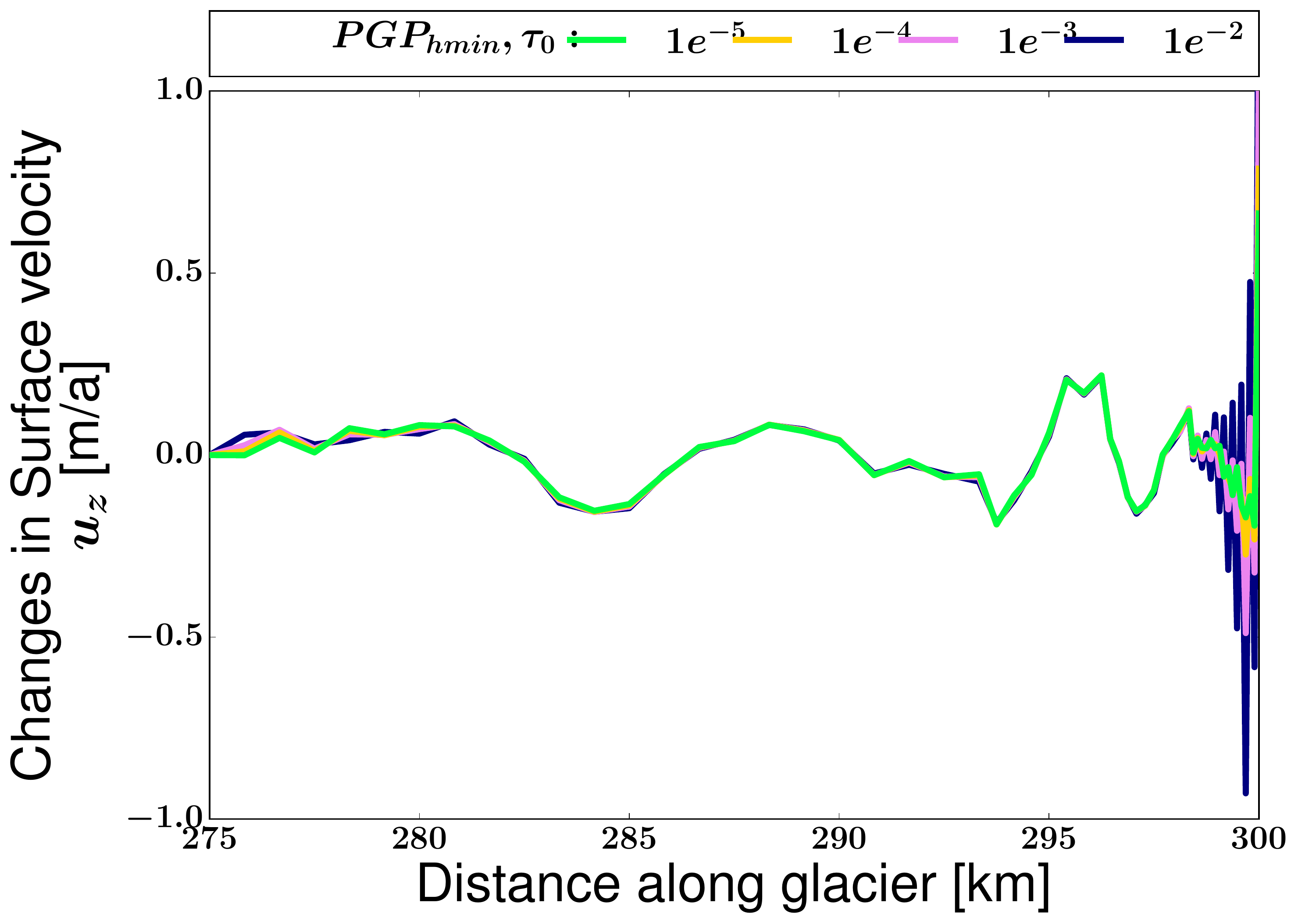}}\\
    \caption{Changes in vertical velocity for IP (top panel) and PGP (bottom panel) with different values of $\tau_0$, compared to an unstabilized velocity solution. The values of $\tau_0$ in each plot have been chosen such that any smaller values produced visible oscillations in the pressure field. For each method the left and right panels show $h_{\mathtiny{max}}$ and $h_{\mathtiny{min}}$ respectively.}
    \label{fig:Vialov_stabilization_alternative}
\end{figure}

The results from previous section indicate that the IP and PGP methods are accurate for a wide span of stabilization parameters. We therefore choose them to make further investigations, and here repeat the outlet glacier experiment from \cref{sec:vialov}. \Cref{fig:Vialov_stabilization_alternative} shows the changes in the vertical surface velocities when compared to an unstabilized, non-oscil\-latory, velocity field, for both IP and PGP, using both  $h_K=h_{\mathtiny{max}}$ and  $h_K=h_{\mathtiny{min}}$.  The stabilization parameters are ranging four orders of magnitude; the smallest stability parameter shown is the smallest value needed to stabilize the pressure.

The range of possible stability parameters for these two methods are substantially narrowed compared to the ISMIP-HOM case of \cref{sec:alternative_stabilization_accuracy}, because of the inlet boundary condition. However, as for the ISMIP-HOM problem, the IP and the PGP methods are more robust than GLS with regards to the stabilization parameter. In the interior the effects of over-stabilization is much less severe than for the GLS method, and also the frontal oscillations are reduced.

\subsection{Poiseuille Flow}
As a final experiment we examine how the IP method compares to the GLS method for the Poiseuille flow experiment. We choose the IP method due to its fairly low computational cost, which is comparable to the GLS method. This is a significant benefit compared to the PGP method. Similarly as for the GLS method, we here choose $h_K$ to be the maximum cell edge.

\begin{figure}[ht!]
  \centering
    \stackinset{l}{0.5cm}{b}{-0.1cm}{a)}{%
      \includegraphics[width=0.46\textwidth]{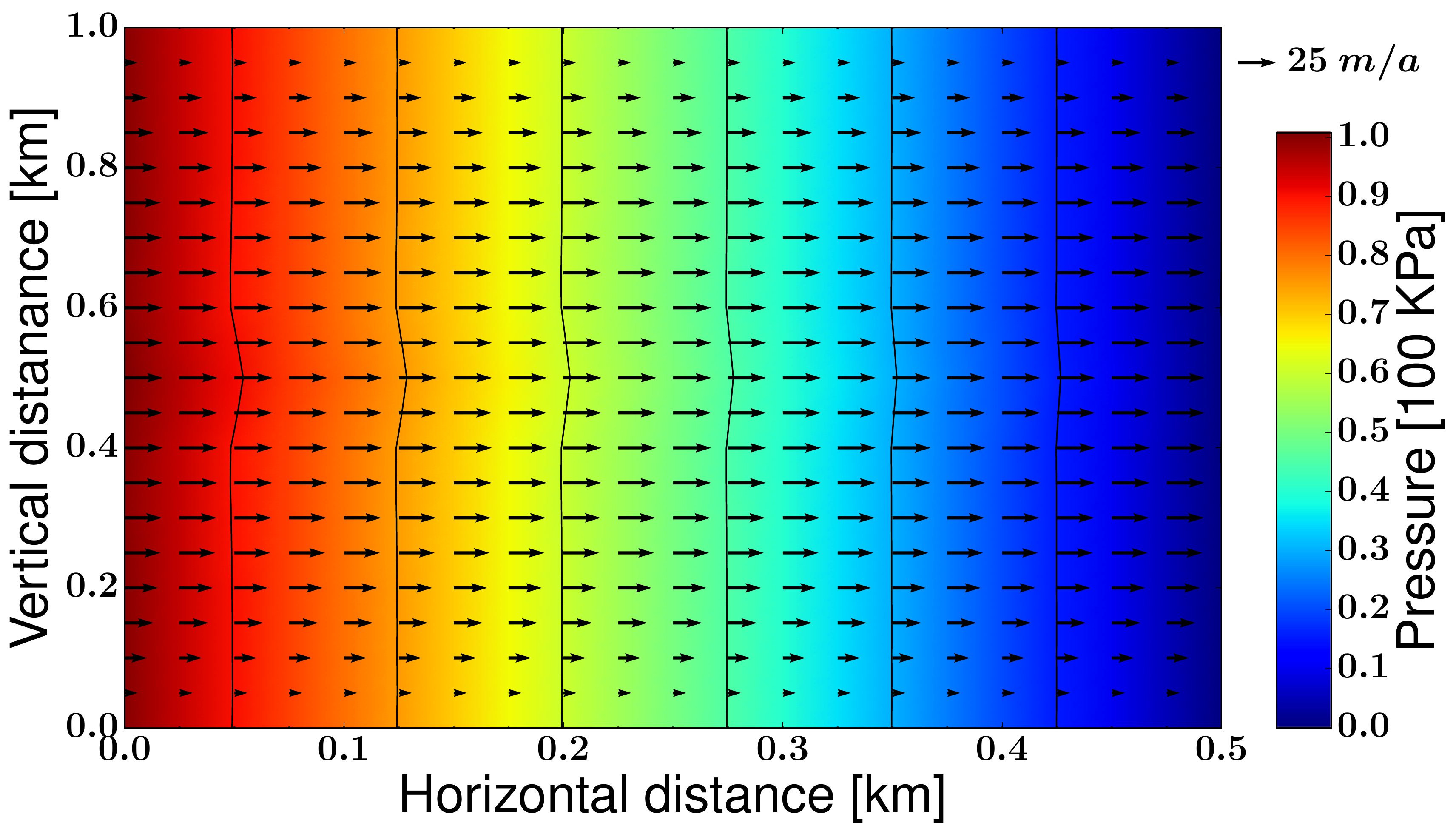}}
    \stackinset{l}{0.5cm}{b}{-0.1cm}{b)}{%
      \includegraphics[width=0.46\textwidth]{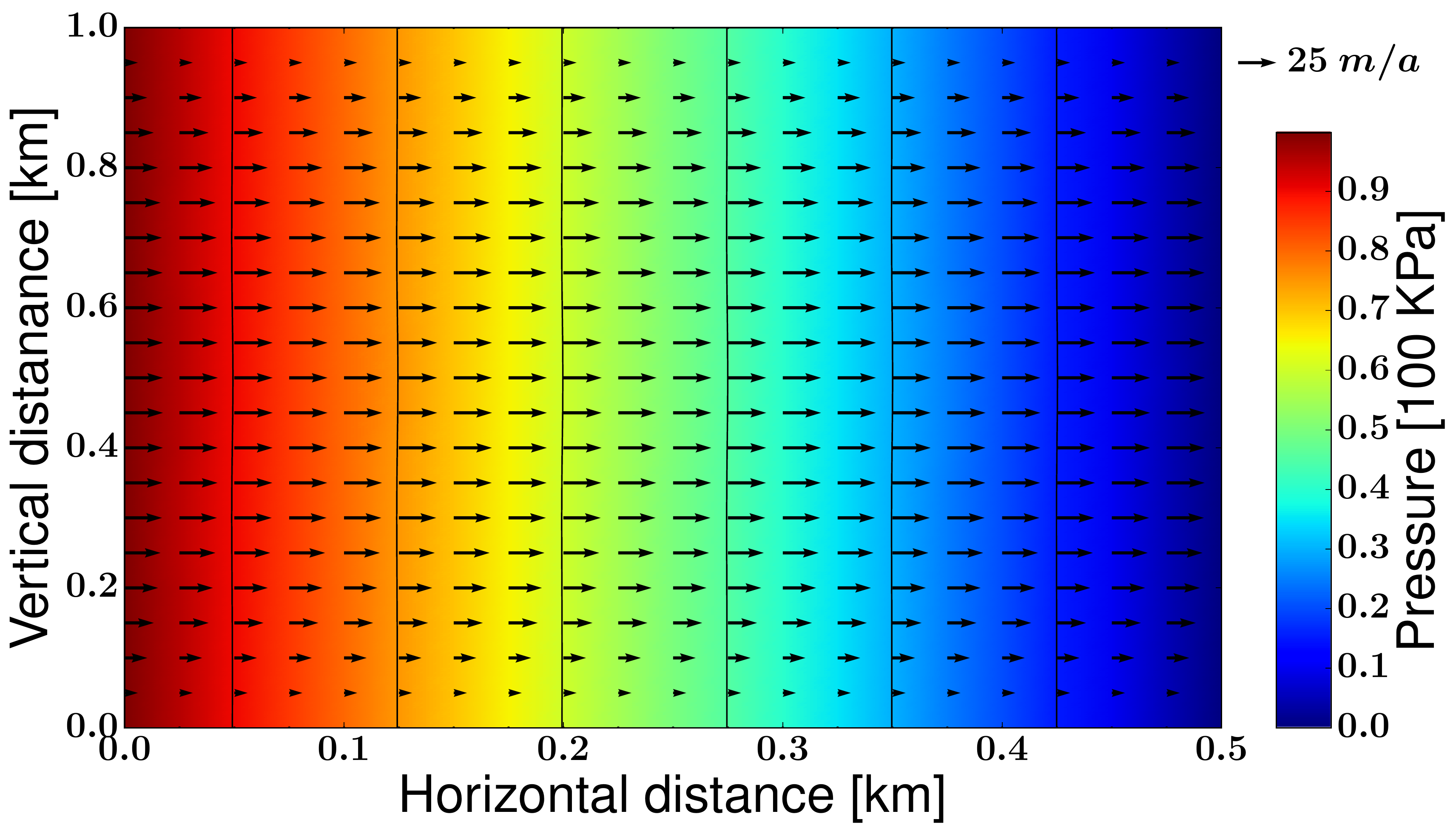}}\\
    \caption{Poiseuille computed with IP stabilization. The stabilization parameter is specified with $\tau_0=1$ (left column) or $\tau_0=100$ (right column). Velocity is depicted as \emph{black arrows} and pressure as \emph{color} with \emph{black lines} as pressure contours.}
    \label{fig:poiseuille_IP}
\end{figure}

Compared to the GLS method in \cref{fig:poiseuille_GLS}a, the curvature of the pressure contours in the upper left and lower right is not as evident in the IP method, see \cref{fig:poiseuille_IP}a.  Like for the ISMIP-HOM problem and the outlet glacier case, the IP method is not prone to over-stabilization (\cref{fig:poiseuille_IP}b). The pressure artifacts due to the singularity is present also for the IP method, although they are much smaller and disappear for a high stabilization parameter, and for large values of $\tau_0$ the method converges to the analytical solution, even at the point of singularity (compare \cref{fig:poiseuille_GLS}b and \cref{fig:poiseuille_IP}b).

\section{Summary and Conclusions}

We have investigated the accuracy and robustness of the commonly used GLS stabilized equal (first) order methods for the non-linear $\mathfrak{p}$-Stokes equations with application to ice sheet modeling. We have done so keeping a common goal in glaciology in mind - accurate ice volume predictions. Furthermore, we have compared the GLS method with other available stabilization techniques, namely a Pressure Penalty method, the Interior Penalty method, the Pressure-Global-Projection method, and the Local Projection Stabilization. We have mainly been working with the FEM code \fenics{} in the numerical experiments, but to reduce the risk of erroneous implementations and measuring techniques, and to utilize some special functionalities, we have complemented some experiments with Elmer/Ice. 

We used both a numerical reference solution and the manufactured solutions presented in \citet{Leng2013} to compute errors. We chose to follow earlier studies and measured the errors in the $L^2$ norm instead of the more natural $L^{\mathfrak{p}/(\mathfrak{p}-1)}$ and $W^{1,\mathfrak{p}}$ norm. When comparing with the numerical reference solution on the two-dimensional ISMIP-HOM B problem, a second order convergence in both \fenics{} and Elmer/Ice was observed. Using the manufactured solutions the convergence rate is two in \fenics{} and three in Elmer/Ice. The surprising third order convergence was found also in \cite{Gagliardini2013}, regardless of using stabilized lower order elements or unstabilized higher order elements. We believe the explanation for this is high artificial surface forces at singularities in the manufactured problem. The manufactured problem is very sensitive to the implementation of the singular viscosity and artificial surface forces. This results in a super-convergence when refining the mesh in the vertical direction (but not in the horizontal plane) in Elmer/Ice. The strong artificial forces also results in a misrepresentation of the distribution of errors in the ice domain, and overestimates the error in the vertical velocity component. \textit{We therefore recommend great caution if using the manufactured solutions for evaluating the accuracy of ice sheet models.}  We worked with the numerical reference solution in the remainder of the experiments of the paper, except for in one experiment.

The errors of the $\mathfrak{p}$-Stokes equations propagate into the calculations of the ice surface position. \textit{Due to the form of the ice surface evolution equation, the relevant errors to control are the errors in the vertical surface velocity component. These errors accumulate in time.}

\textit{Unfortunately, the vertical velocity is the variable that is the most sensitive to over-stabilization}, and it is thus important to choose an appropriate GLS stabilization parameter to accurately calculate the ice surface position. Such over-stabilization may happen if the parameter pre-multiplying the term in the GLS formulation is not defined appropriately. The classical stabilization parameter as suggested in \citet{FrancaFrey1992} depends on the finite element cell size and the implementation of viscosity.   \textit{If the cell size is chosen as the minimum edge length and the viscosity is allowed to vary non-linearly in the stabilization parameter, the value recommended in \citet{FrancaFrey1992} is appropriate for the simple ISMIP-HOM problem}. However, the optimal value of the stability parameter  does depend on the element aspect ratio, and a horizontal mesh refinement may therefore not always give the expected gain in accuracy. \textit{On a more realistic outlet glacier simulation with inlet velocity conditions, the value from \citet{FrancaFrey1992} is unfortunately too large and results in velocity oscillations at the glacier front.} The GLS method also introduce artificial boundary conditions for ice flow. Hypothetically, this could cause inaccuracies in region with hydrodynamic pressure components, such as at the grounding line, but further investigations are needed to determine if this is an important issue in realistic simulations. 

Out of the alternative stabilization methods tested, the simple Pressure Penalty was more sensitive to the stabilization parameter than GLS and was also less accurate. The (linear) Local Projection Stabilization was similar to GLS in terms of sensitivity to the stability parameter and accuracy, but is more complicated to implement. It is possible that the non-linear version presented in \citet{Hirn2012} would improve its properties. Both the Pressure-Global-Projection method  and the Interior Penalty method showed very good stability results on the ISMIP-HOM problem. On the more realistic outlet glacier simulation the range of viable stability parameter values narrowed for these methods, although it was still wider than for the GLS method and less prone to suffer from over-stabilization. Due to its high computational cost we do however not recommend the Pressure-Global-Projection method for ice sheet simulations, and \textit{the most attractive alternative method is thus the Interior Penalty method}. This method was also less prone to introduce artificial boundary conditions and handle singularities in the viscosity.

\textit{In summary, standard equal order GLS stabilized methods perform well on simplistic problems, but for more realistic glaciological problems inaccuracies in the vertical velocities may become large and affect the accuracy of the free surface position and thereby ice volume predictions. Care should therefore be taken in choosing the stability parameter as recommendations from literature may not automatically guarantee good results. A more robust alternative is the Interior Penalty method, although the stability parameter of this method also needs to be chosen mindfully and would benefit of studies of more general ice sheet scenarios.}

\section{Acknowledgments}
Christian Helanow was supported by the nuclear waste management organizations in Sweden (Svensk Kärnbränslehantering AB) and Finland (Posiva Oy) through the Greenland Analogue Project and by Gålöstiftelsen. Josefin Ahlkrona was supported by the Swedish strategic research programme eSSENCE. The computations with Elmer/Ice were performed on resources provided by the Swedish National Infrastructure for Computing (SNIC) at PDC Centre for High Performance Computing (PDC-HPC) and at Uppmax at Uppsala University. Both facilities provided excellent support. We thank Per L\"{o}tstedt and Peter Jansson for valuable comments on the manuscript. 

% references
\clearpage
\bibliographystyle{abbrvnat}
\bibliography{HelanowAhlkrona_arxiv}

\end{document}